\documentclass[twocolumn,review] {autart}

\usepackage[pdftex] {graphicx}
\usepackage{amsmath, amssymb}

\newcommand {\qp} [0] {\partial}
\newcommand {\qol} [1] {\overline{#1}}
\newcommand {\qc} [1] {\check{#1}}
\newcommand {\qR} [0] {\mathbb{R}}
\newcommand {\qS} [0] {\mathbb{S}}

\def\qdetailed{detailed}
\def\qelsaut{elsaut}
\def\qver{detailed}

\newtheorem{rmk}{Remark}[thm]
\let\oldrmk\rmk
\renewcommand{\rmk}{\oldrmk\normalfont}
\renewenvironment {pf} {{\it Proof:}} {}

\begin{document}
\begin{frontmatter}

%\title{Analytic solutions to two quaternion attitude estimation problems\thanksref{footnoteinfo}} 
\title{A unified geometric framework for rigid body attitude estimation\thanksref{footnoteinfo}} 

\ifx\qver\qelsaut
\thanks[footnoteinfo] {The authors would like to gratefully acknowledge partial support from the Air Force Office of Scientific Research, and the National Science Foundation. Corresponding author: Prof Kamran Mohseni. Email: mohseni@ufl.edu, Phone: (352) 273-1834.}
\else
\thanks[footnoteinfo] {The authors would like to gratefully acknowledge partial support from the Air Force Office of Scientific Research, and the National Science Foundation. Corresponding author: Yujendra Mitikiri. Email: yujendra@ufl.edu, Phone: (352) 273-2824.}
\fi
\author[First,Third]{Yujendra Mitikiri} 
\author[First,Second,Third]{Kamran Mohseni} 
\address[First] {Dept of Mechanical and Aerospace Engineering, University of Florida, Gainesville, FL 32611 USA.}
\address[Second] {Dept of Electrical and Computers Engineering, University of Florida, Gainesville, FL 32611 USA.}
\address[Third] {Institute of Networked Autonomous systems, University of Florida, Gainesville, FL 32611 USA.}

\begin{abstract}
This paper presents solutions to the following two common quaternion attitude estimation problems: (i) estimation of attitude using measurement of two reference vectors, and (ii) estimation of attitude using rate measurement and measurement of a single reference vector.
Both of these problems yield to a direct geometric analysis and solution. The former problem already has a well established analytic solution in literature using linear algebraic methods. This paper shows how the solution may also be obtained using geometric methods, which are not only more intuitive, but also amenable to unconventional extensions beyond the traditional least-squares formulations. With respect to the latter problem, existing solutions typically involve filters and observers and use a mix of differential-geometric and control systems methods. Again, this solution may also be derived analytically using the geometric method, which helps improve the estimation accuracy.
In this paper, both the problems are formulated as angle optimization problems, which can be solved to obtain a unique closed-form solution. The proposed approach has the favourable consequences that the estimation is (i) exact, thus overcoming errors in solutions based upon linear methods, (ii) instantaneous with respect to the measurements, thus overcoming the latency inherent in solutions based upon negative feedback upon an error, which can at best show asymptotic convergence, and (iii) geometry-based, thus enabling imposition of geometric inequality constraints.
The geometric approach has been verified in simulations as well as experiments, and its performance compared against existing methods.
\end{abstract}


\end{frontmatter}

\addtolength {\topmargin} {-3pt}
\addtolength {\textheight} {3pt}

\section {Introduction} \label{sec:intro}

\begingroup
\allowdisplaybreaks

The problem of estimating the attitude of a rigid body with respect to a reference coordinate system, by measuring reference vectors in a body-fixed frame, has been treated abundantly in literature. One of the earliest, and arguably simplest, solution was Black's three-axis attitude estimator TRIAD \cite{Black:64a}. A least squares formulation of the attitude estimation problem was posed by Wahba in \cite{Wahba:65a}. Multiple solutions have been reported for Wahba's problem: using polar decomposition \cite{Farrell:66a}, an SVD method, Davenport's $q$-method \cite{Keat:77a}, the Quaternion estimator QUEST \cite{Markley:00a}, \ifx\qver\qelsaut a factored-quaternion algorithm FQA \cite{Yun:08a},\fi {\it etc}.

Although both Davenport's $q$-method and QUEST use the quaternion representation of attitude, they ultimately reduce to an eigenvalue-eigenvector problem. Thus it can be seen that most solutions are linear algebraic in nature, and given the vast array of tools available for linear problems, they are all readily solved. This advantage is, however, associated with the accompanying weakness that it is not straightforward to incorporate nonlinear and nonholonomic constraints in the problem.
For instance, in \cite{Singh:10a}, the authors describe the attitude control of a spaceshuttle during a docking operation, when there is a hard constraint with respect to a nominal pitch angle in order to ensure that a trajectory control sensor is oriented towards the target platform. The attitude guidance module then estimates an optimal pitch attitude that complies with the hard constraint and minimizes the control effort.
Similarly, in \cite{Kalabic:14a}, the authors describe a reference governor with a pointing inclusion constraint such that the spacecraft points towards a fixed target, or an exclusion constraint such that sensitive equipment is not exposed to direct solar radiation. Such inequality constraints are obviously nonholonomic, and while being quite common in practice, are notoriously difficult to incorporate in a linear algebraic solution. Once the guidance or reference module determines an attitude that complies with the constraints, a controller module is used to achieve bounded or asymptotic stability with respect to the reference.

Relatedly, the advent of small unmanned vehicles has motivated the development of solutions that depend upon minimal measurement resources in order to reduce the weight and cost of the sensor payload. In particular, it is of considerable interest to estimate the attitude using a single vector measurement, possibly supplemented by a rate measurement, thus leading us to the second of the stated problems. This interest is partly fueled by the availability of cheap commercial-off-the-shelf inertial measurement units (IMUs) that contain MEMS-based gyroscopes and accelerometers \cite{tdkinvensense:17a}. The research is also partly fueled by the realization that attitude estimation and control is a key challenge in the design of small autonomous aerial robots.% \cite{Mohseni:18n}.

The second problem is most frequently solved using an extended Kalman filter (EKF) \cite{Lefferts:82a}. The EKF provides a point-wise attitude estimate and is instantaneous with respect to the measurements. However, resulting from linearization of an intrinsically nonlinear problem, this solution is not robust to large changes in the attitude state \cite{Baritzhack:96a}.

More recently, some solutions have been reported in literature which use nonlinear observers or filters to solve the single-vector measurement problem \cite{Baritzhack:96a}, \cite{Choukroun:04a}, \cite{Mahony:08a}, \cite{Grip:12a}, \cite{Batista:12a}. These solutions have typically used an appropriate error signal in negative feedback to estimate the attitude. The solutions in \cite{Baritzhack:96a}, and \cite{Mahony:08a} are quite general, and while having been developed for multiple vector measurements, they extend smoothly to the case of a single vector measurement. The solutions presented in \cite{Grip:12a}, and \cite{Batista:12a} are more specific to the availability of single vector measurements.
A common characteristic in this group of solutions is the use of negative feedback from an error signal to estimate the attitude and an (a-priori) unknown gain, that needs to be tuned in order to achieve satisfactory estimator performance. Such a feedback-based estimator is bound to have a finite latency with respect to the input, and cannot instantaneously track abrupt or discontinuous changes in the measurements, and the convergence of the estimate to the true attitude is at best asymptotic.\ifx\qver\qelsaut Even more recently, an optimal algebraic solution has been presented in \cite{Valenti:15a} in the specific case of a gravity vector measurement. However, the solution does not directly extend to arbitrary time-varying reference vectors.\fi

In contrast to the linear algebraic and filter approaches available in literature, this paper analyzes the attitude estimation problems from a geometric perspective. In the process, we obtain solutions that overcome some of the shortcomings in the previous solutions. Firstly, being of a geometric nature, the solutions easily extend to problems involving geometric constraints, irrespective of whether they are holonomic equations or nonholonomic inequality constraints. Secondly, the analytic solutions provide an instantaneous estimate for the attitude which is consistent with respect to the vector measurement at every time step. Besides the mathematical elegance of having an analytic solution, this also has several applications in autonomous guidance, navigation, and control systems: it enables the deployment of frugal single-vector-measurement sensor-suites, and the zero-latency accuracy of the solution is useful in multiple-vector-measurement suites in overcoming sudden failures or intermittent losses in some of the components without leading to large transient errors that could potentially cause system breakdown.

A brief outline of the paper is as follows. We begin by introducing the geometric approach and formulating the stated problems in the language of mathematics in section \ref{sec:prob}. The next section, section \ref{sec:soln1}, presents the solution to the first problem, and relates it to the existing solutions from literature. The next section, Section \ref{sec:soln2}, solves the second problem and also provides results relating to the accuracy of the solution. A filtering method is introduced in section \ref{sec:noifx} to address the issue of measurement noise. This is followed by verification of the theory using simulations and experiment in sections \ref{sec:ressim} and \ref{sec:resexp}.

\section {Notation, definitions, and problem formulation} \label{sec:prob}

In this section, we describe the geometry associated with vector measurements and formulate the attitude estimation problems as well-posed mathematical problems.

The attitude of the rigid body with respect to a reference coordinate system shall be represented using a unit quaternion, denoted using a check accent, {\it e.g.} $\qc p = [p_0\; p_1\; p_2\; p_3]^T$, $\qc q = [q_0\; q_1\; q_2\; q_3]^T \hdots $, such that $\qc p^T\qc p = \qc q^T\qc q = \hdots = 1$, so $\qc p, \qc q \in \qS^3$, the unit 3-sphere. The quaternion components are related to the axis-angle representation of a rotation by the relation $q_0 = \cos\Phi/2$, and $[q_1\; q_2\; q_3]^T = n\sin\Phi/2$, for a rotation through $\Phi$ about the axis $n$. The product of two quaternions $\qc q$ and $\qc p$ shall be denoted as $\qc q\otimes\qc p$. We shall follow the quaternion algebraic conventions described in \cite{PhillipsWF:10a} chapter 11. % Rotations on three dimensional vectors $b$ are accomplished by the operation $\qc q\otimes\qc b\otimes\qc q^{-1}$, where $\qc b = [0\; b^T]^T$.

A reference vector, denoted in bold as $\*h,\, \*k,\, \hdots$, shall be defined as a unit magnitude vector that points in a specified direction. Examples include the direction of fixed stars relative to the body, the Earth's magnetic field, gravitational field {\it etc}. The components of any such vector may be measured in any three-dimensional orthogonal coordinate system. In the context of our problems, two obvious choices for the coordinate system are the reference coordinate system (relative to which the rigid body's attitude is to be determined), and a coordinate system fixed in the body. We assume the availability of measurement apparatus to obtain the vector's components in a three-dimensional orthogonal coordinate system, $g,\,h,\, \hdots ,\, a,\, b,\, \hdots \in \qS^2 \subset \qR^3$ in the reference and body-fixed frames.

A rotation quaternion (or, for that matter, any rotation representation) has three scalar degrees of freedom. A body-referred measurement $b$ of a reference vector has 3 scalar components, that are related to the reference measurement $h$, in terms of the rotation quaternion. However, we also know that the measurement would retain the magnitude of the vector, {\it i.e.}, $h^Th = b^Tb = 1$, so there is one scalar degree of redundancy in our measurement $b$ and only two scalar degrees of information. Reconciling with this redundancy, we can therefore isolate the quaternion from a three-dimensional set of possibilities to a single-dimensional set.

The redundancy can be visualized as shown in figure \ref{fig:snglobscone}. The measurement of a single vector in body-fixed axes confines the body's attitude to form a conical solid of revolution about $\*h$: those and only those attitudes on the cone would yield the same components $b$. We shall refer to the set of attitude quaternions consistent with a measurement as the ``feasibility cone'' $Q_b$ corresponding to that measurement $b$, {\it i.e.}, the measurement confines the attitude quaternion $\qc q$ to lie in $Q_b$. From the previous discussion, $Q_b$ is one-dimensional and $\qc q$ has effectively a single degree of freedom.
We shall repeatedly draw intuition from the geometry in figure \ref{fig:snglobscone} to guide us in the solutions to the stated problems.
\begin {figure} [!htbp] \begin {center}
\includegraphics [scale=0.4, trim={0cm 0cm 0cm 0cm}] {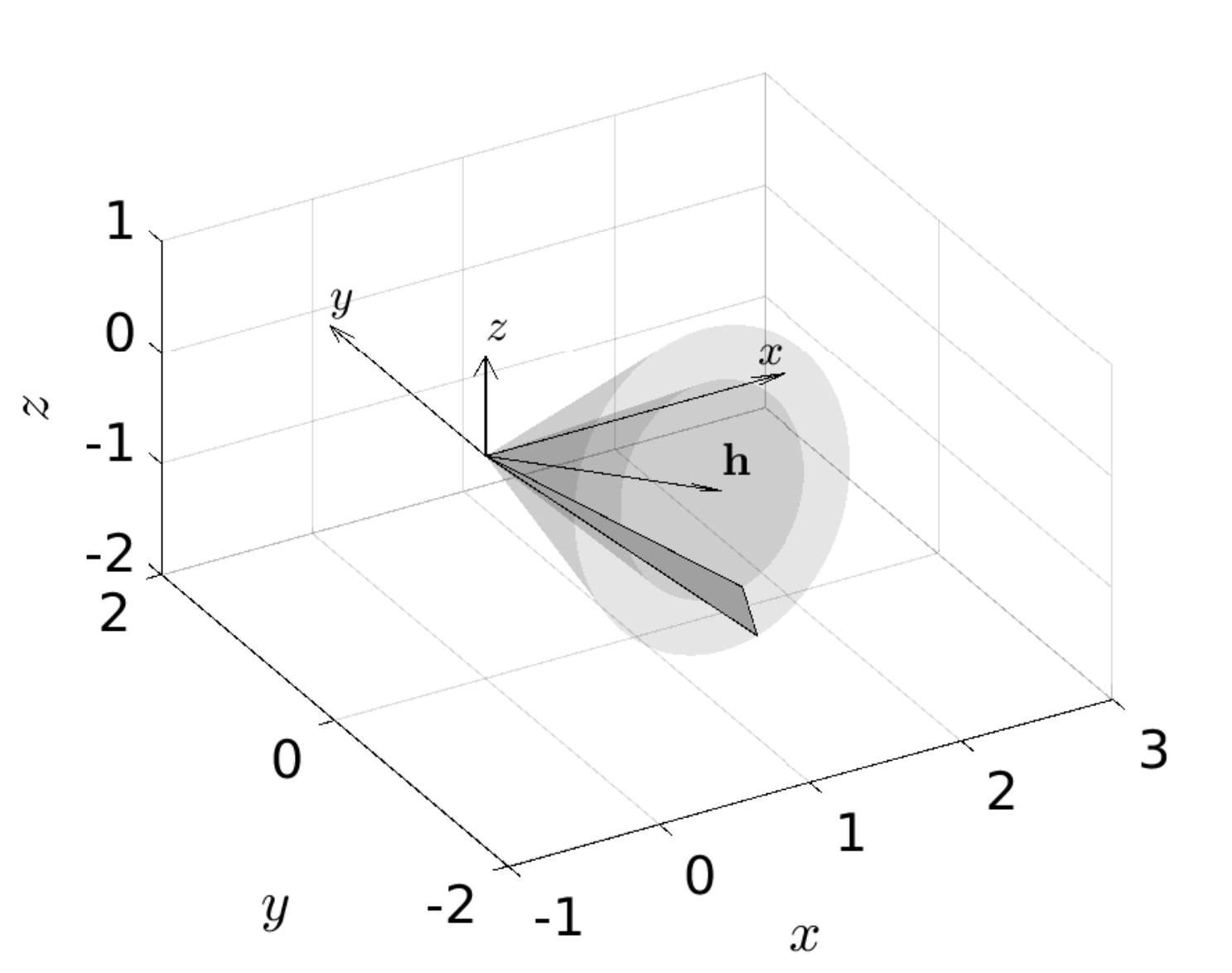}
\caption {Possible attitudes of a minimal rigid body formed out of three non collinear points (represented by the triangular patch) consistent with a measurement of a single vector $\*h$. The subspace is a cone of revolution about the vector being measured.}
\label {fig:snglobscone}
\end {center} \end {figure}

\subsection{Problem 1. Estimation using measurements of two reference vectors}

Let the components of two vectors $\*h$ and $\*k$ be $a = [a_1\; a_2\; a_3]^T$ and $b = [b_1\; b_2\; b_3]^T$ in the body coordinate system, and $h = [h_1\; h_2\; h_3]^T$ and $k = [k_1\; k_2\; k_3]^T$ in the reference coordinate system respectively. As described above, each reference vector measurement provides two scalar degrees of information regarding the attitude of the rigid body. It is immediately clear that the problem is overconstrained, and we have more equations than unknowns. Geometrically, we have two feasibility cones $Q_a$ and $P_b$, with the body-axes intersecting along two lines, but with different roll angles for the body about the body-axis. Thus there is no exact solution to this problem in general, unless some of the measurement information is redundant or discarded.

A trivial means to well-pose the problem is to discard components of one of the vector, say $\*k$, along the second, $\*h$. This is exactly what is done with the TRIAD solution \cite{Black:64a}, where we use the orthogonal vector triad $\*h$, $\*h\times\*k$, and $\*h\times(\*h\times\*k)$ to determine the attitude.
A more sophisticated approach is to use all the measurement information -- four scalar degrees of information with two reference vector measurements --, and frame the problem as a constrained four-dimensional optimization problem in terms of the quaternion components. This leads to Davenport's $q$-method and QUEST solutions to Wahba's problem \cite{Wahba:65a}.

A novel third approach presented in this paper, is to first determine two solutions $\qc q$ and $\qc p$, one each lying on each of the feasibility cones $Q_a$ and $P_b$ corresponding to the measurements $a$ and $b$, and ``closest'' to the other cone in some sense. We then fuse the estimates $\qc q$ and $\qc p$ appropriately to obtain the final attitude estimate. For example, the final estimate could be obtained using linear spherical interpolation, and the weights be chosen to represent the relative significance attached to the individual measurements.

The first problem can therefore be stated as: {\it given the measurements $a$ and $b$ in a rotated coordinate system, of the two reference vectors $h$ and $k$, we would like to estimate the rotated system's two attitude quaternions $\qc q \in Q_a$ closest (in the least squares sense) to $P_b$ and $\qc p \in P_b$ closest (in the least squares sense) to $Q_a$, where $Q_a$ and $P_b$ are the respective feasibility cones}. % such that we minimize the weighted least squares deviation in the cost function $w_1\|\qc h - \qc q\otimes\qc a\otimes\qc q^{-1}\|^2 + w_2\|\qc k - \qc q\otimes\qc b\otimes\qc q^{-1}\|^2$}, where $\otimes$ represents quaternion multiplication \cite{PhillipsWF:10a}.

\subsection{Problem 2. Estimation using rate measurement and measurement of single vector}

Suppose we have a measurement of the components $\omega = [\omega_1\; \omega_2\; \omega_3]^T$ of the angular velocity $\boldsymbol\omega$ of a moving rigid body, and that we also have a measurement of the components $b = [b_1\; b_2\; b_3]^T$ of a reference vector $\*h$, both measurements being made in the body coordinate system. The components of $\*h$ in the reference coordinate system are also known, say $h = [h_1\; h_2\; h_3]^T$. The problem is to make a ``best'' estimate of the body's attitude $\qc q$ on the basis of the pair of measurements $\omega$ and $b$, and knowing $h$.

We shall assume that the initial attitude quaternion is determined using, for {\it e.g.}, a solution to the first problem or by some other means TRIAD, QUEST, FQA, {\it etc}. The angular velocity $\omega$ can be forward integrated to obtain a ``dead-reckoning'' estimate of the rotation quaternion. We start with the attitude, $\qc q(t)$, at time $t$, and then integrate the differential kinematic equation, to obtain the integrated estimate $\qc p(t+dt)$. On account of errors in the measurement of $\omega$, this differs from the actual attitude $\qc q$ of the body. Since we are integrating the errors, the attitude estimates are expected to diverge with time and lead to what is referred to as ``drift'' in the predicted attitude estimate. Constant errors in the measurement lead to a drift that is proportional to the time of integration, while random white wide-sense stationary noise leads to a drift that is proportional to the square-root of time \cite{Woodman:07a}. Let the error in $\omega$ be denoted by the unknown signal $e(t) \in \qR^3$ in the body coordinate system. The integrated estimate also has three scalar degrees of error, though it may depend upon $e$ in some complicated path-dependent form.

The second measurement available is $b$ -- and of course the knowledge of its reference axes components $h$. As described at the beginning of this section, this provides two additional scalar degrees of information besides the three from the rate measurement, and constrains the attitude $\qc q$ to lie in the feasibility cone $Q_b$. In order to determine the six scalar unknowns, three related to the attitude $\qc q$, and three related to the integration of the rate measurement error $e$, we are still lacking one scalar degree of information. In order to specify this degree of freedom and close the problem, we now impose a sixth scalar constraint that uses the attitude $\qc p$ that was obtained by integrating the kinematic differential equation. We choose that particular $\qc q \in Q_b$ which is best in the sense that it deviates the least from $\qc p$.
%What information does the measurement of $b$ (and knowledge of $h$) provide regarding the body's final attitude? A rotation quaternion (or, for that matter, any rotation representation) has three scalar degrees of freedom. We see that we have 3 scalar measurements in $b$. However, we also know that the measurement would retain the magnitude of $\*h$, {\it i.e.}, $\|\*h\|^2 = h^Th = b^Tb$, so there is one scalar degree of redundancy in our measurement and only two scalar degrees of information. Reconciling with this, we can isolate the quaternion from a three-dimensional set of possibilities to a single-dimensional set.

To summarize, the second problem is {\it to estimate the attitude quaternion $\qc q$ which would yield the measurement $b$ in the rotated coordinate system for the reference vector $h$, and closest (in the least squares sense) to the estimate $\qc p$ obtained by integrating the angular velocity measurement $\omega$ as given in the kinematic differential equation}.

\subsection {Nature of measurements of reference vector and angular velocity}

The reference vector measurements are assumed to have random, unbiased noise in each of the components, but that they are subsequently normalized for unit magnitude before being passed on to the attitude estimator. This is the most common situation in practice. Any deterministic errors in the measurement are also assumed to be compensated for, {\it e.g.} acceleration compensation in gravity sense, local field compensation in magnetic field sense.

The angular velocity measurement is also assumed to have random, unbiased noise in each of the components. %, and additionally have an independent, constant (or of negligible variation with time) bias error \cite{Mahony:08a}.
Deterministic errors in this measurement are also assumed to be compensated for. Compensation of a time-varying gyroscopic bias has been addressed by the authors in \cite{Mohseni:19y}. The angular velocity is not of unit magnitude, in general.

Having laid the groundwork for both the problems, the detailed solutions follow in the next section.

\section {Attitude quaternion estimation} \label{sec:soln}

\ifx\qver\qdetailed
We first motivate the use of quaternions for attitude representation by establishing the superiority of the quaternion formalism. Several formalisms exist to represent rotations: 3-component Euler angles, 9-component orthogonal matrices, 4-component axis-angle representations, 4-component Euler-Rodrigues symmetric parameters (quaternions), 4-component Cayley-Klein parameters, and the 3-component modified Rodrigues parameters. Among these, the quaternions and the axis-angle representations are closely related, with simple equations transforming one representation to the other. Note that rotations are accomplished in the axis-angle formalism using Rodrigues rotation formula:
\begin{align}
\*v & = \*n\*n\cdot\*u + (\*u - \*n\*n\cdot \*u)\cos\Phi + \*n\times\*u\sin\Phi , \label{eqn:rotformula1}
\end{align}
where vector $\*u$ is rotated about unit direction $\*n$ through angle $\Phi$ to vector $\*v$. The equivalent matrix equation would be (for a given orthogonal basis coordinate system):
\begin{align}
v & = (nn^T + (1_{3\times 3} - nn^T)\cos\Phi + [n\times]\sin\Phi)u . \label{eqn:rotformula2}
\end{align}
Quaternions are related to the axis-angle formalism as:
\begin{align}
\qc q & = \begin{bmatrix} \cos(\Phi/2) \\ \sin(\Phi/2)n \end{bmatrix} , \label{eqn:axisang2quat}
\end{align}
where we use an angle-like check accent to emphasize that the 4-component quaternion represents a rotation, has unit norm, and satisfies the kinematic equation:
\begin{align}
\dot{\qc q} & = \frac{1}{2}\qc q\otimes\qc\omega = \frac{1}{2}[\qc q\otimes]\qc\omega = \frac{1}{2} [\otimes\qc\omega]\qc q , \label{eqn:qdot}
\end{align}
where $\qc\omega = [0\; \omega^T]^T$ is the angular velocity quaternion.

The quaternion formalism presents exactly the same information as the axis-angle formalism, and leads to an elegant algebra for inverse rotations, the composition of sequential rotations, and interpolation between rotations. We shall henceforth consider both formalisms as equivalent.

We shall now show that the Euler's axis-angle (or equivalently the quaternion) formalism yields the most optimal rotation between two rigid body attitudes. Suppose we wish to minimize the cost in evolving a quaternion from a given initial condition $\qc q(0)$ at time $t = 0$ to a specified final condition $\qc q(t)$ at time $t$. Accordingly, we define the below cost functional with respect to the angular velocity $\qc\omega$ to optimize upon:
\begin{align}
J & = \int_0^t \frac{1}{2}\qc\omega^T\qc\omega dt \Rightarrow L = \frac{1}{2}\qc\omega^T\qc\omega , \label{eqn:Lagrang}\\
\Rightarrow \mathcal H & \overset{\Delta}{=} \frac{1}{2}\qc\omega^T\qc\omega + \frac{\qc\lambda^T}{2} \qc q\otimes\qc\omega = \frac{1}{2}\qc\omega^T\qc\omega + \frac{\qc\lambda^T}{2} [\qc q\otimes]\qc\omega . \label{eqn:Hamilt}
\end{align}
The above Hamiltonian $\mathcal H$ yields the following optimal control $\qc\omega$ using Pontryagin's minimum principle:
\begin{align}
0 & = \qp_\omega \mathcal H = \qc\omega^T + \frac{\qc\lambda^T[\qc q\otimes]}{2} \nonumber\\
\Rightarrow \qc\omega & = -\frac{[\qc q\otimes]^T\qc\lambda}{2} = -\frac{\|\qc q\|^2\qc q^{-1}\otimes \qc\lambda}{2} , \label{eqn:optimctrl}
\end{align}
and the Euler-Lagrange equations for the state $\qc q$ and co-state $\qc\lambda$:
\begin{align}
\begin{bmatrix} \dot{\qc q} \\ \dot{\qc\lambda} \end{bmatrix} & = \begin{bmatrix} [\otimes\qc\omega]\qc q/2 \\ -[\otimes\qc\omega]^T\qc\lambda/2 \end{bmatrix} = \begin{bmatrix} \qc q\otimes\qc\omega/2 \\ -\qc\lambda\otimes\qc\omega^{-1}\|\qc\omega\|^2/2 \end{bmatrix} . \label{eqn:EulerLagrange}
\end{align}

A first integral may be obtained by noticing that $\mathcal H$ has no explicit time dependent:
\begin{align}
\qp_t\mathcal H & = 0 \Rightarrow \dot{\mathcal H} = 0 ,\nonumber\\
\Rightarrow & \mathcal H(0) = \mathcal H(t) = \frac{1}{2} (\qc\omega^T + \qc \lambda^T [\qc q\otimes])\qc\omega = -\frac{\qc\omega^T\qc\omega}{2} , \label{eqn:HteqH0} \\
\Rightarrow \|\qc\omega\|^2 & = \frac{\qc\lambda^T[\qc q\otimes][\qc q\otimes]^T\qc\lambda}{4} = \frac{\|\qc q\|^2\|\qc\lambda\|^2}{4} = -2\mathcal H(0) . \label{eqn:magqmaglam}
\end{align}

A second integral may be obtained by the following observation:
\begin{align}
\qc\lambda^T\dot{\qc q} + \dot{\qc \lambda}^T\qc q & = \frac{1}{2}(\qc\lambda^T[\otimes\qc\omega]\qc q - \qc\lambda^T[\otimes\qc\omega]\qc q) = 0 , \nonumber\\
\Rightarrow \qc\lambda^T\qc q & = \text{ constant } k . \label{eqn:lamTqeqk}
\end{align}

The third and fourth integrals are obtained using (\ref{eqn:optimctrl}), (\ref{eqn:EulerLagrange}), and (\ref{eqn:magqmaglam}) below:
\begin{align}
\dot{\qc q} & = \frac{1}{2}[\qc q\otimes]\qc\omega = -\frac{[\qc q\otimes][\qc q\otimes]^T\qc\lambda}{4} = -\frac{\|\qc q\|^2\qc\lambda}{4} , \nonumber\\
\dot{\qc \lambda} & = -\frac{1}{2}(\qc\lambda\otimes\qc\omega^{-1})\|\qc\omega\|^2 = -\frac{\|\qc\omega\|^2(-\qc q)}{\|\qc q\|^2} = \frac{\|\qc \lambda\|^2\qc q}{4} , \label{eqn:magqmaglam1}\\
\Rightarrow & \frac{d}{dt} \begin{bmatrix} \|\qc q\|^2 \\ \|\qc\lambda\|^2 \end{bmatrix} = \begin{bmatrix} 2\qc q^T\dot{\qc q} \\ 2\qc\lambda^T\dot{\qc\lambda} \end{bmatrix} = \frac{1}{2} \begin{bmatrix} -\|\qc q\|^2\qc q^T\qc\lambda \\ \|\qc\lambda\|^2\qc q^T\qc\lambda \end{bmatrix} = \frac{k}{2} \begin{bmatrix} -\|\qc q\|^2 \\ \|\qc\lambda\|^2 \end{bmatrix} \nonumber\\
\Rightarrow & \begin{bmatrix} \|\qc q\|^2 \\ \|\qc\lambda\|^2 \end{bmatrix} = \begin{bmatrix} e^{-kt/2}A \\ e^{kt/2}B \end{bmatrix} \text {for constants } A, B \in \qR^+ . \label{eqn:magqmaglam2}
\end{align}
Substituting (\ref{eqn:magqmaglam2}) back in (\ref{eqn:magqmaglam1}),
\begin{align}
\begin{bmatrix} \dot{\qc q} \\ \dot{\qc\lambda} \end{bmatrix} & = \frac{1}{4} \begin{bmatrix} -Ae^{-kt/2}\qc\lambda \\ Be^{kt/2}\qc q \end{bmatrix} , \nonumber\\
\Rightarrow \frac{4}{A}e^{kt/2} \left(\ddot{\qc q} + \frac{k}{2}\dot{\qc q} \right) & = -\dot{\qc\lambda} = -\frac{B}{4}e^{kt/2}\qc q \nonumber\\
\Rightarrow \ddot{\qc q} + \frac{k}{2} \dot{\qc q} + \frac{AB}{16}\qc q & = 0 . \label{eqn:qode}
\end{align}
Equation (\ref{eqn:qode}) is a linear ODE in $\qc q$, and may be solved in terms of the constants $A$, $B$, and $k$. Further,
\begin{align}
\ddot{\qc q} & = \frac{1}{4} \qc q\otimes\qc\omega\otimes\qc\omega + \frac{1}{2} \qc q\otimes\dot{\qc\omega} = -\left(\frac{k}{2}\dot{\qc q} + \frac{AB}{16}\qc q\right) \nonumber\\
\Rightarrow \dot{\qc\omega} & = -\frac{1}{2}\qc\omega\otimes\qc\omega - \frac{k}{2}\qc\omega - \frac{AB}{8}. \label{eqn:omegadot}
\end{align}

If $\|\qc q(0)\| = \|\qc q(t)\|$, then we must have $k = 0$. The final solution for the state, co-state, and optimal control when $k = 0$ (unit quaternions) and $\omega_0 = 0$ (vector angular velocities) are given below:
\begin{align}
\begin{bmatrix} \qc q \\ \qc\lambda \end{bmatrix} & = \begin{bmatrix} \cos(\sqrt{AB}t/4)\qc q_c + \sin(\sqrt{AB}t/4)\qc q_s \\ \cos (\sqrt{AB}t/4) \qc\lambda_c + \sin (\sqrt{AB}t/4) \qc\lambda_s \end{bmatrix} , \nonumber\\
\dot{\qc\omega} & = -\frac{4\qc\omega\otimes\qc\omega + AB}{8} = \frac{4\|\omega\|^2 - \|\qc q\|^2\|\qc \lambda\|^2}{8} = 0 . \label{eqn:qlamomega_keq0}
\end{align}
where we have used the fact that $\qc\omega\otimes\qc\omega = -\|\omega\|^2$ when $\omega_0 = 0$.

Thus, the angular velocity $\qc\omega$ must remain constant for a rigid body rotation in 3D Euclidean space. This implies that the rotation must be about a single axis, as represented by the axis-angle formalism.
\fi

\ifx\qver\qelsaut
We shall use the following lemmas on quaternion attitude representation, proved in \cite{Mohseni:18z}:
\begin{enumerate}
\item \label{thm:quatequivangle} The Euclidean distance $\|\qc q - \qc 1\|$ of an attitude quaternion, $\qc q = [c_{\Phi/2}\; s_{\Phi/2}n]^T$, from the identity element, $\qc 1$, is a positive definite and monotonic function of the magnitude of the principal angle of rotation $\Phi$.
\item \label{thm:orthoquat} Two attitude quaternions are orthogonal if and only if they are related by rotations through $\pi$ about some axis $n$.
\item \label{thm:snglvecssoln}
Suppose the components of a reference vector are given by $h$ and $b$ in the reference and body coordinate systems respectively. Then the attitude $\qc q$ must satisfy
\begin{gather}
\qc q\otimes\qc b = \qc h\otimes\qc q\, \label{eqn:bconstr1}
\end{gather}
Let $\Phi = \operatorname{acos} b^Th$, $c = \cos\Phi/2 = \sqrt{(1+b^Th)/2}$ and $s = \sin\Phi/2 = \sqrt{(1-b^Th)/2}$. Then, two particular solutions for the body's attitude are given by:
\begin{gather}
\hspace*{-0.9cm}
\qc r_1 = \begin{bmatrix} c \\ s (b\times h)/\|b\times h\| \end{bmatrix},\;
\qc r_2 = \begin{bmatrix} 0 \\ (b + h)/\|b + h\| \end{bmatrix}.
\end{gather}
\item \label{thm:snglvecfeas} If $\qc q$ lies in the feasibility cone $Q_b$ of the measurement $b$ for the reference vector $h$, then so does any attitude quaternion obtained by rotating $\qc q$ through an arbitrary angle about $h$. Conversely, all attitude quaternions lying on the feasibility cone are related to each other by rotations about $h$.
\item \label{thm:snglvecspan} All elements on the feasibility cone $Q_b$, of the measurement $b$ for the reference vector $h$, are in the norm-constrained linear span of the two special solutions in lemma \ref{thm:snglvecssoln}.
\item \label{thm:twoquatfeas} Any two linearly independent attitude quaternions determine a unique feasibility cone containing them.
\end{enumerate}
\else
We first show the equivalence between quaternion displacements and angles, and characterize quaternion orthogonality in terms of rotations, in the following lemmas.
\begin{lem} \label{thm:quatequivangle}
The Euclidean distance $\|\qc q - \qc 1\|$ of an attitude quaternion, $\qc q = [c_{\Phi/2}\; s_{\Phi/2}n]^T$, from the identity element, $\qc 1$, is a positive definite and monotonic function of the magnitude of the principal angle of rotation $\Phi$.
\end{lem}
\begin{pf}
This is a simple consequence of the trigonometric half-angle identities.
\begin{align}
\|\qc q - \qc 1\|^2 & = (c_{\Phi/2} - 1)^2 + s_{\Phi/2}^2 = 2(1 - c_{\Phi/2}) = 4\sin^2 (\Phi/4), \nonumber
\end{align}
which is a positive definite monotonic function of $\|\Phi\|$ for $\Phi \in [-2\pi, 2\pi]$. A corollary is that the distance $\|\qc q - \qc p\| = \|\qc q^{-1} \otimes \qc p - \qc 1\|$ between two attitude quaternions is a positive definite and monotonic function of the angle corresponding to the quaternion $\qc q^{-1} \otimes \qc p$ that takes $\qc q$ to $\qc p$. \hspace*{\fill} \qed
\end{pf}

\begin{lem} \label{thm:orthoquat}
Two quaternions are orthogonal if and only if they are related by rotations through $\pi$ about some axis $n$.
\begin{align}
\qc p^T \qc q = 0 & \Leftrightarrow \exists n \in \qR^3,\, \qc q = \qc p \otimes \begin{bmatrix} 0 \\ n \end{bmatrix} . \label{eqn:orthoquat}
\end{align}
\end{lem}
\begin{pf} This follows upon noting that a rotation through $\pi$ results in the scalar part being zero.
\begin{align}
\qc p^T \qc q = 0 & \Leftrightarrow p_0q_0 + p_1q_1 + p_2q_2 + p_3q_3 = 0 \nonumber\\
& \Leftrightarrow \operatorname{Re}\{\qc q\otimes\qc p^{-1}\} = 0 \nonumber\\
 & \Leftrightarrow \exists n \in \qR^3,\, \qc q \otimes \qc p^{-1} = \begin{bmatrix} 0 \\ n \end{bmatrix} . \nonumber \hspace*{\fill} \qed
\end{align}
\end{pf}

We next provide two particular solutions for the simpler problem of estimating the attitude quaternion using a single reference vector measurement, in Lemma \ref{thm:snglvecssoln}. We note the algebraic constraint imposed by a vector measurement on the attitude quaternion $\qc q$. The quaternion $\qc q$ represents a rigid body rotation, and it transforms the components of the reference vector from $h$ in the reference coordinate system to $b$ in the body-fixed coordinate system:
\begin{flalign}
&& \qc h & = \qc q\otimes\qc b\otimes\qc q^{-1} & \nonumber\\
\text {or \qquad} && \qc q\otimes\qc b & = \qc h\otimes\qc q\,, & \label{eqn:bconstr1}
\end{flalign}
where the checked quantities $\qc h = [0\; h^T]^T$ and $\qc b = [0\; b^T]^T$ are the quaternions corresponding to the 3-vectors $h$ and $b$. Equation (\ref{eqn:bconstr1}) expresses the vector measurement constraint as a linear equation in $\qc q$ subject to a nonlinear normalization constraint.

\begin{lem} \label{thm:snglvecssoln}
Suppose the components of a reference vector are given by $h$ and $b$ in the reference and body coordinate systems respectively. Let $\Phi = \operatorname{acos} b^Th$, $c = \cos\Phi/2 = \sqrt{(1+b^Th)/2}$ and $s = \sin\Phi/2 = \sqrt{(1-b^Th)/2}$. Then, two particular solutions for the body's attitude are given by:
\begin{gather}
\hspace*{-0.3cm}
\qc r_1 = \begin{bmatrix} c \\ s (b\times h)/\|b\times h\| \end{bmatrix} ,\;
\qc r_2 = \begin{bmatrix} 0 \\ (b + h)/\|b + h\| \end{bmatrix} .
\end{gather}
\end{lem}
\begin{pf}
These two solutions are orthogonal in quaternion space, and correspond to the smallest and largest single axis rotations in $[0, \pi]$ that are consistent with the vector measurement in three-dimensional Euclidean space. Geometrically, the first is a rotation through $\text{acos}(b^Th)$ about $(b\times h)/\|b\times h\|$, the second is a rotation through $\pi$ about $(b + h)/\|b + h\|$. Noting that $\|b\times h\| = \|b\|\|h\|\sin\Phi = \|b\|\|h\|2sc$, and $\|b + h\| = 2c$, we obtain
\begin{align}
& \begin{bmatrix} c \\ (b\times h)/(2c) \end{bmatrix} \otimes \begin{bmatrix} 0 \\ b \end{bmatrix} = \begin{bmatrix} 0 \\ cb + (h - bb^Th)/(2c) \end{bmatrix} \nonumber\\
 & \qquad = \begin{bmatrix} 0 \\ (b + h)/2c \end{bmatrix} = \begin{bmatrix} 0 \\ h \end{bmatrix} \otimes \begin{bmatrix} c \\ (b\times h)/(2c) \end{bmatrix} \;, \nonumber
\intertext{and}
& \begin{bmatrix} 0 \\ (b + h)/(2c) \end{bmatrix} \otimes \begin{bmatrix} 0 \\ b \end{bmatrix} = \begin{bmatrix} -(b^Th)/(2c) \\ (h\times b)/(2c) \end{bmatrix} \nonumber\\
 & \qquad = \begin{bmatrix} 0 \\ h \end{bmatrix} \otimes \begin{bmatrix} 0 \\ (b + h)/(2c) \end{bmatrix} \;, \nonumber
\end{align}
which completes the proof.
As a clarification, when $b \rightarrow h$, $\qc r_1$ and $\qc r_2$ are assumed to take the obvious limits, $\qc 1$ and $\qc h$, and when $b \rightarrow -h$, they are assumed to take the obvious limits, $\qc i = [0\; i]^T$ and $\qc j = [0\; j]^T$, where $[h\;i\;j]$ is an orthogonal vector triplet. In the latter case ($b + h \rightarrow 0$), the orthogonal triad is non-unique, but certain to exist: at least one among the three orthogonal triplets $h, h\times e_x, e_x - h_1h$; $h, h\times e_y, e_y - h_2h$; $h, h\times e_z, e_z - h_3h$ (where $e_x = [1\; 0\; 0]^T, \hdots $) is certain to span $\qR^3$, and would be a valid choice for the orthogonal triad $[h\; i\; j]$ after normalization. \hspace*{\fill} \qed
\end{pf}

The two special solutions can be rotated by any arbitrary angle about the reference vector $h$ and we would still lie within the feasibility cone, as shown in the next lemma.

\begin{lem} \label{thm:snglvecfeas}
If $\qc q$ lies in the feasibility cone $Q_b$ of the measurement $b$ for the reference vector $h$, then so does any attitude quaternion obtained by rotating $\qc q$ through an arbitrary angle about $h$. Conversely, all attitude quaternions lying on the feasibility cone are related to each other by rotations about $h$.
\end{lem}
\begin{pf}
Let $\Phi$ be any angle, and let $\qc p$ be $\qc q$ rotated through $\Phi$ about $h$, {\it i.e.},
\begin{align}
\qc p & = \begin{bmatrix} c \\ sh \end{bmatrix} \otimes \qc q\,, \nonumber
\intertext{where $c = \cos\Phi/2$ and $s = \sin\Phi/2$. Then, }
\qc p\otimes\qc b & = \begin{bmatrix} c \\ sh \end{bmatrix} \otimes \qc q \otimes \qc b = \begin{bmatrix} c \\ sh \end{bmatrix} \otimes \qc h \otimes \qc q \nonumber\\
 & = \qc h \otimes \begin{bmatrix} c \\ sh \end{bmatrix} \otimes \qc q = \qc h \otimes \qc p\,. \nonumber
\end{align}
where we have used the fact that two nonzero rotations commute if and only if they are about the same axis. Conversely, $\qc q^{-1}\otimes\qc h\otimes\qc q = b = \qc p^{-1}\otimes\qc h\otimes\qc p$ implies
\begin{align}
\qc p\otimes\qc q^{-1}\otimes\qc h & = \qc h\otimes\qc p\otimes\qc q^{-1} \nonumber\\
\text{or, } \qquad \qc p\otimes\qc q^{-1} & = \begin{bmatrix} c \\ sh \end{bmatrix} \,, \nonumber
\end{align}
for some $c$ and $s$ satisfying $c^2+s^2=1$, which completes the proof. \hspace*{\fill} \qed
\end{pf}

\begin{lem} \label{thm:snglvecspan}
All elements on the feasibility cone $Q_b$, of the measurement $b$ for the reference vector $h$, are in the norm-constrained linear span of the two special solutions in lemma \ref{thm:snglvecssoln}.
\end{lem}
\begin{pf}
Consider an attitude quaternion $\qc q = c'\qc r_1 + s'\qc r_2$, where $c'^2 + s'^2 = 1$, and $\qc r_1$ and $\qc r_2$ are the special solutions of Lemma \ref{thm:snglvecssoln}. Then:
\begin{align}
& \begin{bmatrix} cc' \\ \dfrac{c'b\times h + s'(b+h)}{2c} \end{bmatrix} \otimes \begin{bmatrix} 0 \\ b \end{bmatrix} \nonumber\\
 & \qquad = \begin{bmatrix} -s'(1 + 2c^2 - 1)/2c \\ cc'b + \dfrac{c'}{2c}(h - (2c^2 - 1)b) + \dfrac{s'}{2c}h\times b \end{bmatrix} \nonumber\\
 & \qquad = \begin{bmatrix} -cs' \\ \dfrac{c'(h+b) + s'h\times b}{2c} \end{bmatrix} \nonumber\\
 & \qquad = \begin{bmatrix} -s'(2c^2 - 1 + 1)/2c \\ cc'h + \dfrac{c'}{2c}(b - (2c^2 - 1)h) + \dfrac{s'}{2c}h\times b \end{bmatrix} \nonumber\\
 & \qquad = \begin{bmatrix} 0 \\ h \end{bmatrix} \otimes \begin{bmatrix} cc' \\ \dfrac{c'b\times h + s'(b+h)}{2c} \end{bmatrix} , \nonumber
\end{align}
that is, $\qc q \otimes \qc b = \qc h \otimes \qc q$, which shows that $\qc q$ is an element on the feasibility cone $Q_b$. Conversely, any element on the feasibility cone, $Q_b$, can be written as the composition of $\qc r$ and a rotation about $h$ through the angle $\Phi'$ from lemma \ref{thm:snglvecfeas}. Hence,
\begin{align}
%& \begin{bmatrix} c' \\ s'h \end{bmatrix} \otimes \begin{bmatrix} 0 \\ (b + h)/2c \end{bmatrix} \nonumber\\
% & \qquad = \begin{bmatrix} -s'(2c^2 - 1 + 1)/2c \\ \dfrac{c'(b + h) + s'h\times b}{2c} \end{bmatrix} \nonumber\\
% & \qquad = c' \begin{bmatrix} 0 \\ \dfrac{b + h}{2c} \end{bmatrix} - s' \begin{bmatrix} c \\ \dfrac{b\times h}{2c} \end{bmatrix} \nonumber
& \begin{bmatrix} c' \\ s'h \end{bmatrix} \otimes \begin{bmatrix} c \\ (b\times h)/2c \end{bmatrix} \nonumber\\
 & \qquad = \begin{bmatrix} c'c \\ \dfrac{c'(b\times h)}{2c} + s'ch + \dfrac{s'}{2c}(b - (2c^2 - 1)h) \end{bmatrix} \nonumber\\
 & \qquad = \begin{bmatrix} c'c \\ \dfrac{c'(b\times h) + s'(b + h)}{2c} \end{bmatrix} = c' \begin{bmatrix} c \\ \dfrac{b\times h}{2c} \end{bmatrix} + s' \begin{bmatrix} 0 \\ \dfrac{b + h}{2c} \end{bmatrix} , \nonumber
\end{align}
which completes the proof. \hspace*{\fill} \qed
\end{pf}

It also follows from Lemma \ref{thm:snglvecspan} that the rotation axis of every rotation on the feasibility cone, $Q_b$, of the measurement $b$ for the reference vector $h$, lies on the unit circle containing the vectors $b\times h/\|b\times h\|$, and $(b + h)/\|b + h\|$ (figure \ref{fig:axisswpsph} left).
\begin{figure} [!ht]
\begin{minipage}{0.49\linewidth} \begin{center}
\includegraphics [width=\linewidth] {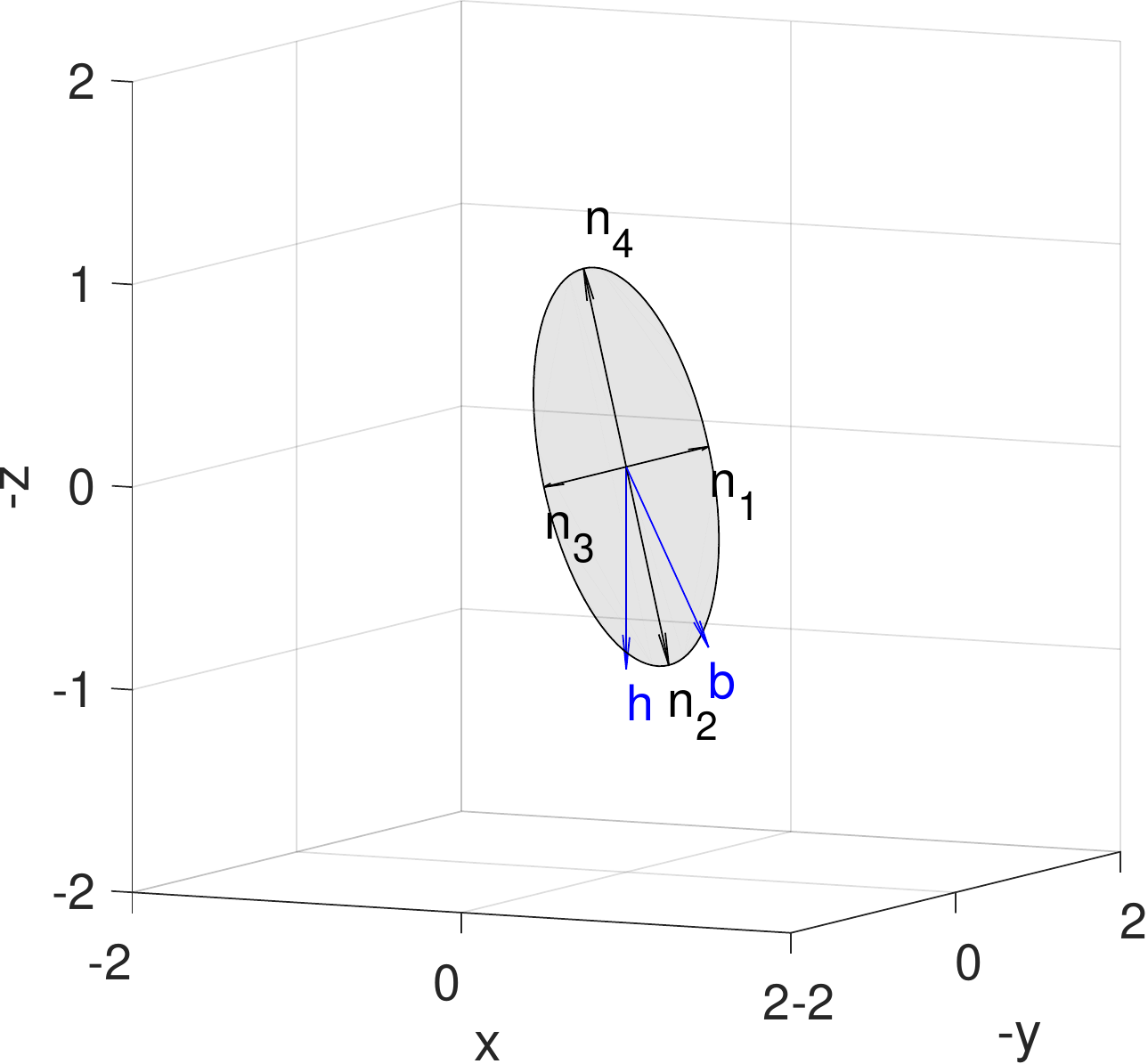}
\end{center}\end{minipage}
\begin{minipage}{0.49\linewidth} \begin{center}
\includegraphics [width=\linewidth] {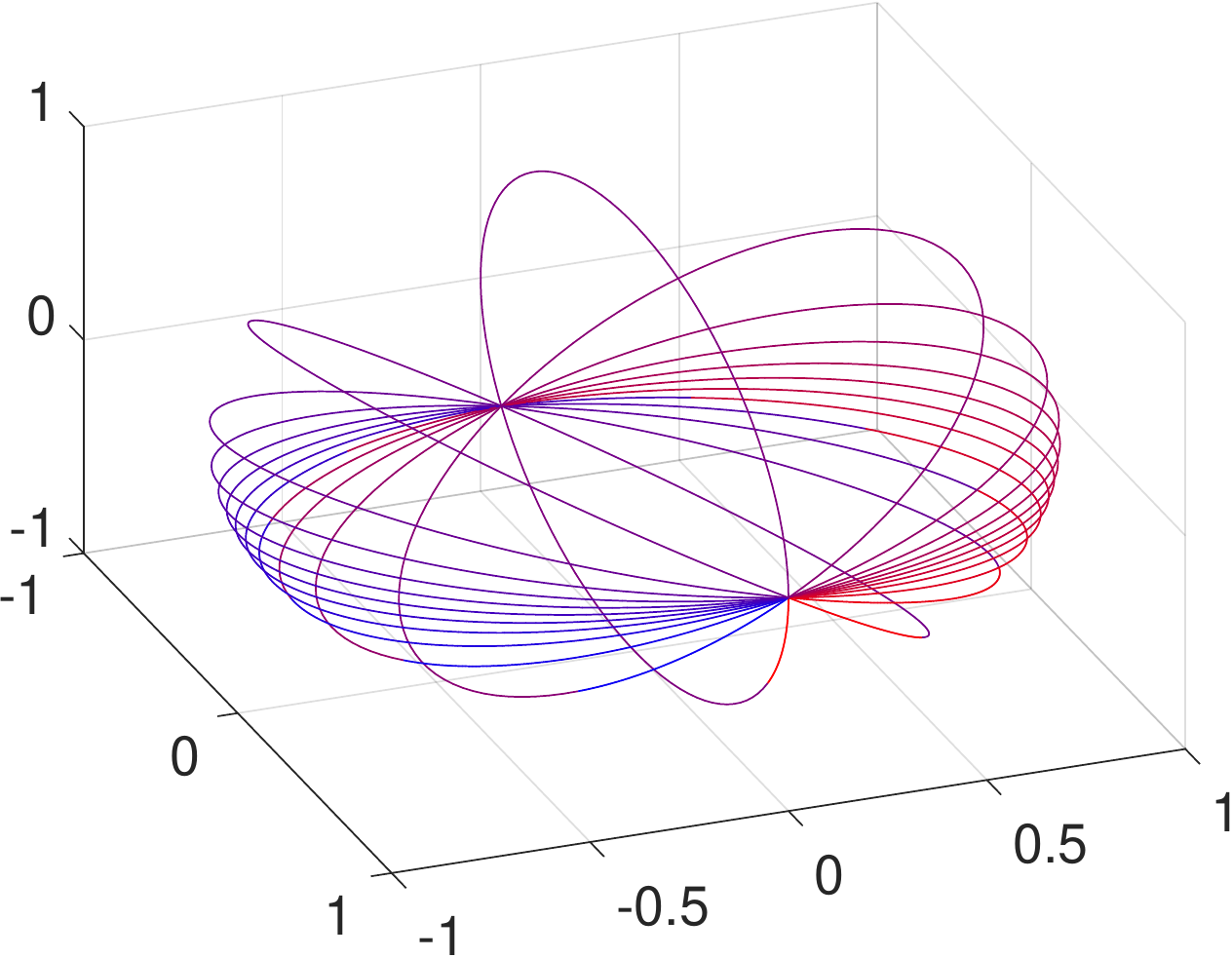}
\end{center}\end{minipage}
\caption{Left: Possible axes to rotate the rigid body about, in order to measure reference vector $h$ as $b$ in the body axes. The rotation axes lie in the unit great circle spanned by $n_1 = b\times h/\|b\times h\|$, $n_2 = (b + h)/\|b + h\|$, $n_3 = -n_1$, $n_4 = -n_2$.
Right: A visual depiction of the covering of the 2-sphere by the body $x$-axis using all rotations on the feasibility cone, $Q_b$. The rigid body is being rotated so as to measure the reference vector $h$ as $b$ in the body frame. In order to obtain this measurement, the body may be rotated (by differing amounts) about the set of unit vectors spanned by $n_1$ and $n_2$. As the rotation axis varies over the unit great circle spanned by these basis elements, the body $x$-axis sweeps great arcs over the 2-sphere that eventually cover all of it. Simultaneously, the yaw angle of the second rotation of the decomposition of a rotation goes from $-2\pi$ to $2\pi$. The color of the great arc is gradually varied from blue to red as the rotation axis begins at $n_1$ and goes through $n_2$, $n_3$, $n_4$, back to $n_1$.}
\label{fig:axisswpsph}
\end{figure}

Thus, we already see that we have a one dimensional infinity of possible solutions for the attitude quaternion if we have a single reference vector measurement. In fact, the two special solutions provided in lemma \ref{thm:snglvecssoln} are rotations of each other about $h$ through $\pi$. In order to obtain a unique solution, we could add either another vector measurement (Wahba's problem), or include an angular velocity measurement (complementary filter).

We note one final trivial result about the feasibility cone subspace.
\begin{lem} \label{thm:twoquatfeascone}
Any two unequal attitude quaternions, $\qc p$ and $\qc q$, define the feasibility cone corresponding to some vector meaurement.
\end{lem}
\begin{pf}
The claim follows trivially upon noting that rotations about the same axis commute, and the axis $n$ of $\qc q\otimes\qc p^{-1}$ is the reference direction whose body frame measurements are the same with both $\qc p$ and $\qc q$:
\begin{align}
\qc q \otimes \qc p^{-1} & = \begin{bmatrix} c \\ sn \end{bmatrix} , \nonumber\\
\Rightarrow \qc q \otimes \qc p^{-1} \otimes \qc n & = \qc n \otimes \qc q \otimes \qc p^{-1} \nonumber\\
\Rightarrow \qc p^{-1} \otimes \qc n \otimes \qc p & = \qc q^{-1} \otimes \qc n \otimes \qc p . \nonumber \hspace*{\fill} \qed
\end{align}
\end{pf}
\fi

%It is now clear why the quaternions have a period of $4\pi$: as the rotation axis sweeps through $2\pi$ beginning from $b\times h$, going through $b+h$, then $-b\times h$, then $-(b+h)$, the decomposition of the rotation quaternion into the minimal rotation $\operatorname{acos}(b^Th)$, followed by a rotation by $\Phi$ about $h$, leads to the yaw angle $\Phi$ sweeping from $-2\pi$ to $+2\pi$. While the rotations through $\Phi$ and $2\pi + \Phi$ result in the same final rigid body attitude, the path followed by the body in reaching the state is different. Moreover, as $\Phi$ sweeps from $-2\pi$ to $2\pi$, every vector attached to the rigid body sweeps the entire 2-sphere (figure \ref{fig:axisswpsph} right).% That also explains why the boundary of a unit 2-sphere is $4\pi$, while the boundary of a unit 1-sphere (circle) is $2\pi$.

\subsection {Attitude estimation using two vector measurements} \label{sec:soln1}

We now derive a unique solution for the attitude quaternion when we have measurements of two reference vectors and would like to incorporate both of them in deriving the attitude estimate. Let $a$ and $b$ be the body-referred components of reference vectors $h$ and $k$ ($h, k \in \qS^2$ contain the components of the two vectors along some reference coordinate axes) respectively. Suppose the rotation quaternion is estimated to be $\qc q = [q_0\;q]^T$ on the basis of $a$, and it is independently estimated to be $\qc p = [p_0\;p]^T$ on the basis of $b$, both estimates being obtained by applying, say, Lemma \ref{thm:snglvecssoln}.

The estimates $\qc q$ and $\qc p$ are each indeterminate to one scalar degree of freedom as shown in lemma \ref{thm:snglvecfeas}: a rotation about the corresponding vectors $h$ and $k$ respectively. Let these rotations be given by the quaternions $\qc r_1 = [c_1\;s_1h]^T$ and $\qc r_2 = [c_2\;s_2k]^T$ respectively where $c_i = \cos\Phi_i/2$ and $s_i = \sin\Phi_i/2$ for $i \in \{1, 2\}$. The problem is to determine the optimal values of $\Phi_1$ and $\Phi_2$ so as to minimize the displacement from the rotated $\qc r_1\otimes\qc q$ to $\qc r_2\otimes\qc p$.
\begin{align}
\hspace*{-0.3cm}\qc r_1\otimes\qc q & = \begin{bmatrix} c_1 \\ s_1h \end{bmatrix} \otimes \begin{bmatrix} q_0 \\ q \end{bmatrix} = \begin{bmatrix} c_1q_0 - s_1q^Th \\ c_1q + s_1q_0h + s_1h\times q \end{bmatrix}, \nonumber\\
\hspace*{-0.3cm}\qc r_2\otimes\qc p & = \begin{bmatrix} c_2 \\ s_2k \end{bmatrix} \otimes \begin{bmatrix} p_0 \\ p \end{bmatrix} = \begin{bmatrix} c_2p_0 - s_2p^Tk \\ c_2p + s_2p_0k + s_2k\times p \end{bmatrix}. \label{eqn:triad1}
\end{align}

We could either minimize $\|\qc r_1\otimes\qc q - \qc r_2\otimes\qc p\|^2$, or equivalently from Lemma \ref{thm:quatequivangle}, maximize the first component of $(\qc r_1\otimes \qc q)^{-1}\otimes\qc r_2\otimes \qc p$. In order to keep the reasoning straightforward, we choose the former. So we need to minimize the cost function
\begin{align}
 & J(\Phi_1, \Phi_2) = (c_1q_0 - s_1q^Th - c_2p_0 + s_2p^Tk)^2 \nonumber\\
 & \qquad + \|c_1q + s_1(q_0h + h\times q) - c_2p - s_2(p_0k + k\times p)\|^2 \,, \nonumber\\
\ifx\qver\qdetailed
 & = c_1^2(q_0^2 + q^Tq) + s_1^2((q^Th)^2 + \|q_0h - q\times h\|^2) \nonumber\\
 & \qquad + c_2^2(p_0^2 + p^Tp) + s_2^2((p^Tk)^2 + \|p_0k - p\times k\|^2) \nonumber\\
 & \qquad - 2c_1c_2(q_0p_0 + q^Tp) \nonumber\\
 & \qquad - 2s_1s_2(q^Thp^Tk + (q_0h - q\times h)^T(p_0k - p\times k)) \nonumber\\
 & \qquad + 2c_1s_1(-q_0q^Th + q_0q^Th - q^Tq\times h) \nonumber\\
 & \qquad + 2c_2s_2(-p_0p^Tk - p_0p^Tk + p^Tp\times k) \nonumber\\
 & \qquad + 2c_1s_2(q_0p^Tk - p_0q^Tk + q^Tp\times k) \nonumber\\
 & \qquad + 2c_2s_1(p_0q^Th - q_0p^Th + p^Tq\times h) \nonumber\\
\fi
 & = 2 + 2l_1c_1c_2 + 2l_2s_1s_2 + 2l_3c_1s_2 + 2l_4s_1c_2 \,, \label{eqn:twoveccost}
\end{align}
where $l_1 = -q_0p_0 - q^Tp$, $l_2 = (-q_0p^T + p_0q^T - (q\times p)^T)h\times k - (q_0p_0 + q^Tp)h^Tk$, $l_3 = k^T(q_0p - p_0q + q\times p)$, and $l_4 = h^T(p_0q - q_0p + p\times q)$, are known quantities.
Now minimizing the cost function with respect to the independent pair of variables $\Phi_1 + \Phi_2$ and $\Phi_1 - \Phi_2$ yields
\begin{align}
\begin{bmatrix} \Phi_1-\Phi_2 \\ \Phi_1+\Phi_2 \end{bmatrix} & = 2\begin{bmatrix} \operatorname{atan2}(l_3-l_4, -(l_1+l_2)) \\ \operatorname{atan2}(-(l_3+l_4), l_2-l_1) \end{bmatrix} . \label{eqn:triad4}
%\begin{bmatrix} c_1c_2 \\ s_1s_2 \\ c_1s_2 \\ s_1c_2 \end{bmatrix} & = \frac{1}{2}\begin{bmatrix} \alpha/\sqrt{\alpha^2 + \beta^2} + \gamma/\sqrt{\gamma^2 + \delta^2} \\ \alpha/\sqrt{\alpha^2 + \beta^2} - \gamma/\sqrt{\gamma^2 + \delta^2} \\ -\beta/\sqrt{\alpha^2 + \beta^2} + \delta/\sqrt{\gamma^2 + \delta^2} \\ \beta/\sqrt{\alpha^2 + \beta^2} + \delta/\sqrt{\gamma^2 + \delta^2} \end{bmatrix} \,. \label{eqn:triad4}
\end{align}
Equation (\ref{eqn:triad4}) can be solved for $\Phi_1$, and $\Phi_2$, and that completes the solution.
The above derivation can be summarized in the form of the following theorem:
\begin{thm} \label{thm:twovecsoln}
If $\qc q$ and $\qc p$ are any two special attitude estimates for a rotated system, derived independently using the measurements $a$ and $b$ in the body-fixed coordinate system of two linearly independent reference vectors $h$ and $k$ respectively, then the optimal estimate incorporating the measurement $b$ in $\qc q$ is $\qc r_1\otimes\qc q$, and the optimal estimate incorporating the measurement $a$ in $\qc p$ is given by $\qc r_2\otimes\qc p$, where $\qc r_1 = [c_1\; s_1h]^T$ and $\qc r_2 = [c_2\; s_2k]^T$, $c_i = \cos\Phi_i$, $s_i = \sin\Phi_i$, and $\Phi_1$ and $\Phi_2$ are given by equation (\ref{eqn:triad4}).
\end{thm}
\begin{pf}
The proof follows from the construction leading to equations (\ref{eqn:triad1}, \ref{eqn:triad4}). Refer figure \ref{fig:solnvis}. \hspace*{\fill} \qed
\end{pf}

\begin{rmk} \label{thm:relnTRIAD}
{\it Relation to the TRIAD attitude estimate \cite{Black:64a}}:
The attitude estimates $\qc r_1\otimes\qc q$ and $\qc r_2\otimes\qc p$, where $\qc r_1 = [c_1\; s_1h]^T$ and $\qc r_2 = [c_2\; s_2k]^T$, are the same as the TRIAD solution in literature \cite{Shuster:81a}. Each of them individually yields an estimate that is competely consistent with one measurement, but only partially consistent with the other.
\end{rmk}

\begin{cor} \label{thm:twovecsoln2}
The rotation from the TRIAD estimate $\qc r_1\otimes \qc q$ to $\qc r_2\otimes\qc p$ in (\ref{eqn:triad4}) is about an axis perpendicular to both $h$ and $k$.
\end{cor}
\begin{pf}
Let $\qc q' = \qc r_1 \otimes \qc q$ and $\qc p' = \qc r_2 \otimes \qc p$ be the optimal TRIAD estimates. Let us now optimize upon these optimal estimates. That should return no required corrections, {\it i.e.} $\qc r_1' = \qc r_2' = \qc 1$. This is equivalent to saying $\Phi_1' = \Phi_2' = 0$. This in turn is equivalent to $l_3' = l_4' = 0$, or $h^T(p_0'q' - q_0'p' + p'\times q') = k^T(q_0'p' - p_0'q' + q'\times p') = 0$. But then $q_0'p' - p_0'q' - p'\times q'$ is just the vector portion of $\qc p'\otimes \qc q'^{-1}$, the rotation taking the optimal TRIAD estimate $\qc q'$ to $\qc p'$ in the reference coordinate system. \hspace*{\fill} \qed
\end{pf}

\begin{rmk} {\it Geometric filtering between the TRIAD estimates}:
In order to filter the noise in the vector measurements, we could now interpolate between the two solutions obtained in equations (\ref{eqn:triad1}, \ref{eqn:triad4}). Let $\qc q, \qc p$ be the TRIAD attitude estimates (denoted as $\qc q'$ and $\qc p'$ in Corollary \ref{thm:twovecsoln2}) using vector measurements $a$ and $b$ of $h$ and $k$ respectively, and $x \in [0, 1] \subset \qR$. The interpolated quaternion, $\qc q_f$, from $\qc q$ to $\qc p$ is given by any of the following four equivalent expressions \cite{DamKochLillholm:98a}:
\begin{align}
\qc q_f & = \qc q\otimes(\qc q^{-1}\otimes\qc p)^x = \qc p\otimes(\qc p^{-1}\otimes\qc q)^{1-x} \nonumber\\
 & = (\qc q\otimes\qc p^{-1})^{1-x}\otimes\qc p = (\qc p\otimes\qc q^{-1})^x\otimes\qc q \,. \label{eqn:interp1}
\end{align}
The interpolation ratio $x$ is now choosen to perform a desired weighting of the two TRIAD estimates $\qc q$ and $\qc p$ in the final result. When the noise in each of the measurements $a$ and $b$ is zero-mean Gaussian with variance $\sigma_i^2$, the appropriate choice for $x$ would be $\sigma_a^2/(\sigma_a^2 + \sigma_b^2)$.
\end{rmk}

%\begin{rmk} {\it Sign indeterminacy}:
%The solution for $c_1$, $s_1$, $c_2$, and $s_2$ in equation (\ref{eqn:triad4}) involves taking a square-root, but once the sign of the square-root is chosen for one of the four quantities, it gets decided for the other three. Both of the resulting attitude quaternions represent the same rotation in three-dimensional Euclidean space.
%\end{rmk}

\begin{rmk} {\it Relation to the solutions of Wahba's problem \cite{Keat:77a}}:
%In order to obtain the solution to Wahba's problem \cite{Keat:77a}, \cite{Shuster:81a}, we could now interpolate between the two solutions obtained in equations (\ref{eqn:triad1}, \ref{eqn:triad4}). Let $\qc q, \qc p$ be unit quaternions and $x \in \qR \in [0, 1]$. The interpolated quaternion, $\qc q_f$, from $\qc q$ to $\qc p$ is given by any of the following four equivalent expressions \cite{DamKochLillholm:98a}:
%\begin{align}
%\qc q_f & =
%\qc q\otimes(\qc q^{-1}\otimes\qc p)^x = \qc p\otimes(\qc p^{-1}\otimes\qc q)^{1-x} \nonumber\\
% & = (\qc q\otimes\qc p^{-1})^{1-x}\otimes\qc p = (\qc p\otimes\qc q^{-1})^x\otimes\qc q \,. \label{eqn:interp1}
%\end{align}
Let the TRIAD estimates again be denoted as $\qc q$ and $\qc p$. Further let $\qc r = \qc p \otimes \qc q^{-1}$ denote the rotation that takes $\qc q$ to $\qc p$ in the reference coordinate system. From Corollary \ref{thm:twovecsoln2}, we know that $\qc r = [c_{\Phi/2}\; s_{\Phi/2} (h\times k)^T/\|h\times k\|]^T$ for some $\Phi$.
Next, let $\qc w$ be the solution to Wahba's problem, that minimizes the loss function $\alpha\|\qc w\otimes\qc a\otimes\qc w^{-1} - \qc h\|^2 + \beta\|\qc w\otimes\qc b\otimes \qc w^{-1} - \qc k\|^2$. Now $\qc w$ must lie on the feasibility cone containing $\qc q$ and $\qc p$. Otherwise, we could move it towards the cone so as to reduce both the errors $\|\qc w\otimes\qc a\otimes\qc w^{-1} - \qc h\|^2$ and $\|\qc w\otimes\qc b\otimes\qc w^{-1} - \qc k\|^2$ in the loss function. So, if $\qc w \otimes \qc q^{-1}$ and $\qc p \otimes \qc w^{-1}$ rotate the body through $\Phi_q$ and $\Phi_p$ about $h\times k$, then we must have $\Phi_q + \Phi_p = \Phi$.
The loss function would be $2\alpha(1 - \cos\Phi_q) + 2\beta(1 - \cos\Phi_p)$. Thus the solution to Wahba's problem maximizes $\alpha\cos\Phi_q + \beta\cos\Phi_p$, subject to $\Phi_p + \Phi_q = \Phi$: $-\alpha\sin\Phi_q + \beta\sin(\Phi - \Phi_q) = 0 \Rightarrow \tan\Phi_q = \sin\Phi/(\alpha/\beta + \cos\Phi)$ and $\tan\Phi_p = \sin\Phi/(\beta/\alpha + \cos\Phi)$.
The filtered estimate $\qc q_f$ may be derived as the rotation through $\Phi_q$ about $h\times k$ from $\qc q$, or $-\Phi_p$ about $h\times k$ from $\qc p$.
\end{rmk}

\begin{rmk} {\it Incorporating hard inequality constraints}:
Since the presented solution is geometric in nature, it is straightforward to include geometric constraints on the solution. For instance, some attitude estimation problems have hard constraints \cite{Singh:10a}, \cite{Kalabic:14a}. In control solutions, such constraints are most often enforced using Barrier Lyapunov functions (BLFs) \cite{Tee:09a} for bounded solutions. Such a strategy can easily be employed in our framework, in contrast with the linear algebraic solutions which are more suitable to handle quadratic forms. Instead of determining the interpolaton factor $x$ using the noise variance, it can be determined as the argument that minimizes a cost function that contains a BLF:
\begin{align}
x = \operatorname*{argmin}_{x\in[0, 1]} (\alpha\operatorname{sec}(x/a) + (1 - x)^2) , \label{eqn:blfinterp}
\end{align}
where $\alpha$ and $a$ are appropriately chosen constants. It may be appreciated that the cost function can be any infinite potential well, and not just the above formulation. This generality is enabled by the simple interpolation of the geometric angle between the two solutions of theorem \ref{thm:twovecsoln}.
\end{rmk}

\begin{figure} [!ht]
\begin{minipage}{0.49\linewidth} \begin{center}
\includegraphics [width=\linewidth] {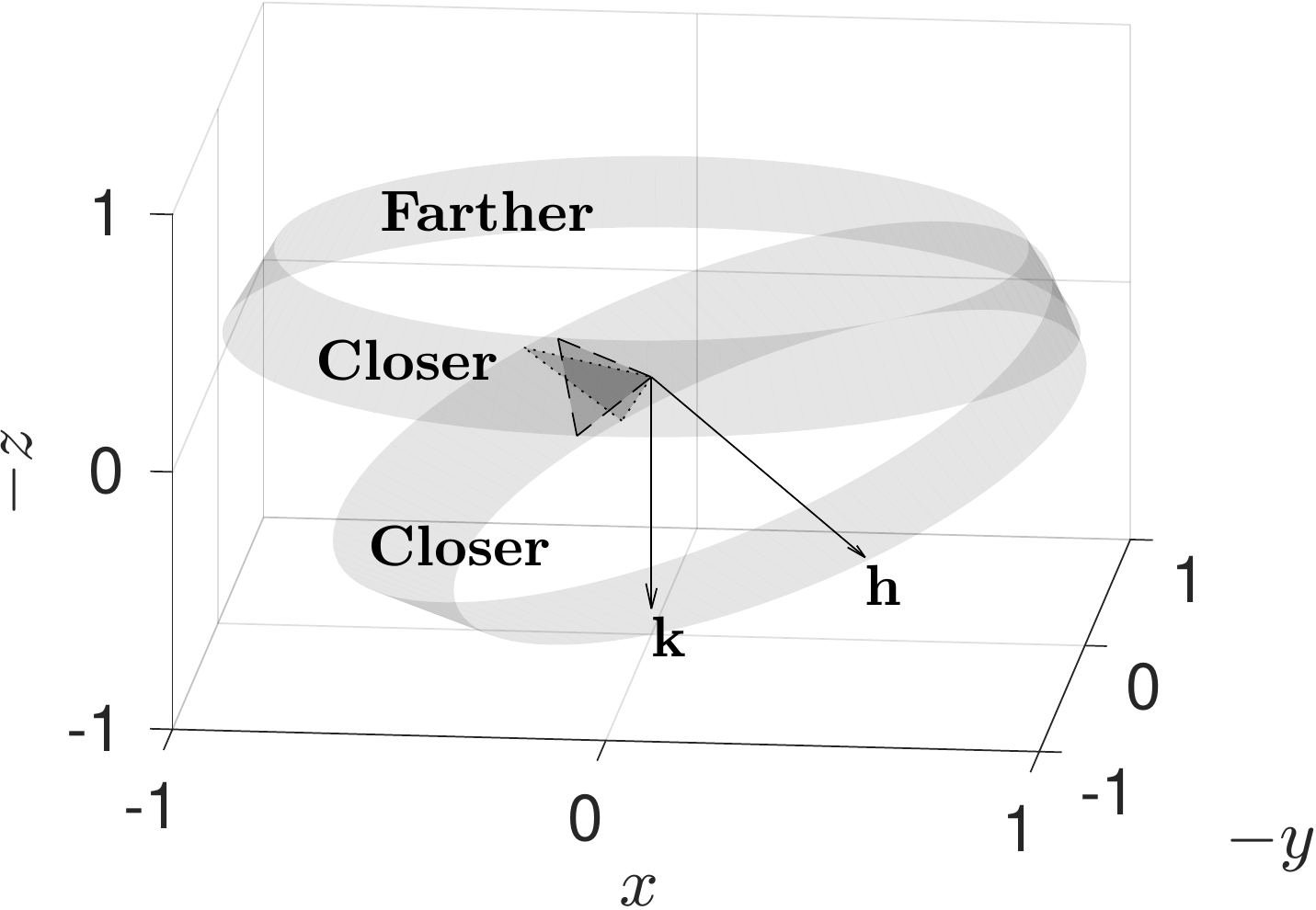}
\end{center}\end{minipage}
\begin{minipage}{0.49\linewidth} \begin{center}
\includegraphics [width=\linewidth] {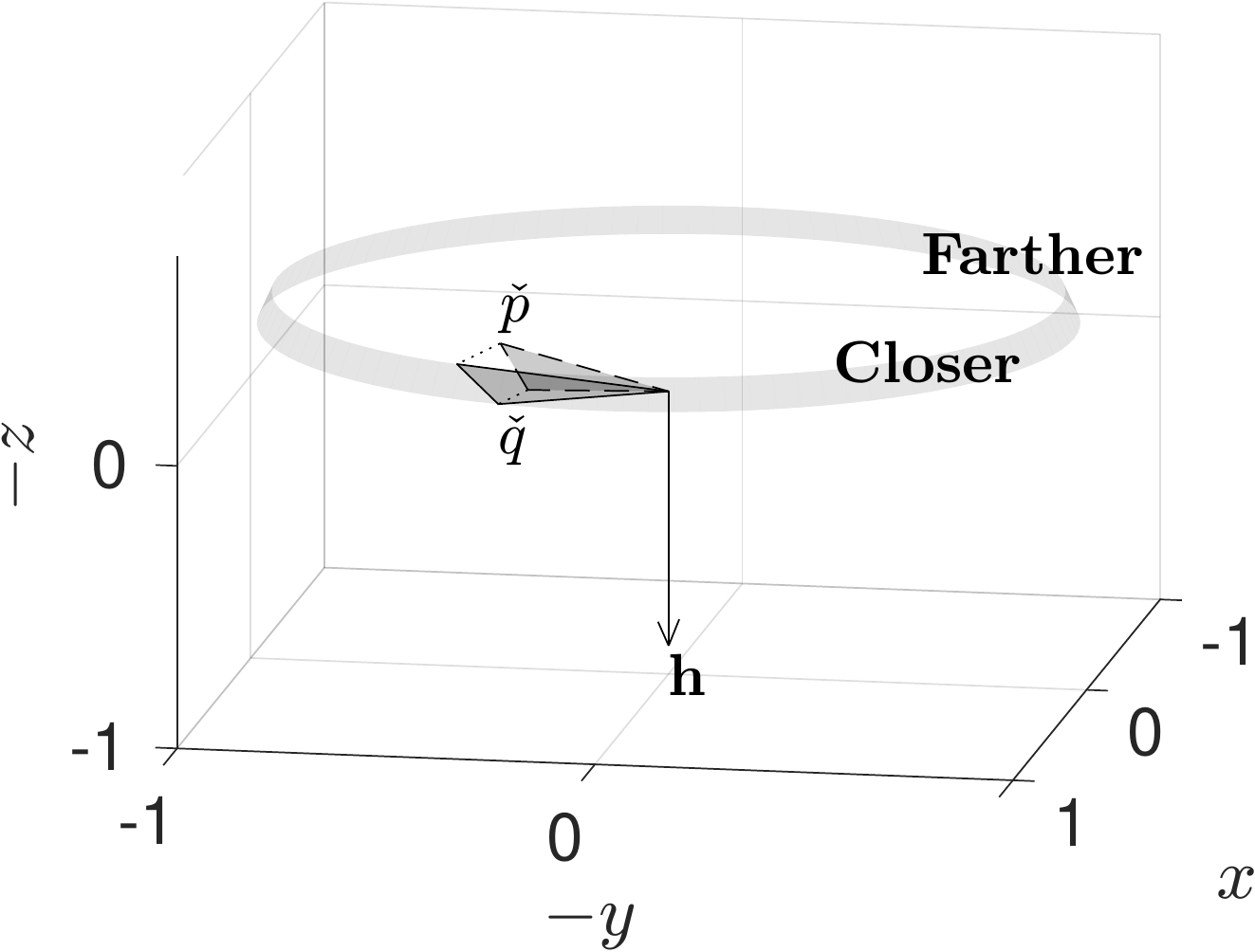}
\end{center}\end{minipage}
\caption{A visual depiction of the solutions presented in Theorems \ref{thm:twovecsoln} and \ref{thm:snglvecratesoln}. The image on the left shows the two solutions $\qc r_1 \otimes \qc q$ (dotted triangle) and $\qc r_2 \otimes \qc p$ (dashed triangle) of theorem 3. The figure on the right shows the solution $\qc q$ (solid triangle) of theorem 4 obtained by projecting the integrated attitude $\qc p$ (dashed triangle) onto the feasibility cone of vector measurement $b$.}
\label{fig:solnvis}
\end{figure}
\subsection {Attitude estimation using single vector measurement and rate measurement} \label{sec:soln2}

We first write down the constraints imposed by the measurement upon the attitude quaternion $\qc q = [c\; s[\*n]]^T = [c\; sn_1\; sn_2\; sn_2]^T$, where $c = \cos(\Phi/2)$ and $s = \sin(\Phi/2)$ are functions of the rotation angle $\Phi$, and $\*n$ is a unit vector along the rotation axis with components $n = [n_1\; n_2\; n_3]^T$ in the reference coordinate system. The constraint is given in equation (\ref{eqn:bconstr1}). Converting the quaternion multiplication to vector notation, equation (\ref{eqn:bconstr1}) can also be written as:
\begin{flalign}
& \begin{bmatrix} -sn^Tb \\ cb + s[n\times]b \end{bmatrix} = \begin{bmatrix} -sh^Tn \\ ch + s[h\times]n \end{bmatrix} \;, & \nonumber\\
\text{\it i.e., } \qquad & \quad \begin{bmatrix} -s(h - b)^Tn \\ c(h - b) + s[(h + b)\times]n \end{bmatrix} = 0 \;, & \nonumber
\end{flalign}
where $[n\times]$ denotes the cross product matrix associated with the 3-vector $n$. Expanding the vectors,
\begin{flalign}
 & \begin{bmatrix} & -f_1 & -f_2 & -f_3 \\ f_1 & & -g_3 & g_2 \\ f_2 & g_3 & & -g_1 \\ f_3 & -g_2 & g_1 & \end{bmatrix} \begin{bmatrix} c \\ sn_1 \\ sn_2 \\ sn_3 \end{bmatrix} = 0 \;, & \label{eqn:bconstr3} \\
\text{where }\;\, \qquad & f = h - b \qquad \text { and } \qquad g = h + b \;, & \nonumber\\
\text{so that } \qquad & f_1g_1 + f_2g_2 + f_3g_3 = f^Tg = h^Th - b^Tb = 0 \;. & \nonumber
\end{flalign}
While it is not obvious, equation (\ref{eqn:bconstr3}) has a double redundancy, so the system of four linear equations actually has rank 2 and nullity 2. This can be seen by considering the solution:
\begin{align}
\qc q & = \begin{bmatrix} c \\ (-cf_2 + sn_3g_1)/g_3 \\ (cf_1 + sn_3g_2)/g_3 \\ sn_3 \end{bmatrix} \;, \label{eqn:quat2dof}
\end{align}
where $sn_1$ and $sn_2$ are solved in terms of $c$ and $sn_3$ using the inner two row equations in equation (\ref{eqn:bconstr3}). Substituting these in the outer two rows of equation (\ref{eqn:bconstr3}) satisfies them trivially, so these two rows do not yield any additional information. This makes sense as we have not yet imposed the normalization constraint that $n_1^2 + n_2^2 + n_3^2 = 1$ ($c$ and $s$, representing $\cos\Phi/2$ and $\sin\Phi/2$, are already assumed to satisfy $c^2 + s^2 = 1$). And we are anyway to end up with one degree of freedom in $\qc q$ if using the vector measurement constraint alone, as discussed earlier.

\ifx\qver\qdetailed
We could apply the normalization constraint,
\begin{gather}
n_1^2 + n_2^2 + n_3^2 = 1 \;, \label{eqn:nconstr1}
\end{gather}
at this point to express $n$ completely in terms of $\Phi$:
\begin{align}
& \frac{c^2f_2^2}{s^2g_3^2} + \frac{g_1^2}{g_3^2}n_3^2 + \frac{c^2f_1^2}{s^2g_3^2} + \frac{g_2^2}{g_3^2}n_3^2 + 2\frac{c}{sg_3^2}n_3(f_1g_2 - f_2g_1) \nonumber\\
& \qquad + n_3^2 = 1 \;, \nonumber
\intertext{or, }
& n_3^2g^Tg + 2\frac{c}{s}n_3(f_1g_2 - f_2g_1) + \frac{c^2}{s^2}(f_1^2 + f_2^2) = g_3^2 . \label{eqn:nconstr2}
\end{align}
The above quadratic equation can be solved for $n_3$ in terms of $c/s = \cot\Phi/2$ to yield:
\begin{align}
n_3 & = -\frac{c(f_1g_2 - f_2g_1)}{sg^Tg} \nonumber\\
& \qquad \pm \sqrt{\frac{c^2((f_1g_2 - f_2g_1)^2 - g^Tg(f_1^2 + f_2^2))}{(sg^Tg)^2} + \frac{g_3^2}{g^Tg}} \;. \label{eqn:nconstr3}
\end{align}
The above equation in conjunction with the inner two rows of equation (\ref{eqn:quat2dof}) expresses all three components of $n$ in terms of $c/s = \cot(\Phi/2)$ and the measured quantities $f$ and $g$. Thus we are left with the single degree of freedom, $\Phi$, in $\qc q$, as expected. However, as shall be seen later, it is easier to retain $n_3$ as a variable in our problem, along with the normalization constraint (\ref{eqn:nconstr2}).
\fi

We now move on to utilizing the angular velocity measurement that determines the differential evolution of the attitude. The kinematic differential equation for the quaternion is the linear first order ODE:
\ifx\qver\qelsaut
\begin{align}
\dot{\qc q} & = \frac{1}{2}\qc q\otimes\qc\omega = \frac{W\qc q}{2}, \label{eqn:pdot}
\end{align}
\else
\begin{align}
\dot{\qc q} & = \frac{1}{2}\qc q\otimes\qc\omega = \frac{1}{2}\,\begin{bmatrix} q_0 & -q_1 & -q_2 & -q_3 \\ q_1 & q_0 & -q_3 & q_2 \\ q_2 & q_3 & q_0 & -q_1 \\ q_3 & -q_2 & q_1 & q_0 \end{bmatrix} \begin{bmatrix} 0 \\ \omega_1 \\ \omega_2 \\ \omega_3 \end{bmatrix} = \frac{W\qc q}{2}, \label{eqn:pdot}
\end{align}
\fi
where $\qc\omega$ is the quaternion form of the 3-vector $\omega$. In continuous time, the integration of (\ref{eqn:pdot}) for a constant $W$ gives an estimate $\qc p(t+T) = \exp(WT/2)\qc q(t)$. For example, if $\omega(t+s) = [0\; (\xi\cos\xi s)\; 0]^T$, then $\qc p = \exp\{j\sin\xi T/2\} \qc q = \cos(\sin(\xi T)/2)\qc q + j\sin(\sin(\xi T)/2)\qc q$, where $j = [\otimes \qc e_2]$, where $\qc e_2 = [0\; 0\; 1\; 0]^T$. For a time-varying $\omega$, the state transition matrix replaces the exponential.
In discrete time, denoting the integrated estimate as $\qc p(i+1)$, the above equation takes the form
\begin{align}
\qc p(i+1) & = \qc q(i) + \frac{T}{2}\qc q(i)\otimes\qc\omega(i) \;, \label{eqn:pdotdisc}
\end{align}
where $T$ is the time step from the previous estimation of $\qc q(i)$ to the current estimation $\qc p(i+1)$. In the subsequent derivation, we shall omit the time argument of $\qc p$, as there is no ambiguity.

The deviation of the vector-aligned quaternion estimate, $\qc q$ in equation (\ref{eqn:quat2dof}), from the integrated estimate, $\qc p$ in equation (\ref{eqn:pdotdisc}), can be expressed as the difference of $\qc p^{-1}\otimes\qc q$ from $\qc 1$. But minimizing the distance of a quaternion from the unit quaternion is the same as minimizing the rotation angle $\Phi$ (Lemma \ref{thm:quatequivangle}), which is, in turn, the same as maximizing the zeroeth component of the quaternion, $\cos(\Phi/2)$.
Note that, the quaternions $\qc p^{-1}\otimes \qc q$ and $-\qc p^{-1}\otimes \qc q$ affect the same rigid body rotation in 3-dimensional Euclidean space, but minimizing the distance of one from $\qc 1$ maximizes the distance of the other in quaternion space. So we just extremize the distance, rather than specifically minimize it. Once we have the solution set, we can check which solutions correspond to a maximum and which to a minimum, and choose the latter.

We therefore need to extremize the zeroeth component of $\qc p^{-1}\qc q$, where $\qc p = [p_0\; p_1\; p_2\; p_3]^T$ is the attitude estimate obtained by integrating the angular velocity $\omega$ as given in equation (\ref{eqn:pdot}) and $\qc q$ is expressed in terms of $c/s$ and $n_3$ as in equation (\ref{eqn:quat2dof}), while enforcing the constraint in equation (\ref{eqn:bconstr1}). This can be accomplished by using the method of Lagrange multipliers to define a cost function that invokes the error norm as well as the constraint. Below, we have multiplied the cost function by the constant $g_3$ and the constraint by $g_3^2$, noting that the solution is unaffected by such a scaling:
\begin{align}
& J(\Phi, n_3) = g_3[\qc p^{-1}\otimes\qc q]_0 + \lambda g_3^2(n_1^2 + n_2^2 + n_3^2 - 1)\nonumber\\
 & = (cp_0 + sn_3p_3)g_3 + (-cf_2 + sn_3g_1)p_1 + (cf_1 + sn_3g_2)p_2 \nonumber\\
 & \qquad + \lambda\left(n_3^2g^Tg + 2\frac{cn_3}{s}(f_1g_2 - f_2g_1) + \frac{c^2}{s^2}(f_1^2 + f_2^2) - g_3^2\right) \nonumber\\
 & = c(g_3p_0  + f_1p_2 - f_2p_1) + sn_3g^Tp \nonumber\\
 & \;\; + \lambda\left(n_3^2g^Tg + 2\frac{cn_3}{s}(f_1g_2 - f_2g_1) + \frac{c^2}{s^2}(f_1^2 + f_2^2) - g_3^2\right) , \nonumber
\end{align}
where $p$ denotes the vector portion of $\qc p$. Now we set the first order partial derivatives of $J$ to 0:
\begin{align}
\hspace*{-0.3cm} 0 & = \qp_\Phi J = -s(g_3p_0  + f_1p_2 - f_2p_1) + cn_3g^Tp \nonumber\\
\hspace*{-0.3cm}  & \; + \left(-\frac{2\lambda}{s^2}\right)\left(\frac{c}{s}(f_1^2 + f_2^2) + n_3(f_1g_2 - f_2g_1)\right) , \label{eqn:dJdPhi} \\
\hspace*{-0.3cm} 0 & = \qp_{n3}J = sg^Tp + 2\lambda g^Tgn_3 + 2\lambda\frac{c}{s}(f_1g_2 - f_2g_1) , \label{eqn:dJdn3} \\
\hspace*{-0.3cm} 0 & = \qp_\lambda J = n_3^2g^Tg - g_3^2 + \frac{2cn_3}{s}(f_1g_2 - f_2g_1) + \frac{c^2}{s^2}(f_1^2 + f_1^2) . \label{eqn:dJdmu}
\end{align}
\ifx\qver\qdetailed
Equation (\ref{eqn:dJdn3}) yields:
\begin{align}
-2\lambda & = \frac{sg^Tp}{g^Tgn_3 + \displaystyle\frac{c}{s}(f_1g_2 - f_2g_1)} \;. \label{eqn:mu}
\end{align}
Substituting this in equation (\ref{eqn:dJdPhi}), we obtain:
\begin{align}
& -s(g_3p_0  + f_1p_2 - f_2p_1)\left(g^Tgn_3 + \frac{c}{s}(f_1g_2 - f_2g_1)\right) + \nonumber\\
 & \qquad g^Tp\left[cn_3\left(g^Tgn_3 + \frac{c}{s}(f_1g_2 - f_2g_1)\right)\right. \nonumber\\
 & \qquad + \left.\frac{c}{s^2}(f_1^2 + f_2^2) + \frac{n_3}{s}(f_1g_2 - f_2g_1)\right] = 0 \;. \label{eqn:dJdphi2}
\end{align}
The factor in the square brackets can be substantially simplified using the constraint equation (\ref{eqn:dJdmu}) as:
\begin{align}
& cn_3\left(g^Tgn_3 + \frac{c}{s}(f_1g_2 - f_2g_1)\right) + \frac{c}{s^2}(f_1^2 + f_2^2) \nonumber\\
 & \qquad + \frac{n_3}{s}(f_1g_2 - f_2g_1) \nonumber\\
& = c\left(g_3^2 -\frac{c^2}{s^2}(f_1^2 + f_2^2) - \frac{cn_3}{s}(f_1g_2 - f_2g_1)\right) \nonumber\\
 & \qquad + \frac{c}{s^2}(f_1^2 + f_2^2) + \frac{n_3}{s}(f_1g_2 - f_2g_1) \nonumber\\
& = c(g_3^2 + f_1^2 + f_2^2) + sn_3(f_1g_2 - f_2g_1) \;. \nonumber
\end{align}
Substituting this back into equation (\ref{eqn:dJdphi2}), we obtain:
\begin{align}
& -(g_3p_0  + f_1p_2 - f_2p_1)\left(sg^Tgn_3 + c(f_1g_2 - f_2g_1)\right) \nonumber\\
& \qquad + g^Tp(c(g_3^2 + f_1^2 + f_2^2) + sn_3(f_1g_2 - f_2g_1)) = 0 \;. \nonumber
\end{align}
Accumulating terms containing $sn_3$ and $c$, we obtain an expression for the ratio $\kappa = c/(sn_3)$ in terms of known quantities as:
\begin{align}
\kappa & = \frac{(g_3p_0 + f_1p_2 - f_2p_1)g^Tg - g^Tp(f_1g_2 - f_2g_1)}{g^Tp(f_1^2 + f_2^2 + g_3^2) - (g_3p_0 + f_1p_2 - f_2p_1)(f_1g_2 - f_2g_1)} \;, \label{eqn:kappa}
\end{align}
where $f = h - b$ and $g = h + b$ were defined in terms of the vector measurements, and $\qc p$ is obtained by integrating the angular velocities. We can simplify the numerator and denominator in equation (\ref{eqn:kappa}) further. First the numerator of $\kappa$:
\begin{align}
& (p_0g_3 - p_1f_2 + p_2f_1)g^Tg + p^Tg(g_1f_2 - g_2f_1) \nonumber\\
& = p_0g^Tgg_3 + p_1(-f_2g^Tg + g_1^2f_2 - g_1f_1g_2) \nonumber\\
 & \qquad + p_2(f_1g^Tg + g_1g_2f_2 - g_2^2f_1) + p_3g_3(g_1f_2 - g_2f_1) \nonumber\\
& = g_3(p_0g^Tg + p_3(g_1f_2 - g_2f_1)) + p_1(-f_2g_2^2 - f_2g_3^2 - g_1f_1g_2) \nonumber\\
 & \qquad + p_2(f_1g_1^2 + f_1g_3^2 + g_1f_2g_2) \nonumber\\
& = g_3(p_0g^Tg + p_3(g_1f_2 - g_2f_1)) + p_1(f_3g_3g_2 - f_2g_3^2) \nonumber\\
 & \qquad + p_2(-f_3g_3g_1 + f_1g_3^2) \nonumber\\
& = g_3(p_0g^Tg + p_1(f_3g_2 - f_2g_3) + p_2(-f_3g_1 + f_1g_3) \nonumber\\
 & \qquad + p_3(g_1f_2 - g_2f_1)) \nonumber\\
& = g_3(p_0g^Tg + p^Tg\times f) . \label{eqn:kappanumer}
\end{align}
Next, the denominator of $\kappa$:
\begin{align}
& (p_0g_3 - p_1f_2 + p_2f_1)(g_1f_2 - g_2f_1) + p^Tg(f_1^2 + f_2^2 + g_3^2) \nonumber\\
& = p_0g_3(g_1f_2 - g_2f_1) + p_1(f_2g_2f_1 + g_1f_1^2 + g_1g_3^2) \nonumber\\
 & \qquad + p_2(f_1g_1f_2 + g_2f_2^2 + g_2g_3^2) + p_3g_3(f_1^2 + f_2^2 + g_3^2) \nonumber\\
& = g_3(p_0(g_1f_2 - g_2f_1) + p_3(f_1^2 + f_2^2 + g_3^2))) \nonumber\\
 & \qquad + p_1(-f_3g_3f_1 + g_1g_3^2) + p_2(-f_3g_3f_2 + g_2g_3^2) \nonumber\\
& = g_3(p_0(g_1f_2 - g_2f_1) + p_1(g_1g_3 - f_1f_3) \nonumber\\
 & \qquad + p_2(g_2g_3 - f_2f_3) + p_3(f_1^2 + f_2^2 + g_3^2)) . \label{eqn:kappadenom}
\end{align}
\fi
This yields, for the ratio $\kappa = c/sn_3$:
\begin{align}
\kappa & = \frac{p_0g^Tg + p^Tg\times f}{p_0(g_1f_2 - g_2f_1) + \sum_{1, 2} p_i(g_ig_3 - f_if_3) + p_3(f_1^2 + f_2^2 + g_3^2)} . \label{eqn:kappa2}
\end{align}

Fortuituously, $c/s = \cot(\Phi/2)$ is therefore just proportional to $n_3$, and upon expressing $c/s$ in terms of $n_3$ in the normalization constraint (equation (\ref{eqn:dJdmu})), the resulting equation becomes extremely simple to solve:
\begin{align}
g_3^2 & = g^Tgn_3^2 + 2\kappa(f_1g_2 - f_2g_1)n_3^2 + \kappa^2n_3^2(f_1^2 + f_2^2) \;, \nonumber
\end{align}
or
\begin{align}
n_3 & = \frac{g_3}{\sqrt{g^Tg + 2\kappa(f_1g_2 - f_2g_1) + \kappa^2(f_1^2 + f_2^2)}} \;, \label{eqn:n3}\\
\frac{c}{s} & = \frac{\kappa g_3}{\sqrt{g^Tg + 2\kappa(f_1g_2 - f_2g_1) + \kappa^2(f_1^2 + f_2^2)}} \;. \label{eqn:cbys}
\end{align}
The other components of the attitude quaternion can be obtained using the inner two rows of equation (\ref{eqn:quat2dof}). Thus we obtain the following theorem.

\begin{thm} \label{thm:snglvecratesoln}
If the angular velocity of a rigid body is integrated to yield a attitude quaternion estimate $\qc p$, then the estimate $\qc q \in Q_b$ lying in the feasibility cone of measurement $b$ which is closest to $\qc p$, is given by equations (\ref{eqn:quat2dof}, \ref{eqn:kappa2}, \ref{eqn:n3}, \ref{eqn:cbys}).
\end{thm}
\begin{pf}
The proof follows from the construction leading to equations (\ref{eqn:quat2dof}, \ref{eqn:kappa2}, \ref{eqn:n3}, $\ref{eqn:cbys}$). Refer figure \ref{fig:solnvis}. \hspace*{\fill} \qed
\end{pf}

\ifx\qver\qdetailed
\begin{rmk}
{\it Sign indeterminacy}:
There are two instances of taking square-roots in the construction of the optimal estimate: one in the denominators in equations (\ref{eqn:n3}, \ref{eqn:cbys}), and a second when determining $s = 1/\sqrt{(c/s)^2+1}$. They multiply all the components, and thus result in a net sign indeterminacy of the complete quaternion. We could choose the sign as yielded by the equations, or such that the zeroeth component is positive. Both choices yield a correct attitude in three-dimensional Euclidean space.
\end{rmk}
\fi

\begin{rmk}
{\it Solution when reference vector is aligned with $z$-axis}:
A common application of the presented solution would be to an aerial robot that uses an accelerometer to measure the gravity vector (after acceleration compensation). Since the reference coordinate system's $z$-axis is aligned with the reference vector $\*h$, we have $f = [(-b_1)\; (-b_2)\; (1-b_3)]^T$ and $g = [b_1\; b_2\; (1+b_3)]^T$. Equations (\ref{eqn:kappa2}, \ref{eqn:cbys}) now simplify to:
\begin{align}
\hspace*{-0.6cm} \kappa & = \frac{c}{sn_3} = \frac{(1 + b_3)p_0 - b_1p_2 + b_2p_1}{b_1p_1 + b_2p_2 + (1 + b_3)p_3} \;, \nonumber\\
\hspace*{-0.6cm} \begin{bmatrix} q_0 \\ q_1 \\ q_2 \\ q_3 \end{bmatrix} & = \begin{bmatrix} c \\ sn_1 \\ sn_2 \\ sn_3 \end{bmatrix} = \frac{1}{\sqrt{2(1+\kappa^2)(1+b_3)}} \begin{bmatrix} \kappa(1 + b_3) \\ \kappa b_2 + b_1 \\ -\kappa b_1 + b_2 \\ (1+b_3) \\ \end{bmatrix} , \label{eqn:hzsoln2}
\end{align}
where we have used the fact that $(1+b_3)^2 + b_1^2 + b_2^2 = 2(1 + b_3)$.
While the introduction of the auxillary variable $\kappa$ in equations (\ref{eqn:kappa2} - \ref{eqn:cbys}) seems adhoc, its role is more clearly visible now -- $\kappa$ parameterizes the feasibility cone $Q_b$ in terms of the two special solutions provided in lemma \ref{thm:snglvecssoln}:
\ifx\qver\qelsaut
\begin{align}
\qc q & = \frac{\kappa \qc r_1 + \qc r_2}{\sqrt{1 + \kappa^2}} = \frac{(\qc r_1\qc r_1^T + \qc r_2\qc r_2^T)\qc p}{\|(\qc r_1\qc r_1^T + \qc r_2\qc r_2^T)\qc p\|} \,. \label{eqn:kappaparam}
\end{align}
\else
\begin{align}
& \sqrt{2(1+\kappa^2)(1+b_3)}\qc q = \kappa \begin{bmatrix} 1 + b_3 \\ b_2 \\ -b_1 \\ 0 \end{bmatrix} + \begin{bmatrix} 0 \\ b_1 \\ b_2 \\ 1 + b_3 \end{bmatrix} \,, \nonumber\\
\text{or, } & \qc q = \frac{\kappa \qc r_1 + \qc r_2}{\sqrt{1 + \kappa^2}} = \frac{(\qc r_1\qc r_1^T + \qc r_2\qc r_2^T)\qc p}{\|(\qc r_1\qc r_1^T + \qc r_2\qc r_2^T)\qc p\|} \,. \label{eqn:kappaparam}
\end{align}
\fi

\ifx\qver\qdetailed
Equation (\ref{eqn:hzsoln2}) may be checked for sanity against the Euler angle solution by using the relations $\sin\theta = 2(q_0q_2 - q_1q_3)$, $\cos\theta\sin\phi = 2(q_0q_1 + q_2q_3)$, and $\cos\theta\cos\phi = q_0^2 - q_1^2 - q_2^2 + q_3^3$. The reduction of the quaternion form to the Euler angle form is straightforward, but the details are long and omitted. The final result is that
\begin{align}
\begin{bmatrix} -\sin\theta \\ \cos\theta\sin\phi \\ \cos\theta\cos\phi \end{bmatrix} & = \begin{bmatrix} b_1 \\ b_2 \\ b_3 \end{bmatrix}. \nonumber
\intertext{So, }
\begin{bmatrix} \tan\phi \\ \sin\theta \end{bmatrix} & = \begin{bmatrix} b_2/b_3 \\ -b_1 \end{bmatrix}\,, \nonumber
\end{align}
as expected.
\fi
\end{rmk}

\begin{rmk} \label{thm:attYvsEKF}
{\it Relation to the EKF \cite{Lefferts:82a}}:
A filtered attitude estimate $\qc q_f$ can be obtained by projecting the integrated estimate, $\qc p$, onto the feasibility cone corresponding to a filtered vector measurement $b_f$, to yield the vector aligned estimate $\qc q$ of Theorem \ref{thm:snglvecratesoln}. The predict-step in Theorem \ref{thm:snglvecratesoln} is identical to that in the EKF: we just integrate the dynamics of the state from the previous time step. Note that the EKF accommodates nonlinearity in the dynamics in the prediction step, and so it is okay for the attitude dynamics to be bilinear in the state (attitude) and input (angular velocity). It is the correction step where the geometric method diverges from the EKF. It may be noted that the projection onto the feasibility cone affects only two degrees of freedom of the attitude. The attitude degree of freedom associated with rotation about the reference vector is completely unaffected by the projection. Thus the filtering may be precisely accomplished by implementing it upon the vector measurement. A detailed derivation of the filtered vector measurement and the propagation of the covariance matrices is given in the next subsection, and the improvement in performance is verified in simulations in section \ref{sec:ressim}.
%\begin{align}
%\qc q_f & = \qc p \otimes (\qc p^{-1} \otimes \qc q)^x. \label{eqn:interp2}
%\end{align}
%The optimal interpolation parameter $x$ in equation \ref{eqn:interp1} is obtained from the measurement noise variances, similar to the Kalman gain in an EKF. An exact expression for $x$ under a standard set of assumptions is derived in the following subsection. In the limit $\omega T \rightarrow 0$, the interpolated estimate is identical to the solution obtained using an optimally tuned EKF. However, for large incremental changes in the attitude, the geometric solution presented in this note is superior to the EKF which suffers from a loss of accuracy on account of the linearization.
\end{rmk}

%\begin{rmk}
%{\it Random-walk bias}: Filtered bias estimation for variable bias with exponential autocorrelation (WIP). Filtering details to be expanded.
%\begin{align}
%A_{i+1} & = (1 - T/\tau)A_i + (1 - b_ib_i^T) \nonumber\\
%B_{i+1} & = (1 - T/\tau)B_i + 2(1 - b_ib_i^T)\delta r_i/ T\nonumber\\
%A\qol e & = B. \label{eqn:e5}
%\end{align}
%\begin {figure} [!htbp] \begin {center}
%\includegraphics [scale=0.6, trim={0cm 0cm 0cm 0cm}] {biasrndwlkestim.pdf}
%\caption {Estimation of an exponentially autocorrelated gyroscopic bias $[e_p\; e_q\; e_r]^T$. The time constant of the exponential autocorrelation is 5s, while the estimator has a time constant of 1s.}
%\label {fig:biasrndwlkestim}
%\end {center} \end {figure}
%\end{rmk}

The following corollary follows from theorem \ref{thm:snglvecratesoln}.
\begin{cor} \label{thm:corrquat}
The correction that takes the integrated estimate $\qc p$ into the feasibility cone $Q_b$ is essentially a rotation about an axis that is orthogonal to the reference vector $h$.
\end{cor}
\begin{pf}
With the simplifying choice for the reference coordinate system's $z$-axis that leads to equation (\ref{eqn:hzsoln2}), the proof is simple.
The correcting rotation in the reference coordinate system is:
\begin{align}
\qc r & = \qc q\otimes \qc p^{-1} = \begin{bmatrix} \kappa(1 + b_3) \\ \kappa b_2 + b_1 \\ -\kappa b_1 + b_2 \\ (1 + b_3) \end{bmatrix} \otimes \begin{bmatrix} p_0 \\ -p_1 \\ -p_2 \\ -p_3 \end{bmatrix} /\sqrt{2(1+\kappa^2)(1+b_3)} \,. \nonumber
\end{align}
So, using the expression for $\kappa$ in equation (\ref{eqn:hzsoln2}), we obtain $r_3 = 0$.
In the general case of an arbitrary $h$, the proof is more tedious, but still valid \cite{Mohseni:18z}.
\ifx\qver\qdetailed
The projected attitude estimate $\qc q$ of theorem \ref{thm:snglvecratesoln} may be written as:
\begin{align}
\gamma \qc q & = \begin{bmatrix} 0 \\ g_1 \\ g_2 \\ g_3 \end{bmatrix} + \kappa \begin{bmatrix} g_3 \\ -f_2 \\ f_1 \\ 0 \end{bmatrix} , \nonumber\\
\end{align}
where $\gamma = \sqrt{(g_1-\kappa f_2)^2 + (g_2 + \kappa f_1)^2 + (1+\kappa^2)g_3^2}$. So the correction quaternion in the reference coordinate system is:
\begin{align}
\gamma\qc q\otimes\qc p^{-1} & = \begin{bmatrix} 0 \\ g_1 \\ g_2 \\ g_3 \end{bmatrix} \otimes \begin{bmatrix} p_0 \\ -p_1 \\ -p_2 \\ -p_3 \end{bmatrix} + \kappa \begin{bmatrix} g_3 \\ -f_2 \\ f_1 \\ 0 \end{bmatrix} \otimes \begin{bmatrix} p_0 \\ -p_1 \\ -p_2 \\ -p_3 \end{bmatrix} \nonumber\\
 & = \begin{bmatrix} p^Tg + \kappa(p_0g_3 - p_1f_2 + p_2f_1) \\ p_0g + p\times g - \kappa pg_3 + \kappa\begin{bmatrix} -p_0f_2 - p_3f_1 \\ p_0f_1 - p_3f_2 \\ p_1f_1 + p_2f_2 \end{bmatrix} \end{bmatrix} . \nonumber
\end{align}
Suppose the vector potion of the correction quaternion is not orthogonal to $h$:
\begin{align}
& 0 \ne h^T(\gamma\qc q\otimes \qc p^{-1}) \nonumber\\
& = (g + f)^T \left(p_0g + p\times g - \kappa pg_3 - \kappa p_3f + \kappa \begin{bmatrix} -p_0f_2 \\ p_0f_1 \\ p^Tf \end{bmatrix} \right) \nonumber\\
& = p_0g^Tg + p^Tg\times f - \kappa g_3p^T(g + f) - \kappa p_3f^Tf \nonumber\\
& \qquad + \kappa (g_3+f_3)p^Tf + \kappa p_0((g_2 + f_2)f_1 - (g_1 + f_1)f_2) . \nonumber
\end{align}
where we use the fact that $g^Tf = 0$. Separating the terms multiplying $\kappa$,
\begin{align}
& p_0g^Tg + p^Tg\times f \ne \kappa p^T(g + f)g_3 + \kappa p_0(g_1f_2 - g_2f_1) \nonumber\\
 & \qquad + \kappa p_3f^Tf - \kappa p^Tf(g_3+f_3) , \nonumber\\
%p_0g^Tg + p^Tg\times f \ne \kappa \left\{ p_0(g_1f_2 - g_2f_1) + p_3f^Tf \right. \nonumber\\
% & \qquad \left. + p^T(gg_3 - ff_3) \right\} , \nonumber\\
& \hphantom{p_0g^Tg + p^Tg\times f}= \kappa \left\{ p_0(g_1f_2 - g_2f_1) + p_1(g_1g_3 - f_1f_3) \right. \nonumber\\
 & \qquad \left. + p_2(g_2g_3 - f_2f_3) + p_3(f_1^2 + f_2^2 + g_3^2) \right\} . \label{eqn:hTvecerr}
\end{align}
Comparing equation (\ref{eqn:kappa2}) with equation (\ref{eqn:hTvecerr}) leads to a contradiction, and the proof is complete.
\fi
The underlying reason for this result is just that a rotation about any other axis would have an unnecessary component about $h$, and that would make the correction to reach $Q_b$ suboptimal. \hspace*{\fill} \qed
\end{pf}

\ifx\qver\qdetailed
Let us further analyze the required reference-frame correction $\qc r = \qc q\otimes\qc p^{-1}$ when the reference vector is along the reference $z$-axis:
\begin{align}
\qc r & = \qc q\otimes \qc p^{-1} \nonumber\\
 & = \frac{1}{\sqrt{2(\alpha^2 + \beta^2)(1 + b_3)}} \begin{bmatrix} \alpha(1+b_3) \\ \alpha b_2+\beta b_1 \\ -\alpha b_1+\beta b_2 \\ \beta(1+b_3) \end{bmatrix} \otimes \begin{bmatrix} p_0 \\ -p_1 \\ -p_2 \\ -p_3 \end{bmatrix} , \nonumber
\end{align}
where  $\alpha = p_0(1+b_3) + p_1b_2 - p_2b_1$, $\beta = p_1b_1 + p_2b_2 + p_3(1+b_3)$. Expanding the quaternion multiplication,
\begin{align}
 & \sqrt{2(1+b_3)(\alpha^2+\beta^2)}\qc r = \nonumber\\
 & \quad \alpha \begin{bmatrix} 1+b_3 \\ b_2 \\ -b_1 \\ 0 \end{bmatrix} \otimes \begin{bmatrix} p_0 \\ -p_1 \\ -p_2 \\ -p_3 \end{bmatrix} + \beta \begin{bmatrix} 0 \\ b_1 \\ b_2 \\ 1+b_3 \end{bmatrix} \otimes \begin{bmatrix} p_0 \\ -p_1 \\ -p_2 \\ -p_3 \end{bmatrix} \nonumber\\
 & = \alpha \begin{bmatrix} p_0(1+b_3)+p_1b_2-p_2b_1 \\ -p_1(1+b_3)+p_0b_2+p_3b_1 \\ -p_2(1+b_3)-p_0b_1+p_3b_2 \\ -p_3(1+b_3)-p_2b_2-p_1b_1 \end{bmatrix} \nonumber\\
 & \qquad + \beta \begin{bmatrix} p_1b_1+p_2b_2+p_3(1+b_3) \\ p_0b_1-p_3b_2+p_2(1+b_3) \\ p_0b_2+p_3b_1-p_1(1+b_3) \\ p_0(1+b_3)-p_2b_1+p_1b_2 \end{bmatrix} . \nonumber
\end{align}

Substituting for $\alpha$ and $\beta$, and simplifying, we obtain:
\begin{align}
\alpha^2 + \beta^2 & = [p_0(1+b_3)+p_1b_2-p_2b_1]^2 \nonumber\\
 & \qquad + [p_1b_1+p_2b_2+p_3(1+b_3)]^2 \nonumber\\
 & = (p_0^2 + p_3^2)(1 + b_3)^2 + (p_1^2 + p_2^2)(1 - b_3^2) \nonumber\\
 & \qquad + 2[p_0(p_1b_2 - p_2b_1) + p_3(p_1b_1+p_2b_2)](1 + b_3) \nonumber\\
 & = (1 + b_3)\left[1 + (p_0^2 - p_1^2 - p_2^2 + p_3^2)b_3 \right. \nonumber\\
 & \qquad \left. + 2(p_0p_1 + p_2p_3)b_1 + 2(-p_0p_2 + p_3p_1)b_2 \right] \nonumber\\
 & = (1 + b_3)(1 + h_{p3}) , \nonumber
\end{align}
where, $\qc h_p = [h_{p0}\; h_{p1}\; h_{p2}\; h_{p3}]^T = \qc p\otimes\qc b\otimes\qc p^{-1}$ is the reference vector that would have yielded $\qc b$ after rotation by $\qc p$, and for the correction:
\begin{align}
(1+b_3)\sqrt{2(1+h_{p3})} \qc r & = (1 + b_3) \begin{bmatrix} 1+h_{p3} \\ h_{p2} \\ -h_{p1} \\ 0 \end{bmatrix} , \nonumber\\
\Rightarrow \sqrt{2(1+h_{p3})} \qc r & = \begin{bmatrix} 1+h_{p3} \\ h_{p2} \\ -h_{p1} \\ 0 \end{bmatrix} . \label{eqn:corr2}
\end{align}
Thus, the correction $\qc r$ is the smallest rotation which takes measurement $\qc h_p$ to $\qc h$: $\qc r\otimes \qc h_p = \qc h \otimes \qc r$. An alternative way to derive this result leading to an elegant expression of the result in Theorem \ref{thm:snglvecratesoln} for the general case (when $h$ need not be $[0\; 0\; 1]^T$) is as follows.

%A geometric filtering mechanism is now described that takes into account the constraints associated with attitude quaternions, and the nonlinear optimization that underlies the filtering between the integrated estimate, $\qc p$, and the vector-aligned estimate, $\qc q$.
We first note that the correction $\qc r = \qc q \otimes\qc p^{-1}$ which takes the integrated estimate $\qc p$ to the vector-aligned estimate $\qc q$ must be a rotation about an axis orthogonal to the reference vector $h$ in the reference coordinate system (Corollary \ref{thm:corrquat}). Then it follows that the correction $\qc r$ must be the smallest rotation which would take a hypothetical body-frame measurement $h_p = \qc p \otimes \qc b \otimes \qc p^{-1}$ to the reference coordinate system measurement $h$, as shown below:
\begin{align}
\qc r \otimes \qc h_p & = \qc q \otimes \qc p^{-1} \otimes \qc h_p \nonumber\\
 & = \qc q \otimes \qc b\otimes \qc p^{-1} = \qc h\otimes \qc q\otimes \qc p^{-1} = \qc h\otimes \qc r . \nonumber
\end{align}
Further, being the smallest rotation implies that $\qc r$ is the first special solution in Lemma \ref{thm:snglvecssoln} for the vector measurement constraint $\qc r\otimes \qc h_p = \qc h \otimes \qc r$, which yields:
\begin{align}
\Rightarrow \sqrt{2(1+h_p^Th)} \qc r & = \begin{bmatrix} 1+h_p^Th \\ h_p\times h \end{bmatrix} . \label{eqn:corr1}
\end{align}
The correction can be written solely in terms of $\qc p$ and $\qc b$ as:
\begin{align}
\sqrt{2(1+h_p^Th)} \qc r & = \qc 1 - \qc h\otimes \qc h_p = \qc 1 - \qc h\otimes\qc p\otimes \qc b\otimes \qc p^{-1} . \label{eqn:corr3}
\end{align}
\fi

\ifx\qver\qelsaut
An elegant expression for the corrected attitude estimate $\qc q$ in terms of the integrated estimate $\qc p$ and vector measurement $b$ of a single vector is:
\else
Equation (\ref{eqn:corr3}) leads to the most elegant form for the corrected attitude estimate $\qc q$ in terms of the integrated estimate $\qc p$ and measurement $b$ of a single vector:
\fi
\begin{align}
\qc q & = \frac{\qc p - \qc h \otimes \qc p \otimes \qc b}{\|\qc p - \qc h \otimes \qc p \otimes \qc b\|} . \label{eqn:qeqphpb}
\end{align}
Equation (\ref{eqn:qeqphpb}) is directly consistent with the measurement constraint $\qc h\otimes\qc q = (\qc h\otimes\qc p + \qc p\otimes\qc b)/\|\qc h\otimes\qc p + \qc p\otimes\qc b\| = \qc q\otimes\qc b$, so it lies on the feasibility cone by definition. At the same time, the correction $\qc q\otimes\qc p^{-1}$ in the reference coordinate system is about an axis perpendicular to $h$ as required by Corollary \ref{thm:corrquat}.

Equation (\ref{eqn:qeqphpb}) may rigorously be derived from equation (\ref{eqn:kappaparam}) as follows:
\ifx\qver\qdetailed
\begin{align}
2(\qc r_1\qc r_1^T + \qc r_2\qc r_2^T) & = \begin{bmatrix} 2c \\ \dfrac{b\times h}{c} \end{bmatrix} \begin{bmatrix} c & \dfrac{(b\times h)^T}{2c} \end{bmatrix} \nonumber \\
 & \qquad + \begin{bmatrix} 0 \\ \dfrac{b+h}{c} \end{bmatrix} \begin{bmatrix} 0 & \dfrac{(b+h)^T}{2c} \end{bmatrix} \nonumber\\
 & = \begin{bmatrix} 2c^2 & (b\times h)^T \\ b\times h & \dfrac{(b\times h)(b\times h)^T}{2c^2} \end{bmatrix} \nonumber\\
 & \qquad + \begin{bmatrix} 0 & 0 \\ 0 & \dfrac{(b+h)(b+h)^T}{2c^2} \end{bmatrix} \nonumber\\
 & = \begin{bmatrix} 1+h^Tb & (b\times h)^T \\ b\times h & 1 - h^Tb + hb^T + bh^T \end{bmatrix} \nonumber\\
 & = 1_{4\times 4} - [h\otimes][\otimes \qc b], \nonumber\\
\Rightarrow (\qc r_1\qc r_1^T + \qc r_2\qc r_2^T)\qc p & = (\qc p - \qc h\otimes \qc p \otimes \qc b)/2 , \nonumber
\end{align}
\else
\begin{align}
2(\qc r_1\qc r_1^T + \qc r_2\qc r_2^T)
 & = 1_{4\times 4} - [h\otimes][\otimes \qc b], \nonumber\\
\Rightarrow (\qc r_1\qc r_1^T + \qc r_2\qc r_2^T)\qc p & = (\qc p - \qc h\otimes \qc p \otimes \qc b)/2 , \nonumber
\end{align}
\fi
which upon normalizing yields the stated equivalence between (\ref{eqn:qeqphpb}) and (\ref{eqn:kappaparam}). Note that
\begin{align}
1 - [h\otimes][\otimes \qc b] & = 1 - \begin{bmatrix} & -h^T \\ h & [h\times] \end{bmatrix} \begin{bmatrix} & -b^T \\ b & -[b\times] \end{bmatrix} \nonumber\\
 & = \begin{bmatrix} 1+h^Tb & (b\times h)^T \\ b\times h & (1 - h^Tb) + hb^T + bh^T \end{bmatrix} . \nonumber
\end{align}

\begin{rmk} \label{thm:attYvsattM}
{\it Relation to the Explicit complementary filter (ECF) \cite{Mahony:08a}}:
The ECF in \cite{Mahony:08a} Theorem 5.1 may be realized out of Theorem \ref{thm:snglvecratesoln} by noting that the correction quaternion in the body frame is given by:
\begin{align}
\hspace*{-0.3cm} \qc p^{-1}\otimes\qc q & = \frac{\qc 1 - \qc p^{-1}\otimes\qc h\otimes\qc p\otimes\qc b}{\|\qc p - \qc h\otimes\qc p\otimes\qc b\|} = \frac{\qc 1 - \qc b_p\otimes\qc b}{\|\qc p - \qc h\otimes\qc p\otimes\qc b\|} , \label{eqn:corrquat2} 
\end{align}
where, $\qc b_p = \qc p^{-1}\otimes\qc h \otimes\qc p$ is the expected measurement of $\*h$ in the body frame, if $\qc p$ was already the correct attitude. On the other hand, the correction from the integrated estimate can be obtained by including a correction term $\omega_c$ in the angular velocity such that:
\begin{align}
\frac{\qc q - \qc p}{T} & = \frac{1}{2}\qc p\otimes\qc \omega_c, \nonumber
\end{align}
where $\qc\omega_c$ is the equivalent correction required in the angular velocity over a time-step $T$.
For small corrections, $\qc q \approx \qc p \approx -\qc h\otimes\qc p\otimes\qc b$, and so the incremental correction angular velocity is given by:
\begin{align}
\qc\omega_c & = \frac{2}{T}\left[\qc p^{-1}\otimes\qc q - \qc 1\right] = \frac{2}{T} \left[ \frac{\qc 1 - \qc b_p\otimes \qc b}{\|\qc p - \qc h\otimes\qc p\otimes\qc b\|} - \qc 1 \right] , \nonumber\\
 & \approx \frac{-\qc b_p\otimes \qc b - \qc 1}{T} \approx \frac{1}{T} \begin{bmatrix} 0 \\ b\times b_p \end{bmatrix} , \nonumber
\end{align}
whose vector portion is exactly the same as that reported in \cite{Mahony:08a} Theorem 5.1, with the gain $k_P$ equal to the time step $1/T$. Note that this also ensures that \cite{Mahony:08a} Theorem 5.1 is dimensionally consistent: $k_P$ must have dimensions of reciprocal time. For values of $k_P$ larger than $1/T$, we obtain a larger correction $\qc\omega_c$, and a larger weightage of $\qc q$ in the final filtered estimate.
\end{rmk}

\if\qver\qdetailed
We finally show that the projection from the integrated estimate $\qc p$ to $\qc q$ on the feasibility cone $Q_b$ leads to progressively smaller angular deviations from all attitudes on the feasibility cone. This result was presumed in establishing the relation between the geometric attitude estimation and Wahba's problem.
\begin{cor} \label{thm:rotsum}
As the attitude $\qc p'$ of a rigid body rotates from $\qc p$ outside the feasibility cone corresponding to a vector measurement $b$ of $h$, onto $\qc q$ on the feasibility cone, the angle to any fixed attitude $\qc r$ on the feasibility cone monotonically reduces.
\end{cor}
\begin{pf}
Since $\qc p'$ is on the path of projection from $\qc p$ to $\qc q$, it must be obtained as a rotation through an angle $x\Phi$, where $\Phi$ is the angle from $\qc p$ to $\qc q$ and $x \in [0, 1]$, about an axis $k$ that is orthogonal to $h$ in the reference coordinate system (Corollary \ref{thm:corrquat}):
\begin{align}
\qc p' & = \begin{bmatrix} c_x \\ s_x k \end{bmatrix} \otimes \qc p , \nonumber
\end{align}
where $c_x = \cos(x\Phi/2)$ and $s_x = \sin(x\Phi/2)$.
Since $\qc r$ lies on the feasibility cone, we must also have:
\begin{align}
\qc r & = \begin{bmatrix} c_r \\ s_r h \end{bmatrix} \otimes \qc q , \nonumber
\end{align}
where $c_r = \cos(\Phi_r/2)$ and $s_r = \sin(\Phi_r/2)$, for some $\Phi_r \in [-\pi, \pi]$. Then,
\begin{align}
\qc r \otimes \qc p'^{-1} & = \begin{bmatrix} c_r \\ s_r h \end{bmatrix} \otimes \qc q \otimes \qc p^{-1} \otimes \begin{bmatrix} c_x \\ -s_x k \end{bmatrix} \nonumber\\
 & =  \begin{bmatrix} c_r \\ s_r h \end{bmatrix} \otimes \begin{bmatrix} c_{1-x} \\ -s_{1-x} k \end{bmatrix} = \begin{bmatrix} c_rc_{1-x} \\ \hdots \end{bmatrix} , \nonumber
\end{align}
since $h^Tk = 0$. So, the scalar part of $\qc r\otimes\qc p'$, $c_rc_{1-x}$, progressively increases as $x$ goes from 0 to 1, and the angle therefore monotonically reduces from $\operatorname{acos}{}(c_rc_{1-x})$ to $\Phi_r$. \hspace*{\fill} \qed
\end{pf}
\fi

\subsection {Effect of noise in measurements on the estimation} \label{sec:noifx}

Let us first describe a filter on the vector measurement $b$ using the angular velocity information. Suppose the angular velocity is integrated to yield the attitude estimate $\qc p = [p_0\; p^T]^T$. This attitude then predicts the body-frame components $b_p$ for the reference vector $h$ through equation (\ref{eqn:bconstr1}). Perturbations in $\qc p$ induce perturbations in the predicted vector measurement $b_p$:
\begin{align}
b_p & = \left[(p_0^2 - p^Tp)1_{3\times 3} + 2pp^T - 2p_0[p\times]\right] h , \nonumber\\
\Rightarrow \frac{\delta b_p}{\delta\qc p} & = 2\begin{bmatrix} (p_0h + h\times p) & (p^Th + ph^T - hp^T  +p_0[h\times]) \end{bmatrix} \nonumber\\
 & = \nabla_p b_p . \label{eqn:dbp}
\end{align}
The above equation yields the covariance $B_p$ of the predicted vector measurement in terms of the covariance $\Pi$ of the predicted attitude estimate.
\begin{align}
B_p & = \nabla_p b_p \Pi \nabla_p^T b_p . \label{eqn:Bp}
\end{align}

\ifx\qver\qdetailed
As an aside, it may be noted that $\nabla_p b_p\nabla_p^T b_p = 1_{3\times 3}$ so that an isotropic noise in $\qc p$ (a diagonal $\Pi$) remains isotropic in $b_p$:
\begin{align}
& \nabla_p b_p\nabla_p^T b_p = \begin{bmatrix} (p_0h+h\times p) & (p^Th+ph^T-hp^T+p_0[h\times]) \end{bmatrix} \nonumber\\
 & \qquad \qquad \qquad \qquad \begin{bmatrix} p_0h^T+(h\times p)^T \\ p^Th+hp^T-ph^T-p_0[h\times] \end{bmatrix} \nonumber\\
 & = p_0^2 \left\{hh^T-[h\times][h\times]\right\} \nonumber\\
 & \qquad + p_0 \left\{h(h\times p)^T + (h\times p)h^T + [h\times](hp^T-ph^T)\right. \nonumber\\
 & \qquad\qquad \left. - (ph^T-hp^T)[h\times]\right\} \nonumber\\
 & \qquad + (h\times p)(h\times p)^T + (p^Th)^2 - (ph^T-hp^T)^2 \nonumber\\
 & = p_0^2 (hh^T-hh^T+1) \nonumber\\
 & \qquad + p_0 \left\{hh^T[p\times] + (h\times p)h^T - [h\times p]h^T + hp^T[h\times]\right\} \nonumber\\
 & \qquad + \|h\times p\|^2 + (p^Th)^2 \nonumber\\
 & = p_0^2 + p_0(0) + p^Tp = 1 , \nonumber
\end{align}
where we have used the vector identity
\begin{align}
(h\times p)(h\times p)^T & = \|h\times p\|^2 + (h^Tp)(hp^T + ph^T) \nonumber\\
 & \qquad - h^Thpp^T - p^Tphh^T . \nonumber
\end{align}
for any 3-vectors $h$ and $p$.
\fi

An expression for the covariance matrix $\Pi$ of the integrated estimate $\qc p$ may be obtained from the kinematic equation (\ref{eqn:pdotdisc}) for small time-steps.
\begin{align}
\Pi & = \Xi + \frac{T^2}{4} \begin{bmatrix} q_0 & -q^T \\ q & q_0+[q\times] \end{bmatrix} W \begin{bmatrix} q_0 & q^T \\ -q & q_0-[q\times] \end{bmatrix} , \label{eqn:Pi}
\end{align}
where $\Xi$ and $W$ are the covariances of the attitude estimate at the previous time step and the angular velocity measurement.

The filtered vector measurement is given by fusing the two estimates:
\begin{align}
b_f & = (B + B_p)^{-1}(Bb_p + B_pb) , \label{eqn:bfilt}
\end{align}
where $B$ and $B_p$ are covariance matrices corresponding to the actual vector measurement $b$ and the predicted vector measurement $b_p$. The covariance matrix of the fused measurement is:
\begin{align}
B_f & = (B + B_p)^{-1}(BB_pB + B_pBB_p)(B + B_p)^{-1} . \label{eqn:Bfilt}
\end{align}

\ifx\qver\qdetailed
For a constant reference vector $\*h$, and constant isotropic noise in the vector and angular velocity measurements, the matrices may be replaced by their scalar equivalents, yielding the asymptotic limit:
\begin{align}
\Pi_\perp & = \Sigma_\perp + \frac{WT^2}{4},\; B_p = 4\Pi_\perp,\; B_f = \frac{B_pB}{B_p+B},\; \Sigma_\perp = \frac{B_f}{4} , \nonumber\\
\Rightarrow \Sigma_\perp & = \frac{B_f}{4} = \sqrt{\frac{W^2T^4}{64}+\frac{BWT^2}{32}} - \frac{WT^2}{8} , \nonumber
\end{align}
where $\Sigma_\perp = (1 - \qc o\qc o^T)\Sigma(1 - \qc o\qc o^T)$ and $\Pi_\perp = (1 - \qc o\qc o^T)\Pi(1 - \qc o\qc o^T)$ are the orthogonal complements of $\Sigma$ and $\Pi$ with respect to the feasibility cone $Q_b$ corresponding to measurement $b$.
\fi

We now analyze the effect of independent, unbiased noise in the angular velocity measurement $\omega$ and vector measurement $b$ on the estimated attitude $\qc q$ \cite{Mohseni:18z}. In particular, we shall assume that there is no bias error in $\omega$. Further, we shall make the reasonable assumption that the errors are small enough relative to the norms of the quantities to consider them as perturbations, and therefore add the effects of individual noise sources to obtain the cumulative effect. The analysis in this section enables the derivation of a filtered projection from the integrated estimate, $\qc p$, onto the feasibility cone $Q_b$ corresponding to the vector measurement $b$, as presented in Theorem \ref{thm:snglvecratesoln}.

We shall introduce some new notation, to avoid lengthy expressions. The quaternion attitude estimate is given by equation (\ref{eqn:hzsoln2}):
\begin{gather}
\hspace*{-0.6cm} \sqrt{2(\alpha^2+\beta^2)(1+b_3)}\,\qc q = \begin{bmatrix} \alpha(1+b_3) \\ \alpha b_2 + \beta b_1 \\ -\alpha b_1 + \beta b_2 \\ \beta(1+b_3) \end{bmatrix} = \alpha\qc u + \beta\qc v \,, \label{eqn:hzsoln3}
\end{gather}
where $\qc u = [(1+b_3)\; b_2\; {-b_1}\; 0]^T$ and $\qc v = [0\; b_1\; b_2\; (1+b_3)]^T$ are scaled versions of the two special solutions from Lemma \ref{thm:snglvecssoln}, $\alpha = p_0(1+b_3) + p_1b_2 - p_2b_1 = \qc p^T\qc u$, and $\beta = p_1b_1 + p_2b_2 + p_3(1+b_3) = \qc p^T\qc v$.

Let us first consider the effect of noise in $\omega$ alone. Suppose the noise in $\omega$ leads to a small error in the integrated estimate $\delta \qc p = (T/2)\qc q\otimes\delta\qc \omega$ (refer equation (\ref{eqn:pdotdisc}) for a small $T$). The errors in $\alpha$, $\beta$ would then be:
\begin{align}
\begin{bmatrix} \delta\alpha \\ \delta\beta \end{bmatrix} & = \begin{bmatrix} 1+b_3 & b_2 & -b_1 & \\ & b_1 & b_2 & 1+b_3 \end{bmatrix} \delta\qc p = \begin{bmatrix} \qc u^T \\ \qc v^T \end{bmatrix} \delta \qc p \,. \nonumber
\end{align}

\begin{thm} \label{thm:pnoi}
In the absence of any other errors, a perturbation error $\delta\qc p$ in the integrated attitude estimate $\qc p$ leads to a perturbation in the vector-aligned attitude estimate $\qc q$ (equation (\ref{eqn:hzsoln2})) equal to the projection of $\delta\qc p$ onto the feasibility cone, {\it i.e.}, the subspace spanned by the two special solutions in lemma \ref{thm:snglvecssoln}, and orthogonal to the nominal attitude estimate.
\end{thm}
\begin{pf}
Taking differentials of equation (\ref{eqn:hzsoln3}):
\begin{gather}
\sqrt{2(\alpha^2+\beta^2)(1+b_3)}\,\delta \qc q = \left\{\begin{aligned} - \frac{(\alpha\delta\alpha+\beta\delta\beta)\sqrt{2(1+b_3)}} {\sqrt{\alpha^2+\beta^2}} \qc q \\ \quad + \qc u\delta\alpha + \qc v\delta\beta \end{aligned}\right. \,, \nonumber\\
\qquad = (1 - \sqrt{\frac{2(1+b_3)}{\alpha^2+\beta^2}} \qc q\qc p^T)(\qc u\delta\alpha + \qc v\delta\beta) \,. \label{eqn:noi1}
\end{gather}
Once we have expressed the error as the sum of first order differentials, the multiplying coefficients may now be approximated to their nominal values -- any error on account of the approximation would be multiplied by the differentials and therefore be of higher order. Specifically, we may approximate $\qc p \approx \qc q$, so $\qc p\otimes\qc b \approx \qc h\otimes\qc p$, so $\alpha \approx 2p_0 \approx 2q_0$, and $\beta \approx 2p_3 \approx 2q_3$, in the coefficients, to obtain
\ifx\qver\qdetailed
\begin{align}
\alpha^2 + \beta^2 & = 4q_0^2 + 4q_3^3 = 2(1 + b_3) \,, \nonumber\\
2(1+b_3)\delta\qc q & = (1 - \qc q\qc q^T) \begin{bmatrix} \qc u & \qc v \end{bmatrix} \begin{bmatrix} \delta\alpha \\ \delta\alpha \end{bmatrix} \,, \nonumber\\
 & = (1 - \qc q\qc q^T)\begin{bmatrix} \qc u & \qc v \end{bmatrix} \begin{bmatrix} \qc u^T \\ \qc v^T \end{bmatrix} \delta\qc p \,, \nonumber\\
 & = (1-\qc q\qc q^T)(\qc u\qc u^T + \qc v\qc v^T)\delta\qc p \nonumber\\
\delta\qc q & = (1-\qc q\qc q^T)(\qc r\qc r^T + \qc s\qc s^T)\delta\qc p = \qc o\qc o^T\delta \qc p \,, \label{eqn:noi2}
\end{align}
\else
\begin{align}
\delta\qc q & = (1-\qc q\qc q^T)(\qc r\qc r^T + \qc s\qc s^T)\delta\qc p = \qc o\qc o^T\delta \qc p \,, \label{eqn:noi2}
\end{align}
\fi
where $\qc o \in Q_b$, and $\qc o = (-\qc r + \kappa \qc s)/\sqrt{1 + \kappa^2} = \qc h \otimes \qc q$. \hspace*{\fill} \qed
\end{pf}

A similar but tedious derivation in \cite{Mohseni:18z} yields the following theorem for noise in the vector measurement $b$. We shall reuse some of the previous notation leading to theorem \ref{thm:pnoi} and equation (\ref{eqn:hzsoln3}).
\begin{thm} \label{thm:bnoi}
In the absence of any other errors, a perturbation error $\delta\qc b$ in the vector measurement $\qc b$ leads to a perturbation in the vector-aligned attitude estimate $\qc q$ (equation (\ref{eqn:hzsoln2})) equal to a rotation through the angle $-b\times\delta b$, which is the smallest angle rotation that takes $b$ to $b+\delta b$.
\end{thm}
\begin{pf}
%\begin{align}
%\begin{bmatrix} -q_1b_1 - q_2b_2 - q_3b_3 \\ q_0b_1 - q_3b_2 + q_2b_3 \\ q_0b_2 + q_3b_1 - q_1b_3 \\ q_0b_3 - q_2b_1 + q_1b_2 \end{bmatrix} & = \begin{bmatrix} -q_3 \\ -q_2 \\ q_1 \\ q_0 \end{bmatrix} \,. \nonumber
%\end{align}
Taking differentials of equation (\ref{eqn:hzsoln3}):
\begin{align}
\left.\begin{aligned} \sqrt{2(\alpha^2+\beta^2)(1+b_3)}\delta \qc q \\ + \qc q\sqrt{2(1+b_3)}\frac{\alpha\delta\alpha + \beta\delta\beta}{\sqrt{\alpha^2+\beta^2}} \\ + \qc q\sqrt{2(\alpha^2+\beta^2)}\frac{\delta b_3}{2\sqrt{1+b_3}} \end{aligned}\right\} & = \left\{\begin{aligned} \delta\alpha\qc u + \delta\beta\qc v \\ + \alpha\delta\qc u + \beta\delta\qc v \end{aligned}\right. . \label{eqn:noi3}
\end{align}
\ifx\qver\qelsaut
Similar to the proof of theorem \ref{thm:pnoi}, the coefficients multiplying the first order differentials are approximated to their nominal values, ultimately yielding
\begin{align}
\delta\qc q & = -\frac{1}{2}\qc q\otimes\qc b\otimes\delta\qc b = -\frac{1}{2}\qc o\otimes\delta\qc b \,, \label{eqn:noi8}
\end{align}
where $\qc o = \qc h\otimes\qc q = [(-q_3)\; (-q_2)\; q_1\; q_0]^T = \qc q \otimes \qc b$, and $\qc q\otimes\qc b$ is already orthogonal to $\qc q$. \hspace*{\fill} \qed
\else
Similar to the proof of theorem \ref{thm:pnoi}, the coefficients multiplying the first order differentials are approximated to their nominal values, ultimately yielding
\begin{align}
\alpha^2 + \beta^2 & = 4q_0^2 + 4q_3^2 = 2(1+b_3) \,, \nonumber\\
%\begin{bmatrix} \delta\alpha \\ \delta\beta \end{bmatrix} & = \begin{bmatrix} -p_2 & p_1 & p_0 \\ p_1 & p_2 & p_3 \end{bmatrix}\delta b = \begin{bmatrix} -q_2 & q_1 & q_0 \\ q_1 & q_2 & q_3 \end{bmatrix}\delta b \,, \nonumber\\
\begin{bmatrix} \alpha & \beta \end{bmatrix} \begin{bmatrix} \delta\alpha \\ \delta\beta \end{bmatrix} & = \qc p^T\begin{bmatrix} \qc u & \qc v \end{bmatrix} \begin{bmatrix} \delta\alpha \\ \delta\beta \end{bmatrix} = \qc q^T\begin{bmatrix} \qc u & \qc v \end{bmatrix} \begin{bmatrix} \delta\alpha \\ \delta\beta \end{bmatrix} \,. \label{eqn:noi4}
\end{align}
Working on the $\delta\qc u$ and $\delta\qc v$ terms,
\begin{align}
\qc q^T(\alpha\delta\qc u + \beta\delta\qc v) & = \qc q^T(2q_0\delta\qc u + 2q_3\delta\qc v) \,, \nonumber\\
 & = 2\qc q^T\left\{q_0 \begin{bmatrix} & & 1 \\ & 1 & \\ -1 & & \\ & & \end{bmatrix} + q_3 \begin{bmatrix} & & \\ 1 & & \\ & 1 & \\ & & 1 \end{bmatrix} \right\}\delta b \,, \nonumber\\
 & = 2(q_0\begin{bmatrix} -q_2 & q_1 & q_0 \end{bmatrix} + q_3\begin{bmatrix} q_1 & q_2 & q_3 \end{bmatrix})\delta b \,, \nonumber\\
 & = \begin{bmatrix} b_1 & b_2 & (1+b_3) \end{bmatrix}\delta b = \delta b_3 \,. \label{eqn:noi5}
\end{align}
Substituting from equations (\ref{eqn:noi4}, \ref{eqn:noi5}) back in equation (\ref{eqn:noi3}),
\begin{align}
\left.\begin{aligned} && 2(1+b_3)\delta\qc q + \qc q\qc q^T(\qc u\delta\alpha + \qc v\delta\beta) \\ && + \qc q\qc q^T(\alpha\delta\qc u + \beta\delta\qc v) \end{aligned}\right\} & = \left\{\begin{aligned} & \qc u\delta\alpha + \qc v\delta\beta \\ & \; + \alpha \delta\qc u + \beta\delta\qc v \end{aligned}\right. \,, \nonumber
\end{align}
It can be seen that the terms on the RHS are projected onto $\qc q$ and the projection appears on the LHS. This is just a consequence of the fact that $\qc q$ has unit magnitude, and therefore $\delta\qc q$ must be orthogonal to $\qc q$:
\begin{align}
2(1+b_3)\delta\qc q & = (1 - \qc q\qc q^T)\left(\qc u\delta\alpha + \qc v\delta\beta + \alpha \delta\qc u + \beta\delta\qc v\right) \,. \label{eqn:noi6}
\end{align}
We now simplify the terms within the parantheses on the RHS using the relations that $b^T\delta b = 0$ and $\qc q\otimes \qc b = \qc h \otimes \qc q$:
\begin{align}
& \qc u\delta\alpha + \qc v\delta\beta + \alpha\delta\qc u + \beta\delta\qc v \nonumber\\
 & = \begin{bmatrix} -q_2(1+b_3) & q_1(1+b_3) & q_0(1+b_3) \\ (-q_2b_2 + q_1b_1) & (q_1b_2 + q_2b_1) & (q_0b_2 + q_3b_1) \\ (q_2b_1 + q_1b_2) & (-q_1b_1 + q_2b_2) & (-q_0b_1 + q_3b_2) \\ q_1(1+b_3) & q_2(1+b_3) & q_3(1+b_3) \end{bmatrix} \delta b \nonumber\\
 & \quad + 2q_0 \begin{bmatrix} & & 1 \\ & 1 & \\ -1 & & \\ & & \end{bmatrix}\delta b + 2q_3 \begin{bmatrix} & & \\ 1 & & \\ & 1 & \\ & & 1 \end{bmatrix}\delta b \,, \nonumber\\
 & = \begin{bmatrix} -q_2(1+b_3) & q_1(1+b_3) & 0 \\ (2q_3 - q_2b_2 + q_1b_1) & (2q_0 + q_1b_2 + q_2b_1) & 0 \\ (-2q_0 + q_2b_1 + q_1b_2) & (2q_3 - q_1b_1 + q_2b_2) & 0 \\ q_1(1+b_3) & q_2(1+b_3) & 0 \end{bmatrix} \delta b \nonumber\\
 & \quad + \begin{bmatrix} 0 & 0 & q_0(1+b_3) \\ 0 & 0 & (q_0b_2 + q_3b_1) \\ 0 & 0 & (-q_0b_1 + q_3b_2) \\ 0 & 0 & q_3(1+b_3) \end{bmatrix} \delta b + 2\begin{bmatrix} q_0 \\ \\ \\ q_3 \end{bmatrix}\delta b_3 \,, \nonumber
\intertext{(Using $\qc q\otimes\qc b = \qc h\otimes\qc q$)}
 & = \begin{bmatrix} -q_2(1+b_3) & q_1(1+b_3) & 0 \\ (2q_1b_1 + q_3(1 + b_3)) & (2q_1b_2 + q_0(1 + b_3)) & 0 \\ (2q_2b_1 - q_0(1 + b_3)) & (2q_2b_2 + q_3(1 + b_3)) & 0 \\ q_1(1+b_3) & q_2(1+b_3) & 0 \end{bmatrix} \delta b \nonumber\\
 & \quad + \begin{bmatrix} 0 & 0 & q_0(1+b_3) \\ 0 & 0 & (q_0b_2 + q_3b_1) \\ 0 & 0 & (-q_0b_1 + q_3b_2) \\ 0 & 0 & q_3(1+b_3) \end{bmatrix} \delta b + 2\begin{bmatrix} q_0 \\ \\ \\ q_3 \end{bmatrix}\delta b_3 \,, \nonumber
\intertext{(Using $b^T\delta b = 0$)}
 & = (1+b_3)\begin{bmatrix} -q_2 & q_1 & q_0 \\ q_3 & q_0 & -q_1 \\ -q_0 & q_3 &-q_2  \\ q_1 & q_2 & q_3 \end{bmatrix} \delta b \nonumber\\
 & \quad + \begin{bmatrix} 0 \\ q_1 - q_1b_3 + q_0b_2 + q_3b_1 \\ q_2 - q_2b_3 - q_0b_1 + q_3b_2 \\ 0 \end{bmatrix} \delta b_3 + 2\begin{bmatrix} q_0 \\ \\ \\ q_3 \end{bmatrix}\delta b_3 \,, \nonumber
\intertext{(Again using $\qc q\otimes\qc b = \qc h\otimes\qc q$)}
 & = (1+b_3)\begin{bmatrix} -q_2 & q_1 & q_0 \\ q_3 & q_0 & -q_1 \\ -q_0 & q_3 &-q_2  \\ q_1 & q_2 & q_3 \end{bmatrix} \delta b + 2\begin{bmatrix} q_0 \\ q_1 \\ q_2 \\ q_3 \end{bmatrix}\delta b_3 \,, \nonumber\\
 & = (1+b_3) \begin{bmatrix} -q_2 & q_1 & q_0 \\ q_3 & q_0 & -q_1 \\ -q_0 & q_3 &-q_2  \\ q_1 & q_2 & q_3 \end{bmatrix} \delta b + 2\qc q\delta b_3 \,. \label{eqn:noi7}
\end{align}
Substituting from equation (\ref{eqn:noi7}) back in equation (\ref{eqn:noi6}), we obtain:
\begin{align}
\delta\qc q & = -\frac{1}{2}\qc q\otimes\qc b\otimes\delta\qc b = -\frac{1}{2}\qc o\otimes\delta\qc b \,, \label{eqn:noi8}
\end{align}
where $\qc o = \qc h\otimes\qc q = [(-q_3)\; (-q_2)\; q_1\; q_0]^T = \qc q \otimes \qc b$, and $\qc q\otimes\qc b$ is already orthogonal to $\qc q$. \hspace*{\fill} \qed
\fi
\end{pf}

\ifx\qver\qdetailed
A quick consistency check may be obtained using equation \ref{eqn:bconstr1}:
\begin{align}
\qc q\otimes \qc b & = \qc h \otimes \qc q , \nonumber\\
\Rightarrow \delta\qc q \otimes \qc b + \qc q \otimes \delta \qc b & = \qc h \otimes \delta \qc q . \nonumber
\end{align}
Checking equation \ref{eqn:noi8},
\begin{align}
-(1/2)\qc q\otimes \qc b \otimes \delta \qc b \otimes \qc b + \qc q \otimes \delta \qc b + (1/2)\qc h \otimes \qc q \otimes \qc b \otimes \delta \qc b & \overset{?}{=} 0 , \nonumber\\
\Leftarrow (1/2)\qc q\otimes \qc b \otimes \qc b \otimes \delta \qc b + \qc q \otimes \delta \qc b + (1/2)\qc q \otimes \qc b \otimes \qc b \otimes \delta \qc b & \overset{?}{=} 0 , \nonumber\\
\Leftarrow -(1/2)\qc q \otimes \delta \qc b + \qc q \otimes \delta \qc b - (1/2)\qc q \otimes \delta \qc b & \overset{?}{=} 0 . \checkmark \nonumber
\end{align}
\fi

Equations (\ref{eqn:pdotdisc}), (\ref{eqn:noi2}), (\ref{eqn:noi8}) can be used to derive an equation for the evolution of noise in the integrated and vector-aligned estimates.
\begin{align}
\delta\qc p_{i+1} & = \delta \qc q_i \otimes \left(\qc 1 + \frac{\qc\omega_i T}{2}\right) + \qc q_i \otimes \frac{\delta \qc\omega_i T}{2} = P_{i+1}\begin{bmatrix} \delta \qc q_i \\ \delta\omega_i \end{bmatrix} , \nonumber\\
%\delta_b\qc q & = -(1/2)\qc q\otimes \qc b \otimes \delta \qc b = -(1/2)\qc h \otimes \qc q \otimes \delta \qc b , \nonumber\\
%\delta_p\qc q & = \qc o\qc o^T \delta\qc p , \nonumber\\
\Rightarrow \delta\qc q_{i+1} & = \qc o_{i+1}\qc o_{i+1}^T\left[\delta\qc q_i \otimes\left(\qc 1 + \frac{\qc\omega_i T}{2}\right) + \qc q_i\otimes\frac{\delta\qc\omega_i T}{2}\right] \nonumber\\
 & \qquad - \frac{1}{2}\qc o_{i+1}\otimes \delta\qc b_{i+1} = Q_{i+1} \begin{bmatrix} \delta\qc q_i \\ \delta\omega_i \\ \delta b_{i+1} \end{bmatrix} , \label{eqn:noi9}
\end{align}
where $\delta \qc q_i$ is the noise in the attitude estimate at the previous time-step.
Equation (\ref{eqn:noi9}) can be used to derive expressions for the covariance matrices corresponding to $\qc p$ and $\qc q$, $\Pi$ and $\Xi$:%, which can be used in deriving an MMSE filtered estimate as done next.
\begin{align}
\Pi_{i+1} & = P_{i+1}\begin{bmatrix} \Xi_i & \\ & W_i \end{bmatrix}P_{i+1}^T , \nonumber\\
\Xi_{i+1} & = Q_{i+1}\begin{bmatrix} \Xi_i & & \\ & W_i & \\ & & B_{f,i+1} \end{bmatrix}Q_{i+1}^T , \label{eqn:noi10}
%\qc q_{f,i+1} & = (\Pi_{i+1}^{-1} + \Xi_{i+1}^{-1})^{-1}(\Pi_{i+1}^{-1}\qc p_{i+1} + \Xi_{i+1}^{-1}\qc q_{i+1}) , \label{eqn:noi10}
\end{align}
where $\Xi$, $W$, and $B_f$ are the covariance matrices corresponding to the attitude estimate $\qc q$, angular velocity measurement noise $\delta\omega$, and filtered vector measurement noise $\delta b_f$ respectively.

%It may be noted at this point that the errors in $\qc p$ and $\qc q$ are not completely independent. %, {\it i.e.}, the matrices $\Pi_{i+1}$ and $\Xi_{i+1}$ are not orthogonal to each other.
%The common portion of the errors arises on account of equation (\ref{eqn:noi2}), which shows that the projection of $\delta\qc p$ on $\qc o$, $\qc o\qc o^T\delta \qc p$, also appears in $\delta\qc q$, and cannot be filtered out. Another way to see this is by noting that the vector observation $\hat b$ provided no information for the attitude estimate within its feasibility cone $Q_b$, and the feasibility cone is accessed from $\qc q$ along $\qc o$. This common error, the orthogonal complement of $\delta\qc p$ with respect to $\qc o$, $(1 - \qc o\qc o^T)\delta\qc p$, and the error in $\qc q$ on account of $\delta b$, $-\qc o\otimes\delta \qc b/2$, are three independent errors in the two attitude estimates, $\qc p$ and $\qc q$. Of these, the latter two may be complementarily filtered in order to obtain an unbiased filtered estimate $\qc q_f$:
%\begin{align}
%\qc q_{f,i+1} & = (\Sigma_{i+1}^{-1} + B_{i+1}^{-1})^{-1} (\Sigma_{i+1}^{-1}\qc p_{i+1} + B_{i+1}^{-1}\qc q_{i+1}) , \label{eqn:noi11}
%\end{align}
%where, $\Sigma = (1 - \qc o\qc o^T)\Pi(1 - \qc o\qc o^T)$ is the covariance matrix of $(1 - \qc o\qc o^T)\qc p$.

% this section is maintained in the ICRA 2019 folder
\section {Observability and estimation of gyroscopic bias} \label{sec:bias}

The angular velocity of the body is measured to have components $\hat\omega = [\hat\omega_1\; \hat\omega_2\; \hat\omega_3]^T$ in the body coordinate system. This is the typical scenario in most applications, where the gyroscope is part of an Inertial measurement unit (IMU) that is fixed with respect to the body. However, the measured angular velocity $\hat\omega$ has an error with respect to the true quantity $\omega$. The angular velocity measurement error is assumed to be an Ornstein-Uhlenbeck process, with mean $\qol\omega$, time-constant $\tau$, and random-walk increments $\tilde\omega$:
\begin{align}
\hat\omega & = \omega + \qol\omega + \tilde\omega . \label{eqn:pnoi}
\end{align}
In this section, we consider the effects of gyroscopic bias on the geometric attitude estimation. Since the gyroscopic bias is exponentially autocorrelated with a time constant that is much larger than the time-step between measurements, this error manifests as a relatively low frequency source in comparison to the Gaussian noise considered in the previous section. The slow variation with time enables the design of an observer that could estimate the noise as well as compensate for it.

First, consider a bias error $\qol\omega$ that is constant with time. At each time-step, the estimate obtained by integrating the angular velocity is projected onto the feasibility cone corresponding to the vector measurement.
\begin{thm} \label{thm:biasestim}
In the absence of any other measurement errors, a fixed bias error in the angular velocity measurement can be completely estimated by applying theorem \ref{thm:snglvecratesoln} on two linearly independent vector measurements.
\end{thm}
\begin{pf}
The incremental change from the integrated attitude quaternion estimate, $\qc p$, to the vector-aligned estimate, $\qc q$, is essentially the correction to the integrated error in the angular velocity measurement $\qol\omega$. Denoting the increment by $\qc r$ in the body-fixed coordinate system (since $\hat\omega$ is available only in this system), for a constant $\qol\omega$ over a small integration time $T$, we must have:
\begin{align}
\qc r = \begin{bmatrix} 1 \\ \delta r \end{bmatrix} = \qc p^{-1}\otimes\qc q & = \begin{bmatrix} 1 \\ (\qol\omega T)/2 \end{bmatrix} + \delta\mu\qc b\,, \label{eqn:e1}
\end{align}
where $\delta\mu$ is an unknown infinitesimal rotation about $\hat b$ in the body system. We have assumed that we start on the feasibility cone, and integrate the angular velocity measurement over a small time, so $\qc r$ is close to $\qc 1$, and its scalar portion is approximately 1.
However, with a single vector measurement, a correction is possible only in the subspace orthogonal to the measured vector $\hat b$. Therefore, we have an unknown term proportional to $\qc b$ in equation (\ref{eqn:e1}). Projecting onto the subspace orthogonal to $\qc b$, we obtain $(1 - \hat b\hat b^T)\qol\omega = 2(1 - \hat b\hat b^T)\delta r/ T$ in the case of a correction onto the feasibility cone of a single measurement $\hat b$. Since $\hat b$ and $\delta r$ are known, this may be used to estimate the portion of $\qol\omega$ normal to $\hat b$. With two or more linearly independent measurements $\hat b_j$ and corrections $\delta r_j$ at a constant $\qol\omega$, the matrix $\sum_j(1 - \hat b_j\hat b_j^T)$ becomes invertible, and we can actually determine $\qol\omega$ completely:
\begin{align}
\sum_j(1 - \hat b_j\hat b_j^T)\qol\omega & = \sum_j 2(1 - \hat b_j\hat b_j^T)\delta r_j/ T,. \label{eqn:e3}
\end{align}
Thus, in the absence of any other measurement errors, a fixed bias error in the angular velocity measurement can be completely estimated using equation (\ref{eqn:e3}) on two linearly independent vector measurements. \hspace*{\fill} \qed
\end{pf}

\begin{rmk}
{\it Observability condition}:
The condition for invertibility of $\sum_i(1 - b_ib_i^T)$ is the same as the full-rank condition in literature, and for sequential measurements of a single vector observation, it is equivalent to the persistently non-parallel and sufficient excitation conditions. The condition can easily be checked by evaluating $\hat b_i^T \sum_{j=1}^i (1 - \hat b_j\hat b_j^T)$, since each of the terms in the summation is positive semi definite, and the inner product of $\hat b_i$ with the last term returns zero. Therefore, the summation is invertible if and only if its inner product with $\hat b_i$ is non-zero.
\end{rmk}

\begin{rmk}
{\it Non constant bias}:
If only measurements of a single constant vector are available, the body would have to rotate faster than the variation in $e$, if any such variation exists, for this estimation to be accurate. If $e$ does happen to vary significantly, we would only be estimating the weighted average of the error, $\qol e$, during the time over which the measurements were taken and the corrections determined:
\begin{align}
\sum_i(1 - b_ib_i^T)\qol e & = \sum_i2(1 - b_ib_i^T)\delta r_i/ T,. \label{eqn:e4}
\end{align}
\end{rmk}

Let us now allow variation in the bias error through the $\tilde\omega$ term (equation (\ref{eqn:pnoi})). If only measurements of a single constant vector are available, the variation in $\tilde\omega = \hat\omega - \omega - \qol\omega$ can cause the bias estimation of equation (\ref{eqn:e3}) to be inaccurate. In this case of time-varying bias $\tilde\omega$, we would be estimating the weighted average of the error, $\qol\omega + \tilde\omega$, during the time over which the measurements were taken and the corrections determined. For a uniformly distributed attitude, that would just be the constant bias error $\qol\omega$ in equation (\ref{eqn:e3}).
Equation (\ref{eqn:e3}) assigns equal weightage to all past measurements and corrections. This suggests a mechanism for estimating a slowly varying bias. Rather than weigh all past measurements equally, their influence on the current bias estimation may be progressively and gradually reduced (analogous to an infinite impulse response filter). This simulates a low pass filter on the attitude corrections whose bandwidth may be determined by the time constant $\tau$ of the autocorrelation of the bias error. For {\it e.g.}, if $\tau/T = 100$, then we could reduce the influence of past measurements by $1 - 1/100 = 0.99$ in each successive measurement. Increasing the attenuation factor towards 1 reduces the bandwidth of the bias estimator and lowers the noise in the estimation. Contrarily, reducing the attenuation factor towards 0 increases the bandwidth of the bias estimator, but at the cost of higher noise.
Such an estimator may be expressed in terms of the matrices $A_i$ and $B_i$, defined inductively, as shown below:
\begin{align}
A_{i+1} & = (1 - T/\tau)A_i + (T/\tau)(1 - \hat b_i\hat b_i^T) , \nonumber\\
B_{i+1} & = (1 - T/\tau)B_i + (1 - \hat b_i\hat b_i^T)2\delta r_i/\tau, \nonumber\\
& A_{i+1}(\qol\omega + \tilde\omega_{i+1}) = B_{i+1}, \label{eqn:e5}
\end{align}
with the initial conditions $A_0 = 0, B_0 = 0$. Note that $\qol\omega$ is a constant across the time-steps, and is the output of the estimator in the special limiting case when $\tau$ goes to infinity.

While equation (\ref{eqn:e5}) is sufficient to estimate the bias when the persistency-of-excitation condition is met, it may fail when the body stops rotating. The failure upon loss of excitation occurs as $\hat b_i$ approaches a limit, and the matrix $A_i$ gradually approaches the now constant $1 - \hat b_i\hat b_i^T$ over time, thus becoming singular. Failure may be avoided under such circumstances by updating only the components of $A_i$ and $B_i$ that have additional information in the new measurements, as done in the following estimator design:
\begin{align}
A_{i+1} & = (\hat b_i\hat b_i^T)A_i + (1 - \hat b_i\hat b_i^T)((1 - T/\tau)A_i + (T/\tau)) , \nonumber\\
B_{i+1} & = (\hat b_i\hat b_i^T)B_i + (1 - \hat b_i\hat b_i^T)((1 - T/\tau)B_i + 2\delta r_i/\tau) , \nonumber\\
& A_{i+1}(\qol\omega + \tilde\omega_{i+1}) = B_{i+1}, \label{eqn:e6}
\end{align}
The estimator of equation (\ref{eqn:e6}) tracks a time-varying bias equally as well as that in equation (\ref{eqn:e5}) under persistant excitation. However, it does not fail when excitation is lost. It provides the best estimate of the bias it could under the circumstances: tracking the components of the bias orthogonal to $\hat b_i$, while retaining the last best estimate for the component of bias along $\hat b_i$. The first order filtering can easily be extended to higher orders by including additional terms on the right hand side of equation (\ref{eqn:e6}) that invoke $A_{i-1}$, $B_{i-1}$ {\it etc}.

\section {Simulation results} \label{sec:ressim}

In this section, we use Matlab simulations to verify the key theoretical results derived in the previous section. The first group of simulations correspond to verifying the solution for the first problem -- attitude estimation using two vector measurements. We assume that the directions of two linearly independent vectors, $h$ and $k$, are measured at 100Hz in the body-fixed coordinate system as $a$ and $b$.
%For purposes of illustration, we choose $h$ to be along the reference coordinate system's $z$-axis, and $k$ to lie in the $xz$-plane at an angle of elevation of $-\pi/4$.
Measurements $a$ and $b$ are assumed to have random, unbiased noise of 0.01 and 0.02 normalized units respectively. The body is prescribed an oscillatory roll and pitch motion, and a constant yaw angle.

Figure \ref{fig:attGattT} shows the attitude estimated using theorem \ref{thm:twovecsoln}, $\qc q_G$, in comparison with the attitude derived by using the TRIAD method, $\qc q_T$, when reference vector $h$ is of greater significance. Both the solutions are identical upto machine precision. The figure on the left shows that the attitude follows a high-amplitude trajectory while the two solutions maintain equivalence.
\begin{figure} [!ht] \begin{center}
\begin{minipage}{0.49\linewidth} \begin{center}
\includegraphics [width=\linewidth] {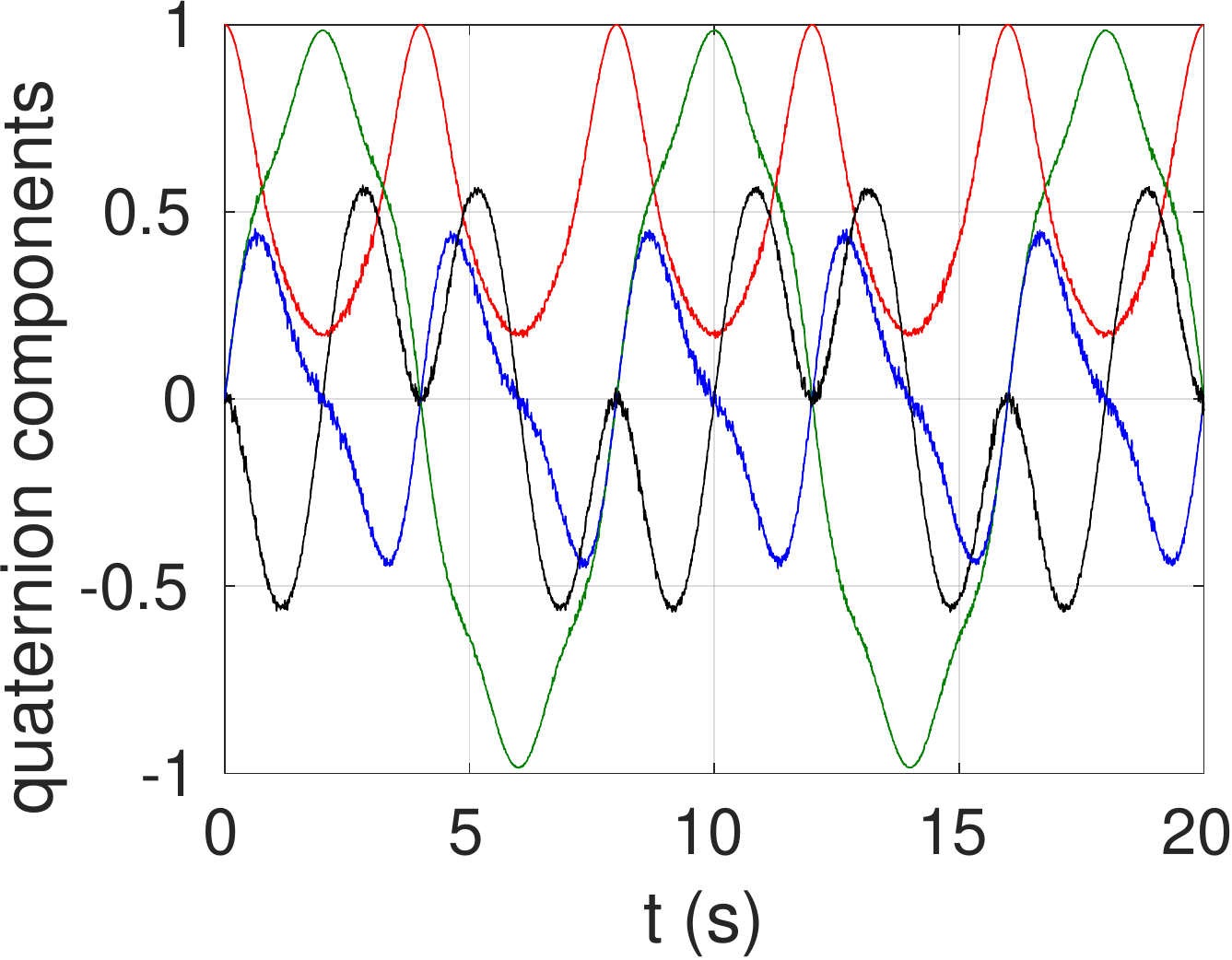}
\end{center}\end{minipage}
\begin{minipage}{0.49\linewidth} \begin{center}
\includegraphics [width=\linewidth] {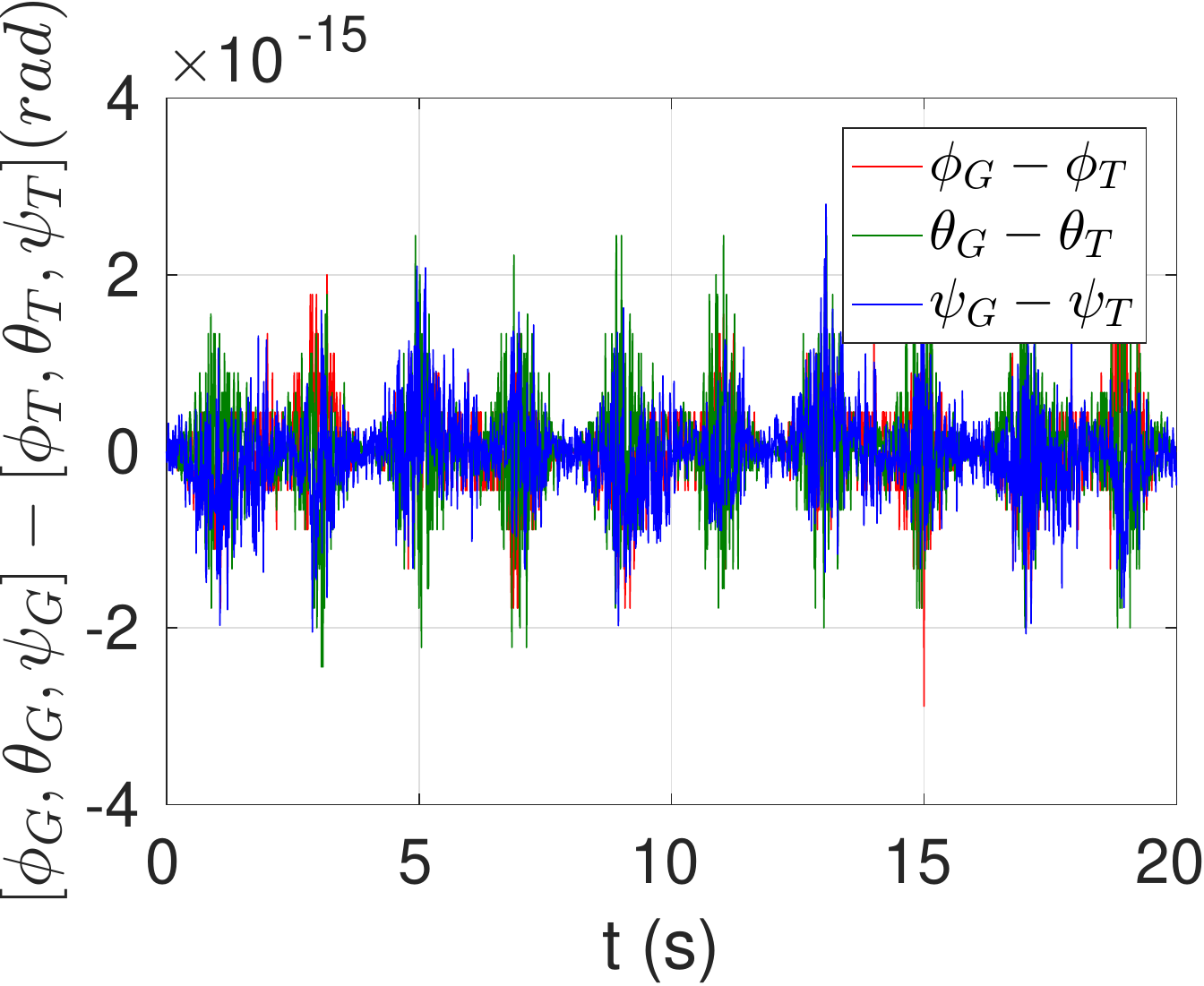}
\end{center}\end{minipage}
\caption{Matlab simulations of full attitude estimation using two vector measurements. The left figure shows the results of applying the TRIAD solution and using the geometric method of Theorem \ref{thm:twovecsoln}. The figure on the right shows that the two solutions are equal upto machine precision.}
\label{fig:attGattT}
\end{center} \end{figure}
%Figure \ref{fig:attGattT} (right) shows the attitude estimated using theorem \ref{thm:twovecsoln}, $\qc q_G$, in comparison with the attitude derived by using the TRIAD method, $\qc q_T$, when reference vector $h$ is of greater significance. Both the solutions are identical upto machine precision.
\begin{figure} [!ht] \begin{center}
\begin{minipage}{0.49\linewidth} \begin{center}
\includegraphics [width=\linewidth] {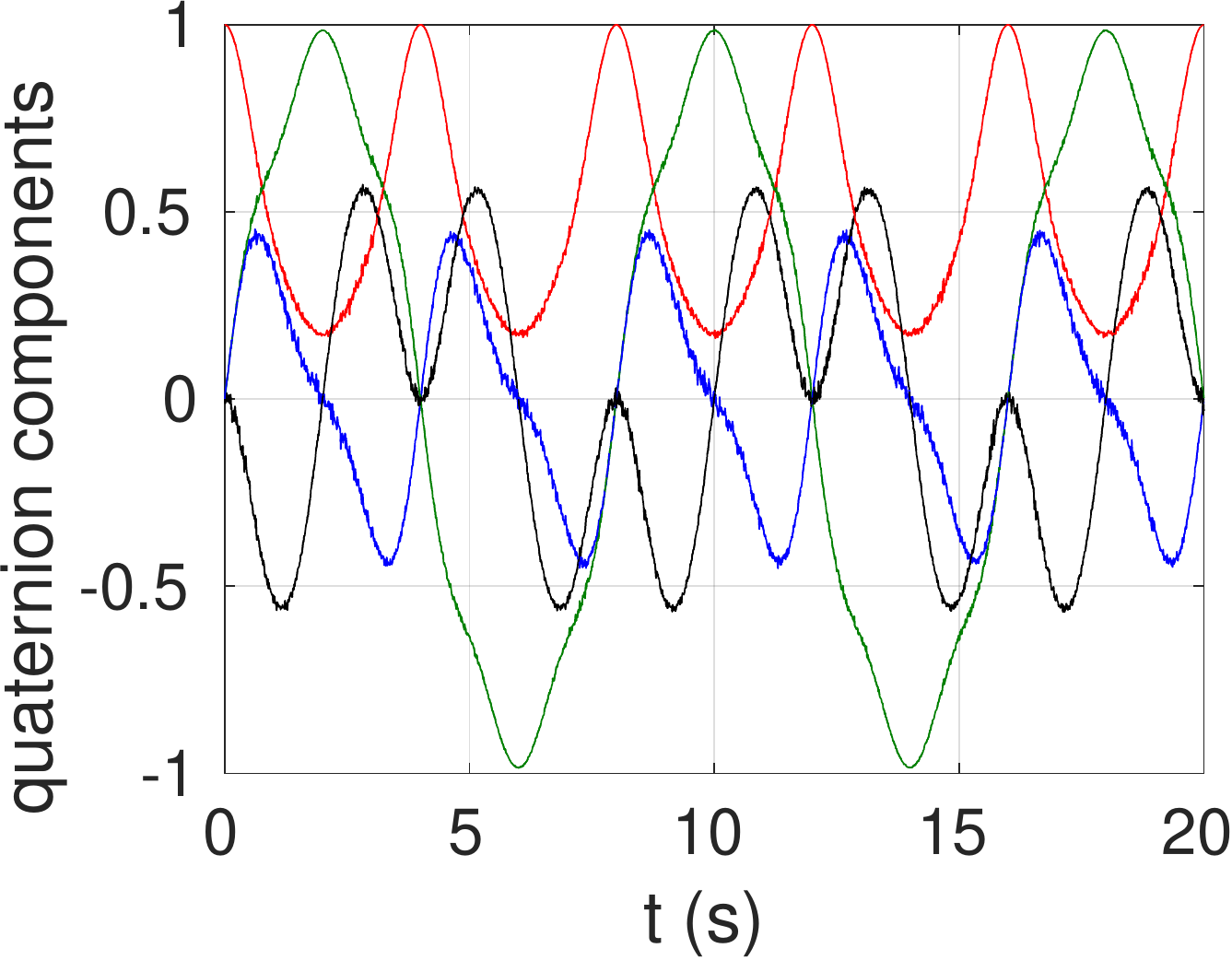}
\end{center}\end{minipage}
\begin{minipage}{0.49\linewidth} \begin{center}
\includegraphics [width=\linewidth] {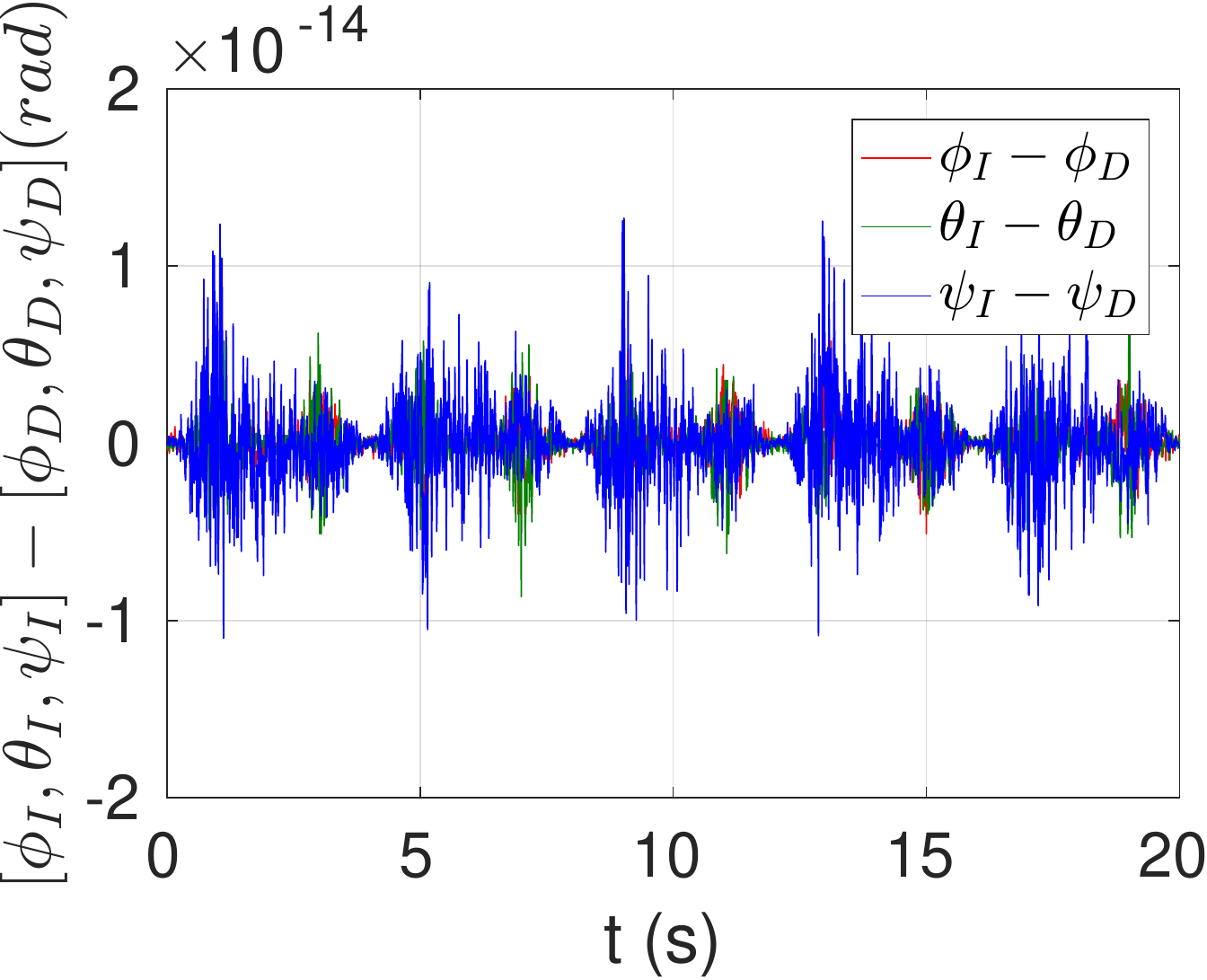}
\end{center}\end{minipage}
\caption{Matlab simulations of full attitude estimation using two vector measurements. The left figure shows the results of applying Davenport's $q$-method and an appropriate geometric filter using (\ref{eqn:interp1}). The figure on the right shows that the two solutions are equal upto machine precision.}
\label{fig:attIattD}
\end{center} \end{figure}

By using equation (\ref{eqn:interp1}) to interpolate between the two solutions obtained from theorem \ref{thm:twovecsoln}, we obtain the solution to Wahba's problem. The interpolation parameter $x$ is chosen to be $2^2/(1^2 + 2^2) = 0.8$, as the rms noise of the two vector measurements have a ratio of 2. Figure \ref{fig:attIattD} (right) shows the equivalence between the result obtained by interpolating (equation (\ref{eqn:interp1})) on the two estimates of theorem \ref{thm:twovecsoln}, $\qc q_I$, and that obtained by using Davenport's $q$-method, $\qc q_D$. The figure on the left shows that the attitude follows a high-amplitude trajectory while the two solutions maintain equivalence.

The next group of simulations verify the result of theorem \ref{thm:snglvecratesoln}.
In these simulations, we assume a constant gyroscopic bias of $[-0.32\;0.16\;{-0.08}]^T$rad/s along the three axes, and a random, unbiased noise of 0.04rad/s in each component. The reference vector components are assumed to be $h = [0\; 0\; 1]^T$. The vector measurement is also assumed to have a random, unbiased noise of 0.01 normalized units, but we assume any constant biases in this measurement have been eliminated. The vector measurement is then normalized before being passed on to the attitude estimator.% All error sources are exaggerated for illustrative purposes. The exact magnitude of the noise would vary depending upon the particular vector being measured, and the equipment being used to perform the measurement.

\begin{figure} [!ht] \begin{center}
\begin{minipage}{0.49\linewidth} \begin{center}
\includegraphics [width=\linewidth] {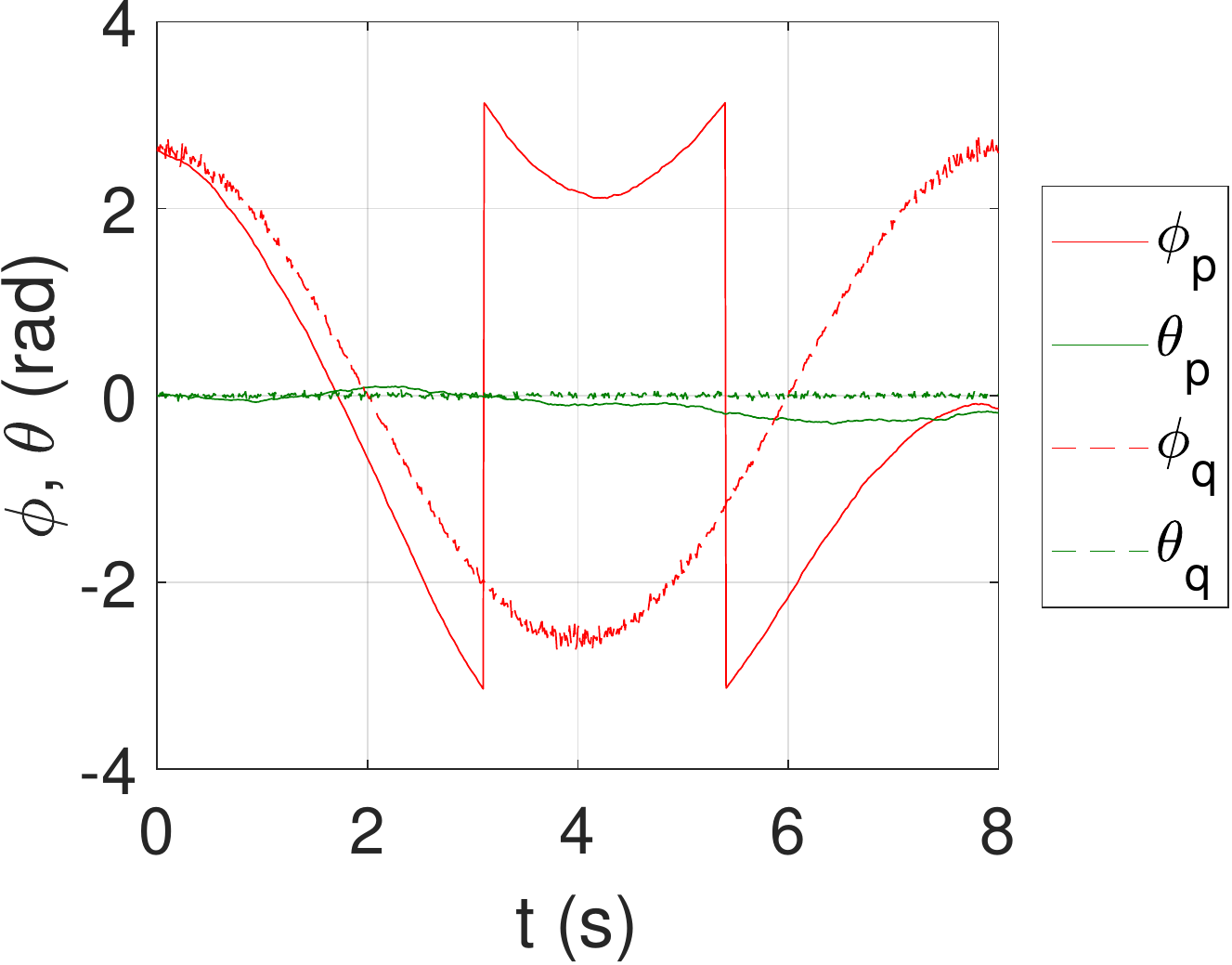}
\end{center}\end{minipage}
\begin{minipage}{0.49\linewidth} \begin{center}
\includegraphics [width=\linewidth] {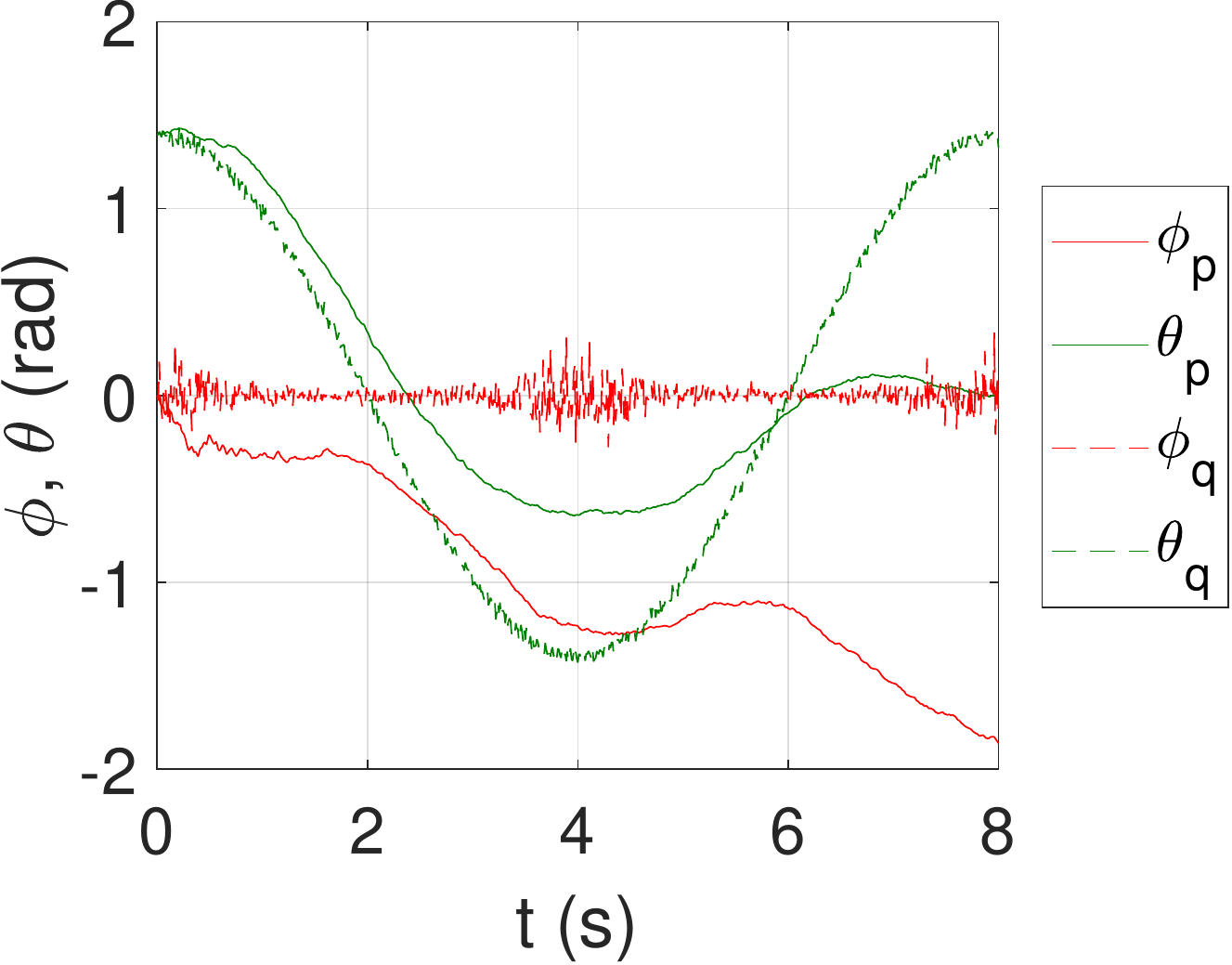}
\end{center}\end{minipage}
\caption{Simulated attitude estimation for pure sinusoidal roll (left) and pitch (right) manoeuvres. While the gyro integrated estimate drifts with time, the vector measurement correction (equation (\ref{eqn:hzsoln2})) realigns the roll and pitch angles at every time step.}
\label{fig:rollestim}
\end{center} \end{figure}
The quaternion output of the attitude estimator is converted to 3-2-1 Euler angles for ease of readability. The angular velocity is measured at 100Hz and integrated (along with the bias and noise errors) to return an integrated estimate for the attitude ($\phi_p$ and $\theta_p$ after conversion to roll and pitch Euler angles). Then, a corrected attitude is determined that is consistent with the noisy vector measurement, also at 100Hz, to yield the vector-aligned estimate ($\phi_q$ and $\theta_q$ respectively).

In this case of the reference vector being aligned with the $z$-axis, the attitude estimator cannot correct for errors on account of yaw drift in the integrated estimate $\qc p$. Therefore, we can evaluate the estimator's performance after isolating the roll and pitch angles from the estimate. The first plot (figure \ref{fig:rollestim} left) considers the case of a sinusoidal roll manoeuvre of amplitude $\pm5\pi/6$rad and frequency 0.25Hz. The second plot (figure \ref{fig:rollestim} right) repeats the simulation with a fixed roll angle and a sinusoidal pitch manoeuvre of amplitude $\pm4\pi/9$rad and frequency 0.25Hz. It can be seen that the integrated estimates drift with time, but the vector-aligned estimates, while having more noise, stay true to the actual values.

\begin{figure} [!ht] \begin{center}
\begin{minipage}{0.49\linewidth} \begin{center}
\includegraphics [width=\linewidth] {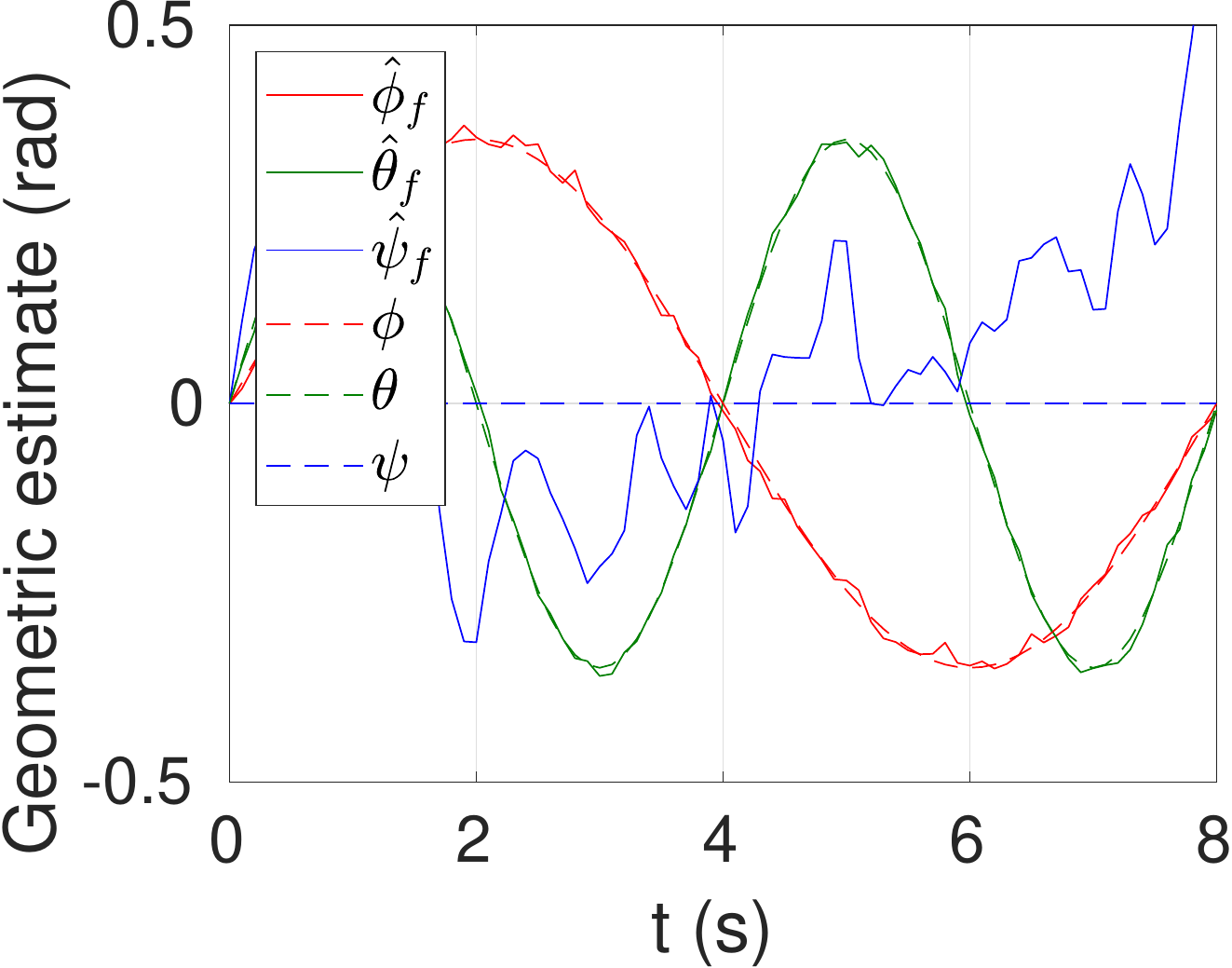}
\end{center}\end{minipage}
\begin{minipage}{0.49\linewidth} \begin{center}
\includegraphics [width=\linewidth] {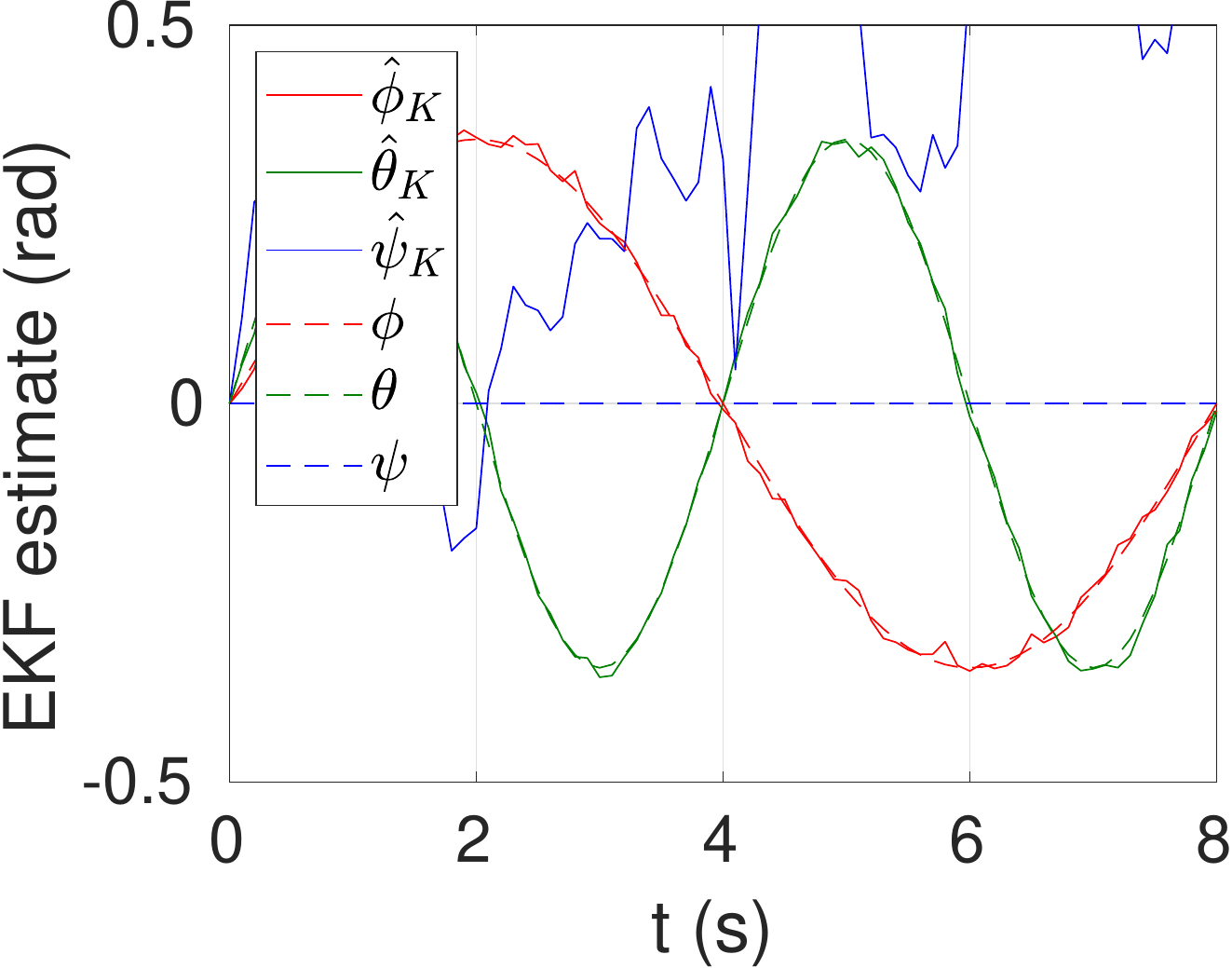}
\end{center}\end{minipage}
\begin{minipage}{0.49\linewidth} \begin{center}
\includegraphics [width=\linewidth] {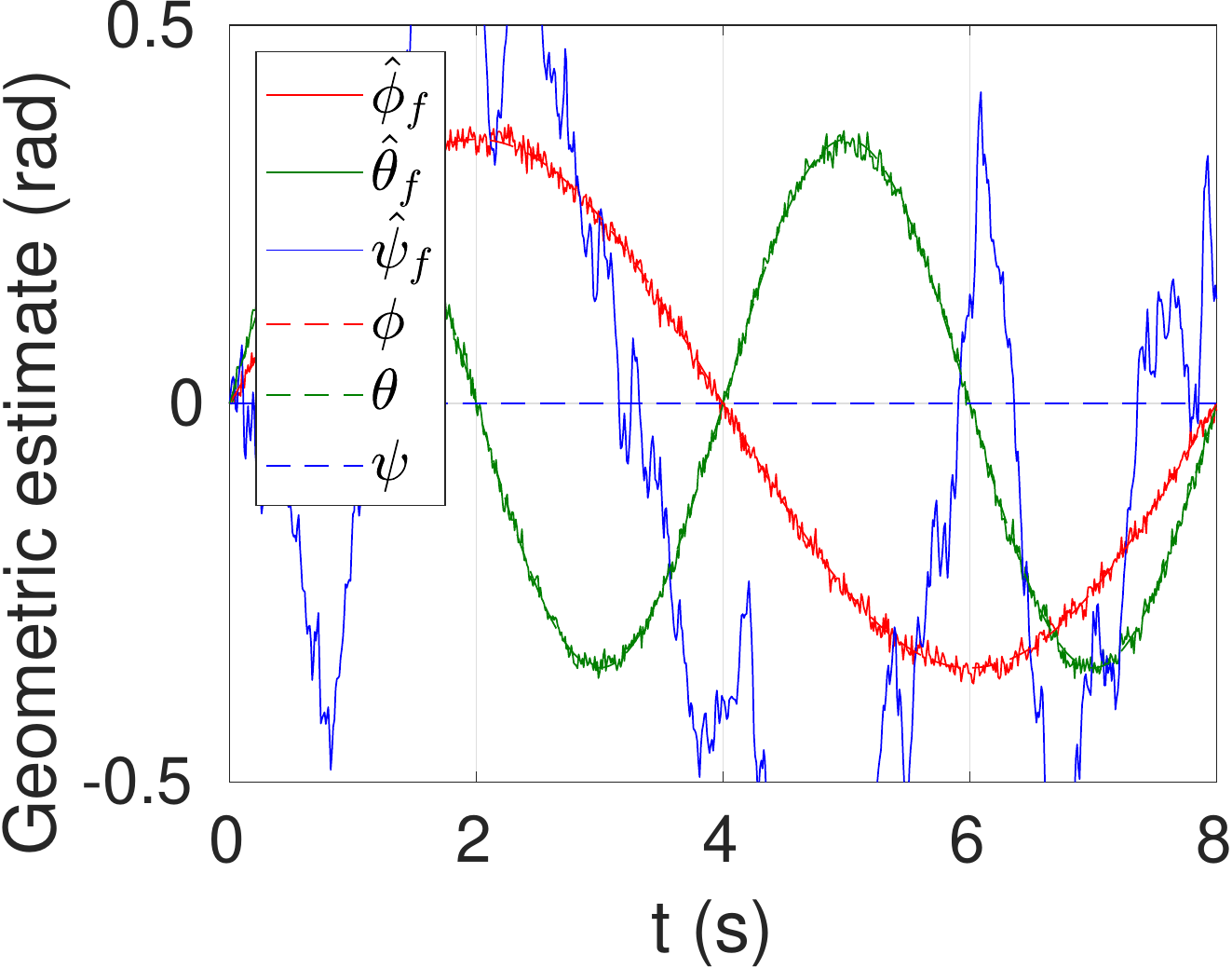}
\end{center}\end{minipage}
\begin{minipage}{0.49\linewidth} \begin{center}
\includegraphics [width=\linewidth] {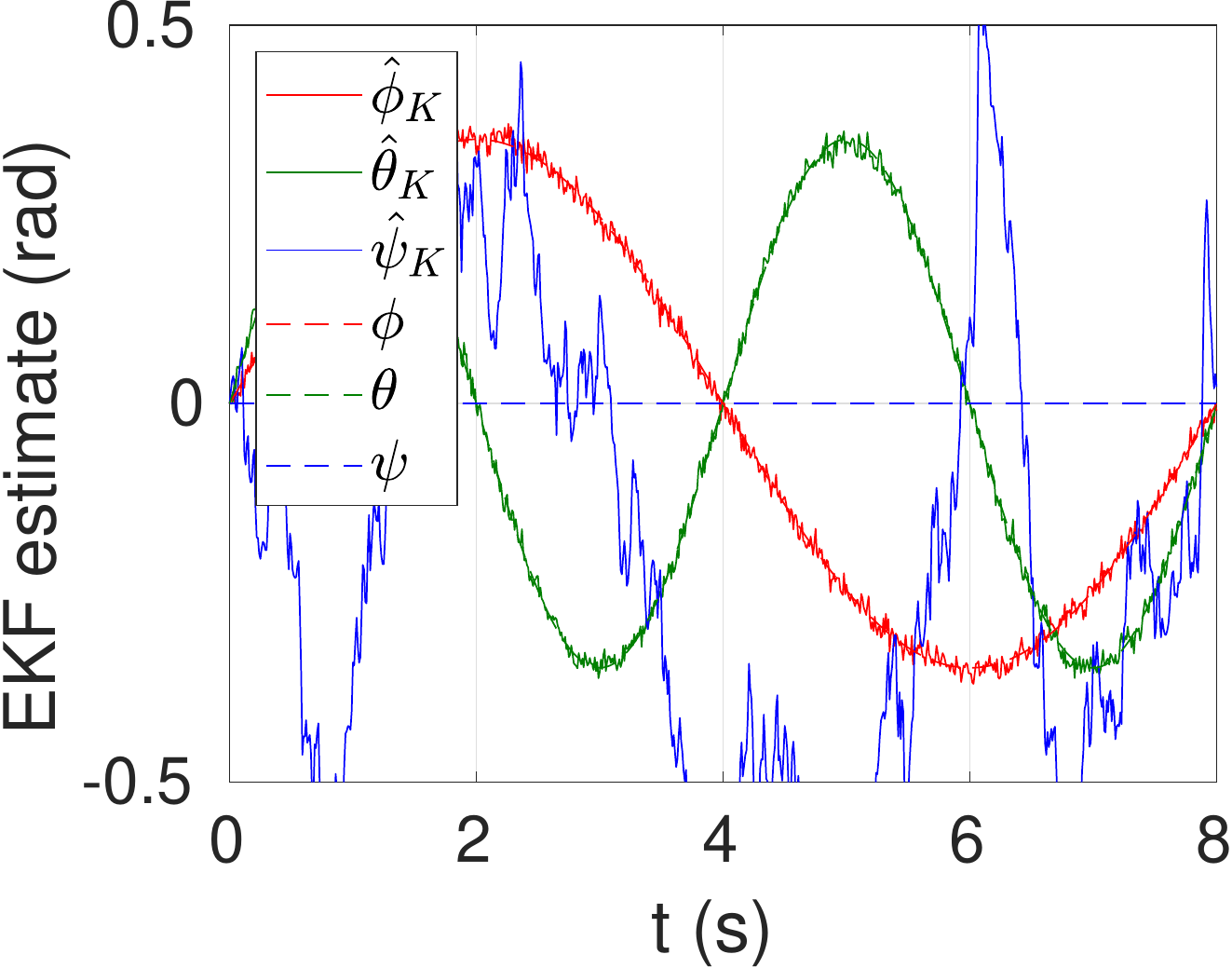}
\end{center}\end{minipage}
\caption{Filtering using equations (\ref{eqn:bfilt}), (\ref{eqn:Bfilt}), (\ref{eqn:noi10}) to obtain a filtered attitude estimate. The roll and pitch angles are prescribes to be sinusoids of amplitude $\pi/9$ rad.The filtered solution (top left) has lower errors than an optimally tuned EKF (top right) for large attitude corrections ($\approx 0.04$ units rms vector measurement noise) at each time-step. In the limit of smaller attitude increments ($\approx 0.01$ units rms noise), the EKF (bottom right) approaches the more accurate interpolated solution of equations (bottom left). The yaw estimates may be ignored for this comparison.}
\label{fig:rollpitchestim}
\end{center} \end{figure}
The attitude estimate $\qc q$ of Theorem \ref{thm:snglvecratesoln} can be filtered to reduce the noise, as decribed in equations (\ref{eqn:bfilt}), (\ref{eqn:Bfilt}), (\ref{eqn:noi10}). For small noise in the vector measurement and consequently small attitude corrections at each time-step, the filtered estimate is the same as that obtained using the traditional EKF, but the linearization inherent in the EKF begins to introduce significant errors for large corrections (figure \ref{fig:rollpitchestim}). In the top panel the vector measurement has a noise of rms $0.04$ units, while the noise is $0.01$ units in the bottom panel. While the variance of the error is similar with both the methods in the bottom panel (0.436e-4 sq-units with the EKF and 0.410e-4 sq-units with the geometric filter), it is 18\% lower with the geometric filter in the top panel (5.43e-4 sq-units with the EKF and 4.58e-4 sq-units with the geometric filter).

\begin{figure} [!ht] \centering
\begin{minipage}{0.49\linewidth}
\includegraphics [width=\linewidth] {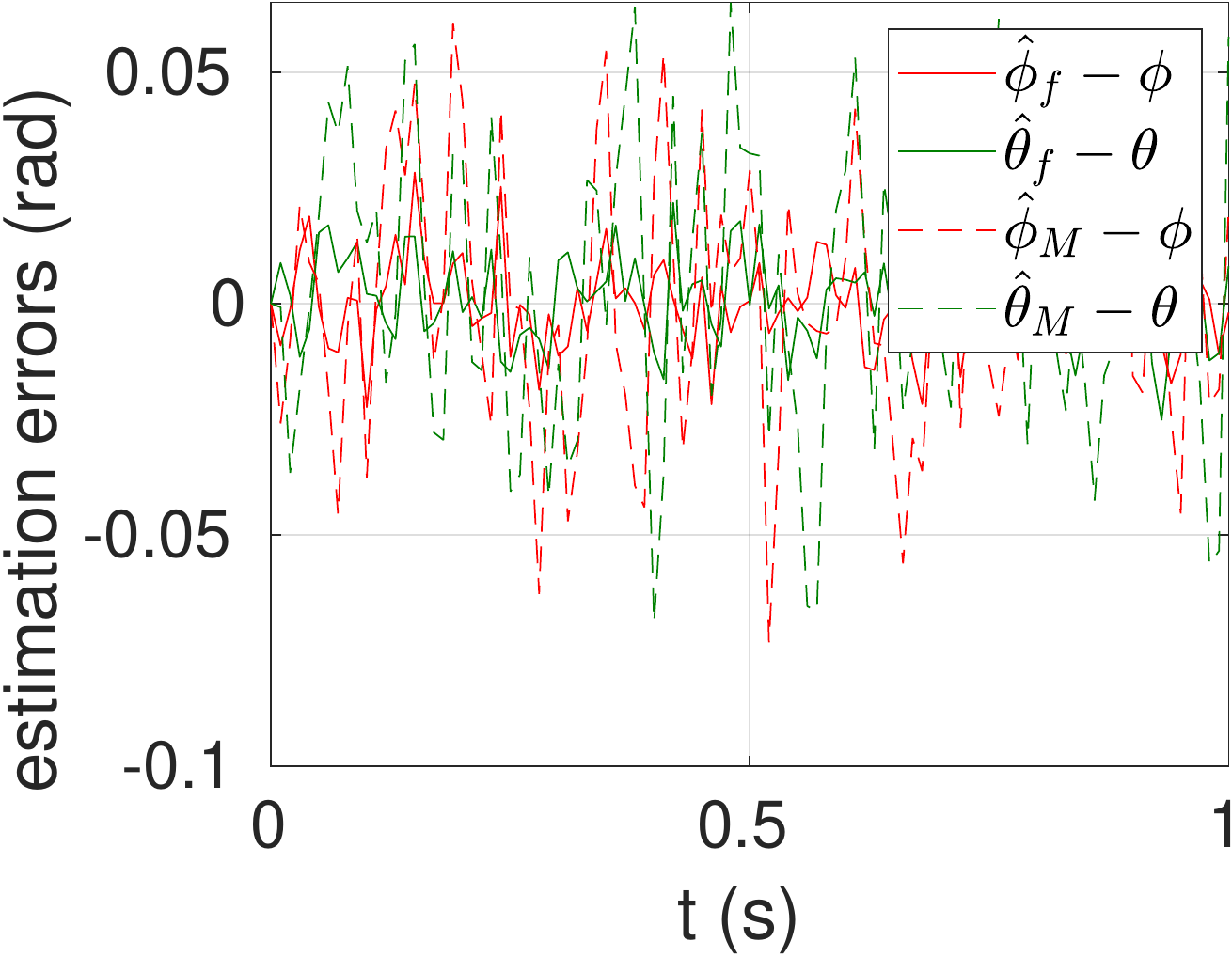}
\end{minipage}
\begin{minipage}{0.49\linewidth}
\includegraphics [width=\linewidth] {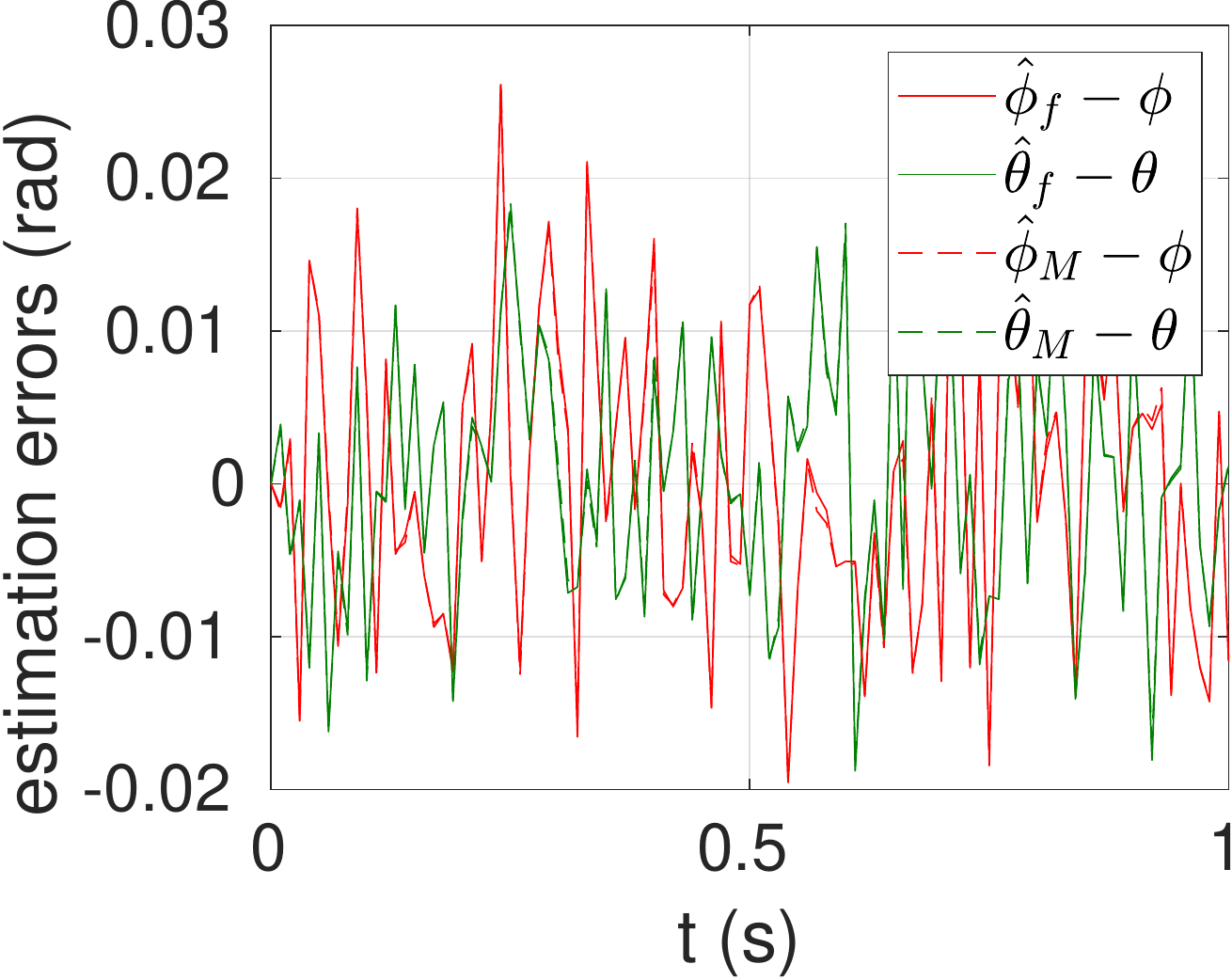}
\end{minipage}
\caption{A comparison of the estimator in Theorem \ref{thm:snglvecratesoln} against the ECF in \cite{Mahony:08a}. The time scale had to be zoomed in, in order to discern the differences between the two estimators. The ECF has larger residual errors unless we use the optimal gain suggested in this paper in a two-step estimation. Left: The ECF with gains recommended in \cite{Mahony:08a}. Right: the ECF using the gain derived in Remark \ref{thm:attYvsattM} in two-step estimation.}
\label{fig:attGvsattM}
\end{figure}

The attitude estimator in Theorem \ref{thm:snglvecratesoln} ($\phi_f$ and $\theta_f$) is compared with the ECF of \cite{Mahony:08a} ($\phi_M$ and $\theta_M$) in figure \ref{fig:attGvsattM}. The true attitude angles are denoted $\phi$ and $\theta$. The geometric filter provides superior accuracy to the ECF with the gains recommended in \cite{Mahony:08a}. Equivalent performance may be obtained with both the solutions only upon following a two-step attitude estimation in the ECF, and using the gains suggested in Remark \ref{thm:attYvsattM}. The two-step estimation is essential so as to ensure that the angular velocity correction $\omega_c$ is with respect to the filtered vector measurement $b_f$ obtained from the first step, and that the subsequent vector-measurement based correction is expressed in the body-frame obtained after integrating the angular velocity in the first step.

\section{Experimental validation of geometric attitude estimation using rate and single vector measurement} \label{sec:resexp}

This section provides experimental verification for the geometric attitude estimator by using a recently developed autopilot in our group, which is equipped with an IMU, the MPU9250, and is described in \cite{Mohseni:17a}. The autopilot is mounted on an inhouse designed model positioning system (MPS) that can independently prescribe roll, pitch, plunge and yaw manoeuvres on a test module.
\ifx\qdetailed
The 4 degree-of-freedom MPS is described by Linehan and Mohseni in \cite{Mohseni:14t}. A key enabling feature of the MPS is that it provides for both static and dynamic positioning of a mounted model, which is required to generate and measure a non-zero angular velocity.
\fi
\begin{figure} [!ht] \begin{center}
\begin{minipage}{0.49\linewidth} \begin{center}
\includegraphics [width=1.0\linewidth] {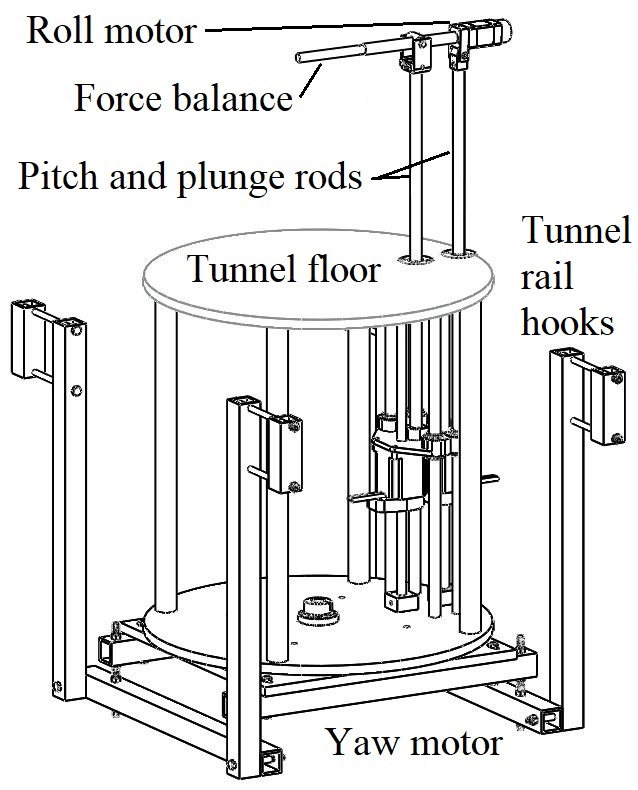}
\end{center}\end{minipage}
\begin{minipage}{0.49\linewidth} \begin{center}
\includegraphics [width=0.8\linewidth] {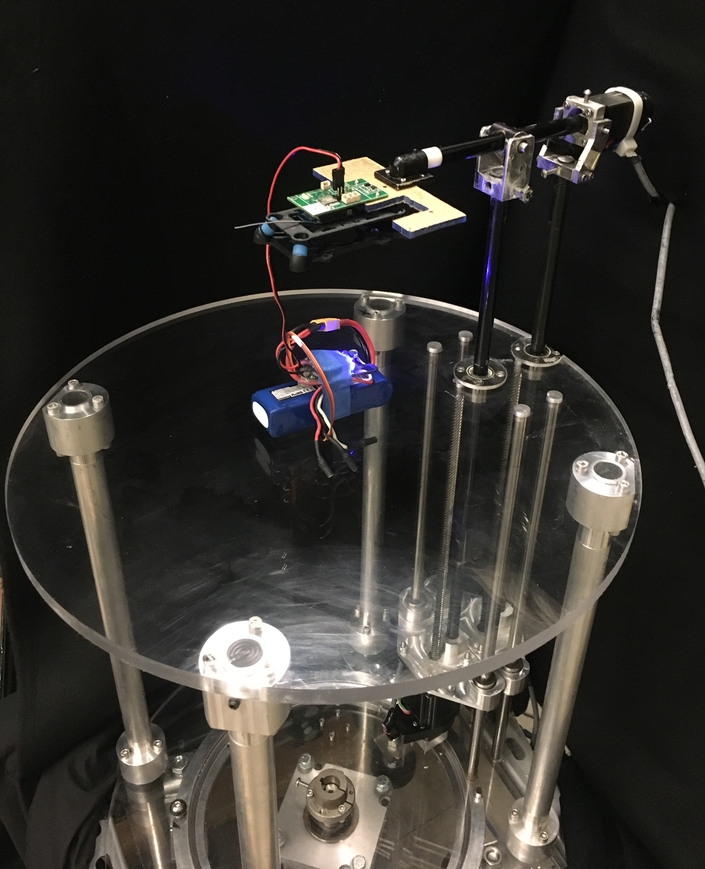}
\end{center}\end{minipage}
\caption{On the left, a schematic of the 4 Degree of freedom Model Positioning System (MPS) described in \cite{Mohseni:14t}. The MPU9250 mounted on the PCB (green in the picture on the right) and being tested on the MPS.}
\label{fig:mpsphoto}
\end{center} \end{figure}

\begin{figure} [!ht] \begin{center}
\ifx\qver\qelsaut
\begin{minipage}{0.49\linewidth}
\hspace*{-0.5cm}
\includegraphics [width=1.2\linewidth] {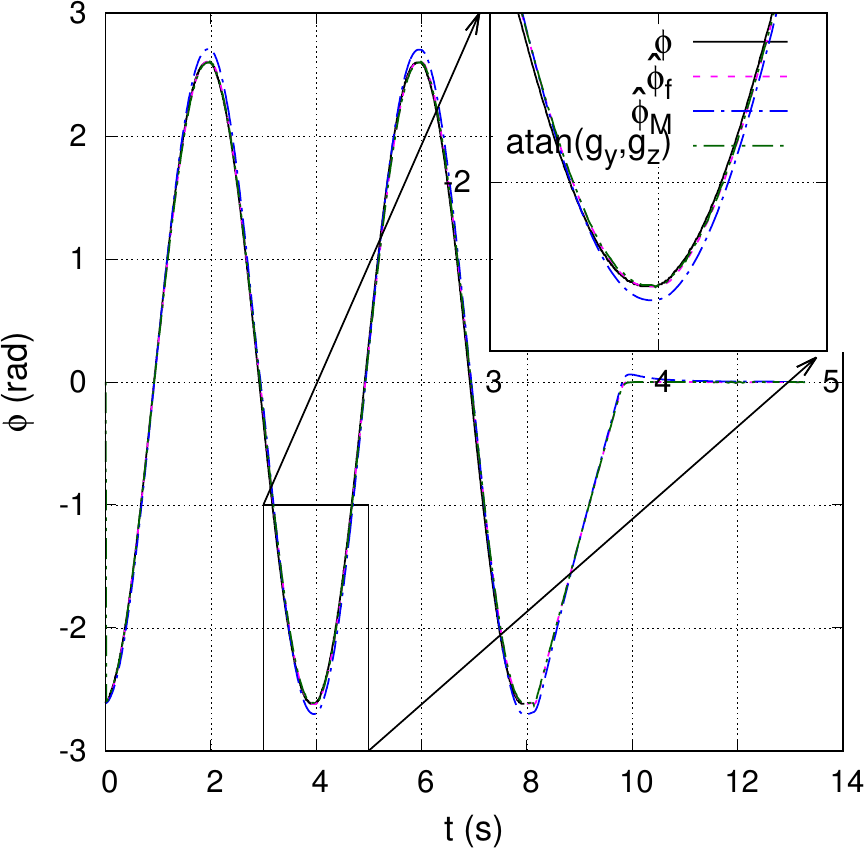}
\end{minipage}
\begin{minipage}{0.49\linewidth}
\hspace*{0.2cm}
\includegraphics [width=1.2\linewidth] {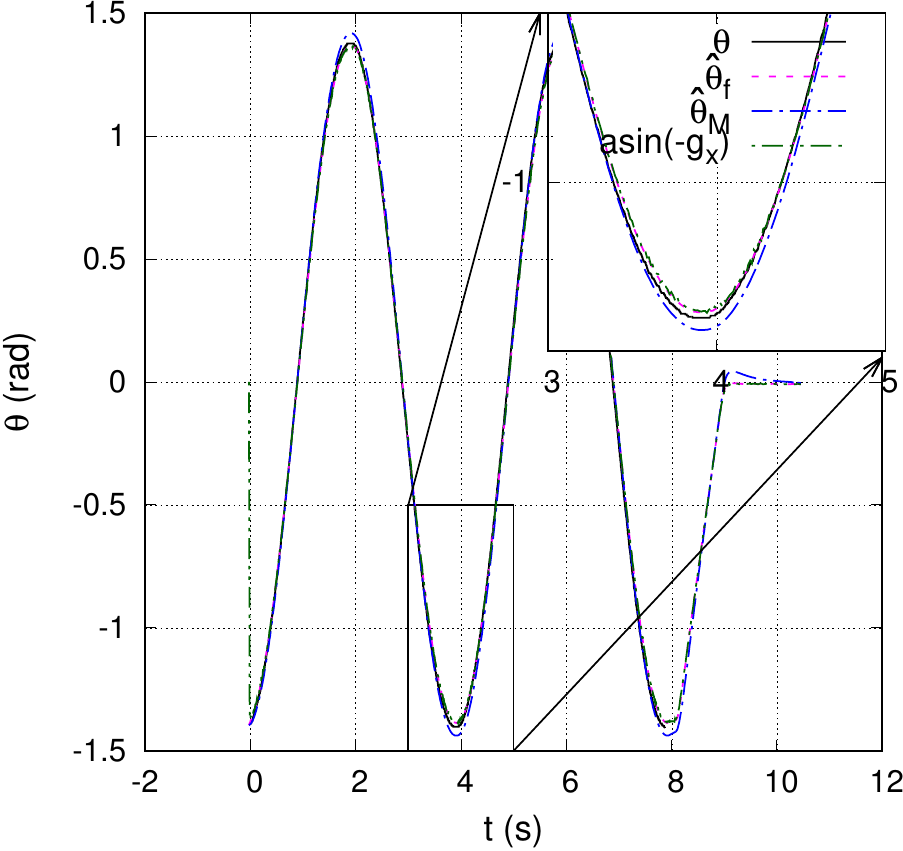}
\end{minipage}
\else
\begin{minipage}{0.6\linewidth}
\includegraphics [width=1.2\linewidth] {mpsattYattMroll.pdf}
\end{minipage}
\begin{minipage}{0.6\linewidth}
\includegraphics [width=1.2\linewidth] {mpsattYattMpitch.pdf}
\end{minipage}
\fi
\caption{Left: Attitude estimation for a pure sinusoidal roll manoeuvre on a real system. Right: Attitude estimation for a pure sinusoidal pitch manoeuvre on a real system. The solid black lines are the true roll and pitch angles returned by the encoder, and the dashed magenta curves are their estimates using Theorem \ref{thm:snglvecratesoln} presented in this paper after the filtering described in section \ref{sec:noifx}. The dash-dot blue curve shows the attitude estimate obtained using the ECF \cite{Mahony:08a}. The dash-dot-dot green line is the attitude consistent with the gravity vector measurement.}% The vector measurements, $g_x$, $g_y$ and $g_z$, are also included to show the instantaneous response from the measurement to the estimation.}
\label{fig:mpsroll}
\end{center} \end{figure}

\ifx\qdetailed
The roll and yaw motions are generated using stepper motors which can be programmed to rotate the model according to a prescribed trajectory. As the motors rotate, real-time measurement is provided using 517 counts-per-revolution differential encoders which provide feedback to the motor controller and also a record of the actual position of the tested model.
Pitch and plunge motions are generated by actuating the pitch plunge rods using linear accelerators. The output of the rods is again measured using encoders to provide real-time feedback and a record of the actual position of the model.

The autopilot is then separately subjected to oscillatory roll and pitch motions.
\fi

The roll motion has an amplitude of $5\pi/6$ and a period of 4s. The pitch motion has the same period, and an amplitude of $4\pi/9$. The encoder on the MPS provides the true angles at 1kHz, while the attitude estimator on the MPU9250 provides estimates at 90Hz. The estimated roll and pitch angles are plotted along with the true values in figure \ref{fig:mpsroll}. The residual errors in estimating the roll and pitch angles can be attributed to experimental errors. Also shown in the zoomed insets is the high-accuracy, zero latency tracking from the vector measurements to the attitude estimation. This may be compared with the larger errors using the ECF. As shown in Remark \ref{thm:attYvsattM}, the ECF is an approximation of the exact geometric estimation that is associated with latency on account of a feedback based correction mechanism. In this experiment, the ECF was used with a gain $k_P$ equal to 1, as suggested in \cite{Mahony:08a}. Using lower vales for $k_P$ introduces greater latency for a gradual improvement in the asymptotic accuracy.

%\FloatBarrier

\section {Conclusion} \label{sec:end}

We have reported a geometry-based analytic solution for the problem of attitude estimation using two reference vector measurements, and using a rate measurement and a measurement of a single reference vector. The estimated attitude is analytically derived, so that the need to tune gains does not arise. The estimate also has no latency and is available at the same timestep when the measurement is available. The estimator is verified using Matlab simulations and also by experiments for accuracy and responsiveness.

The presented approach also leads to a unified framework to derive, as special cases, the most significant among previously reported solutions: namely, the TRIAD solution \cite{Black:64a}, Wahba's formulation \cite{Wahba:65a}, the extended Kalman filter \cite{Lefferts:82a}, and the ECF \cite{Mahony:08a}. These four works represent the four most common approaches for attitude estimation: the former two for estimation using vector observations, the EKF for estimation using a linearized complementary filter, the ECF for estimation using a nonlinear complementary filter. Beyond the optimality metrics of these formulations, the proposed solution can also handle nonlinear and non-holonomic optimization.

%A concluding remark is that the presented approach can be extended in principle to problems involving more than two vector measurements. For instance, with three measurements, the problem reduces to one of determining the centroid with respect to three feasibility cones. On account of the nonlinear nature of the problem, it is not easy to obtain closed-form solutions to such problems by extending the presented solutions. Notwithstanding the algebraic difficulties involved, the conceptual extension to such problems is straightforward.

\endgroup

\ifx\qver\qdetailed
\section {Acknowledgements} \label{sec:ack}

The authors gratefully acknowledge the help of Thomas I Linehan in setting up and performing the MPS experiment that was used to verify the geometric attitude estimator.
\fi

\bibliographystyle{plain}
{\footnotesize
\bibliography{ref3.bib}
}
\end{document}